\documentclass[twocolumn,tighten]{aastex62}
\usepackage{mathtools}
\usepackage{txfonts} 
\usepackage{hyperref}
\usepackage{color}
\usepackage{soul}
\usepackage[utf8]{inputenc}
\usepackage{newunicodechar}
\usepackage[encapsulated]{CJK}
\usepackage{ucs}
\newcommand{\cntext}[1]{\begin{CJK}{UTF8}{gbsn}#1\end{CJK}}

\defcitealias{PaperI}{EHTC~I}
\defcitealias{PaperII}{EHTC~II}
\defcitealias{PaperIII}{EHTC~III}
\defcitealias{PaperIV}{EHTC~IV}
\defcitealias{PaperV}{EHTC~V}
\defcitealias{PaperVI}{EHTC~VI}
\defcitealias{PaperVII}{EHTC~VII}
\defcitealias{PaperVIII}{EHTC~VIII}

\def\lsim{\mathrel{\raise.3ex\hbox{$<$\kern-.75em\lower1ex\hbox{$\sim$}}}}
\def\gsim{\mathrel{\raise.3ex\hbox{$>$\kern-.75em\lower1ex\hbox{$\sim$}}}}

\begin{document}                                              

\title{
The Polarized Image of a Synchrotron Emitting Ring of Gas Orbiting a Black Hole  
}
\shorttitle{Polarized Image of a Ring Orbiting a Black Hole}

\author[0000-0002-1919-2730]{Ramesh Narayan}
\affil{Center for Astrophysics $\vert$ Harvard \& Smithsonian, 60 Garden Street, Cambridge, MA 02138, USA}
\affiliation{Black Hole Initiative at Harvard University, 20 Garden Street, Cambridge, MA 02138, USA}
\author[0000-0002-7179-3816]{Daniel~C.~M.~Palumbo}
\affil{Center for Astrophysics $\vert$ Harvard \& Smithsonian, 60 Garden Street, Cambridge, MA 02138, USA}
\affiliation{Black Hole Initiative at Harvard University, 20 Garden Street, Cambridge, MA 02138, USA}
\author[0000-0002-4120-3029]{Michael~D.~Johnson}
\affil{Center for Astrophysics $\vert$ Harvard \& Smithsonian, 60 Garden Street, Cambridge, MA 02138, USA}
\affiliation{Black Hole Initiative at Harvard University, 20 Garden Street, Cambridge, MA 02138, USA}
\author[0000-0001-8053-4392]{Zachary Gelles}
\affil{Center for Astrophysics $\vert$ Harvard \& Smithsonian, 60 Garden Street, Cambridge, MA 02138, USA}
\affiliation{Black Hole Initiative at Harvard University, 20 Garden Street, Cambridge, MA 02138, USA}
\author{Elizabeth Himwich}
\affil{Center for the Fundamental Laws of Nature, Harvard University, Cambridge, MA 02138, USA}
\affiliation{Black Hole Initiative at Harvard University, 20 Garden Street, Cambridge, MA 02138, USA}
\author{Dominic O.~Chang}
\affil{Center for Astrophysics $\vert$ Harvard \& Smithsonian, 60 Garden Street, Cambridge, MA 02138, USA}
\affiliation{Black Hole Initiative at Harvard University, 20 Garden Street, Cambridge, MA 02138, USA}
\author[0000-0001-5287-0452]{Angelo Ricarte}
\affil{Center for Astrophysics $\vert$ Harvard \& Smithsonian, 60 Garden Street, Cambridge, MA 02138, USA}
\affiliation{Black Hole Initiative at Harvard University, 20 Garden Street, Cambridge, MA 02138, USA}
\author[0000-0003-3903-0373]{Jason Dexter}
\affil{JILA and Department of Astrophysical and Planetary Sciences, University of Colorado, Boulder, CO 80309, USA}
\author[0000-0001-7451-8935]{Charles F. Gammie}
\affil{Department of Physics, University of Illinois at Urbana-Champaign, 1110 West Green Street, Urbana, IL 61801, USA}
\affil{Department of Astronomy, University of Illinois at Urbana-Champaign, 1002 West Green Street, Urbana, IL 61801, USA}
\author[0000-0003-2966-6220]{Andrew A. Chael}
\affil{Princeton Center for Theoretical Science, Jadwin Hall, Princeton University, Princeton, NJ 08544, USA}
\affil{NASA Hubble Fellowship Program, Einstein Fellow}






\collaboration{The Event Horizon Telescope Collaboration}

\author[0000-0002-9475-4254]{Kazunori Akiyama}
\affiliation{Massachusetts Institute of Technology Haystack Observatory, 99 Millstone Road, Westford, MA 01886, USA}
\affiliation{National Astronomical Observatory of Japan, 2-21-1 Osawa, Mitaka, Tokyo 181-8588, Japan}
\affiliation{Black Hole Initiative at Harvard University, 20 Garden Street, Cambridge, MA 02138, USA}

\author[0000-0002-9371-1033]{Antxon Alberdi}
\affiliation{Instituto de Astrof\'{\i}sica de Andaluc\'{\i}a-CSIC, Glorieta de la Astronom\'{\i}a s/n, E-18008 Granada, Spain}

\author{Walter Alef}
\affiliation{Max-Planck-Institut f\"ur Radioastronomie, Auf dem H\"ugel 69, D-53121 Bonn, Germany}

\author[0000-0001-6993-1696]{Juan Carlos Algaba}
\affiliation{Department of Physics, Faculty of Science, University of Malaya, 50603 Kuala Lumpur, Malaysia}

\author[0000-0003-3457-7660]{Richard Anantua}
\affiliation{Black Hole Initiative at Harvard University, 20 Garden Street, Cambridge, MA 02138, USA}
\affiliation{Center for Astrophysics | Harvard \& Smithsonian, 60 Garden Street, Cambridge, MA 02138, USA}
\affiliation{Center for Computational Astrophysics, Flatiron Institute, 162 Fifth Avenue, New York, NY 10010, USA}

\author{Keiichi Asada}
\affiliation{Institute of Astronomy and Astrophysics, Academia Sinica, 11F of Astronomy-Mathematics Building, AS/NTU No. 1, Sec. 4, Roosevelt Rd, Taipei 10617, Taiwan, R.O.C.}

\author[0000-0002-2200-5393]{Rebecca Azulay}
\affiliation{Departament d'Astronomia i Astrof\'{\i}sica, Universitat de Val\`encia, C. Dr. Moliner 50, E-46100 Burjassot, Val\`encia, Spain}
\affiliation{Observatori Astronòmic, Universitat de Val\`encia, C. Catedr\'atico Jos\'e Beltr\'an 2, E-46980 Paterna, Val\`encia, Spain}
\affiliation{Max-Planck-Institut f\"ur Radioastronomie, Auf dem H\"ugel 69, D-53121 Bonn, Germany}

\author[0000-0003-3090-3975]{Anne-Kathrin Baczko}
\affiliation{Max-Planck-Institut f\"ur Radioastronomie, Auf dem H\"ugel 69, D-53121 Bonn, Germany}

\author{David Ball}
\affiliation{Steward Observatory and Department of Astronomy, University of Arizona, 933 N. Cherry Ave., Tucson, AZ 85721, USA}

\author[0000-0003-0476-6647]{Mislav Balokovi\'c}
\affiliation{Black Hole Initiative at Harvard University, 20 Garden Street, Cambridge, MA 02138, USA}
\affiliation{Center for Astrophysics | Harvard \& Smithsonian, 60 Garden Street, Cambridge, MA 02138, USA}

% ...





\begin{abstract} Synchrotron radiation from hot gas near a black hole results in a polarized image. The image polarization is determined by effects including the orientation of the magnetic field in the emitting region, relativistic motion of the gas, strong gravitational lensing by the black hole, and parallel transport in the curved spacetime. We explore these effects using a simple model of an axisymmetric, equatorial accretion disk around a Schwarzschild black hole. By using an approximate expression for the null geodesics derived by \citet{Beloborodov_2002} and conservation of the Walker-Penrose constant, we provide analytic estimates for the image polarization. We test this model using currently favored general relativistic magnetohydrodynamic simulations of M87*, using ring parameters given by the simulations. For a subset of these with modest Faraday effects, we show that the ring model broadly reproduces the polarimetric image morphology. 
Our model also predicts the polarization evolution for compact flaring regions, such as those observed from Sgr~A* with GRAVITY. With suitably chosen parameters, our simple model can reproduce the EVPA pattern and relative polarized intensity in Event Horizon Telescope images of M87*. Under the physically motivated assumption that the magnetic field trails the fluid velocity, this comparison is consistent with the clockwise rotation inferred from total intensity images. 
\end{abstract}


\section{Introduction}\label{sec:intro}

The Event Horizon Telescope (EHT) Collaboration has recently published the first images of a black hole (\citealt{PaperI,PaperII,PaperIII,PaperIV,PaperV,PaperVI,PaperVII,PaperVIII}; hereafter EHTC I-VIII). These images achieve a diffraction-limited angular resolution that corresponds to approximately $5 G M/c^2$, where $M$ is the mass of the black hole. They reveal a bright ring of emission with a twisting polarization pattern and a prominent rotationally symmetric mode. 

The polarization structure in the EHT images depends on details of the emitting plasma, principally the magnetic field geometry. However, it is also affected by the strongly curved spacetime near the black hole. Over the past few decades, simulated polarimetric images of black holes have been studied as a means to understand astrophysical properties of their surrounding accretion flows \citep[e.g.,][]{Bromley_2001,Shcherbakov_2012,Moscibrodzka_2017,Jimenez_2018,PWP_2020} and to infer the disk inclination and black hole spin through the effects of parallel transport  \citep[e.g.,][]{Connors_1980,Broderick_Loeb_2006,Li_et_al_2009,Schnittman_Krolik_2009,Gold_2017,Marin_et_al_2018}. 

While they are becoming increasingly realistic, these simulations are generally difficult to use for broad parameter surveys because of their computational cost, and they often provide little insight into how to decouple astrophysical and relativistic effects. 

In this article, we develop a simple toy model to understand polarimetric images of black holes. This model consists of a ring of magnetized fluid orbiting a Schwarzschild black hole. Our model allows arbitrary emission radius, magnetic field geometry, equatorial fluid velocity, and observer inclination. With a single approximation, described in \autoref{sec:model}, we can analytically compute the resulting polarimetric image and can assess its dependence on the input parameters.  

In \autoref{sec:model}, we describe the toy ring model and work out the relevant relativistic transformations from the frame of a radiating fluid element in the ring to the image as seen on the sky by an observer. In \autoref{sec:examples}, we present a series of examples to illustrate the primary model features. In \autoref{sec:analytic}, we provide analytic estimates of image diagnostics -- the apparent shape of the ring, the vector polarization, and the coefficient of rotational symmetry \citep[$\beta_2$;][]{PWP_2020}. In \autoref{sec:observation_comparisons}, we discuss the suitability of our model for comparisons with observations, focusing on the EHT images of M87* and polarization ``loops'' seen during flares of Sagittarius A* (Sgr~A*). In \autoref{sec:summary}, we summarize our results.

\section{The Model}\label{sec:model}

We consider an accretion disk around a Schwarzschild black
hole of mass $M$. We use standard geometrized units: $G=c=1$. The fluid radiates from the equatorial plane within a narrow range of radii centered on a dimensionless radius $R$, measured in units of $M$ (or $GM/c^2$, including the physical constants). With
respect to a distant observer, the ring is tilted from a face-on
orientation by an angle $\theta_{\rm o}$. We assume that the tilt is towards
the North, so that the line-of-nodes between the ring orbital plane
and the observer's sky plane is in the East-West direction. We take
the sky angular coordinate $x$ to be oriented towards the West
(i.e., to the right), and the coordinate $y$ towards the North
(i.e., towards the top). The fluid has radial and tangential components of velocity in the plane of the ring, but no vertical velocity. In the comoving frame of the fluid, the magnetic field has radial, azimuthal and vertical components. For simplicity, we assume that both the velocity field and the magnetic field are axisymmetric, though the equations developed in this section are valid even without this assumption.

We wish to compute the following primary observables: (1) the shape of the ring as viewed by the distant observer, (2) the variation of
the polarized intensity around the observed ring, and (3) the orientation and
pattern of the polarization vectors around the ring.
An exact calculation requires integrating the geodesic equation, which
has to be done numerically. However, with one simplification,
described below, it is possible to do all the calculations
analytically. This simplified model provides a convenient method for
investigating polarization properties of idealized models.

\subsection{Geometry, Lensing and Special Relativity}

\begin{figure}[t]
\includegraphics[width=\columnwidth]{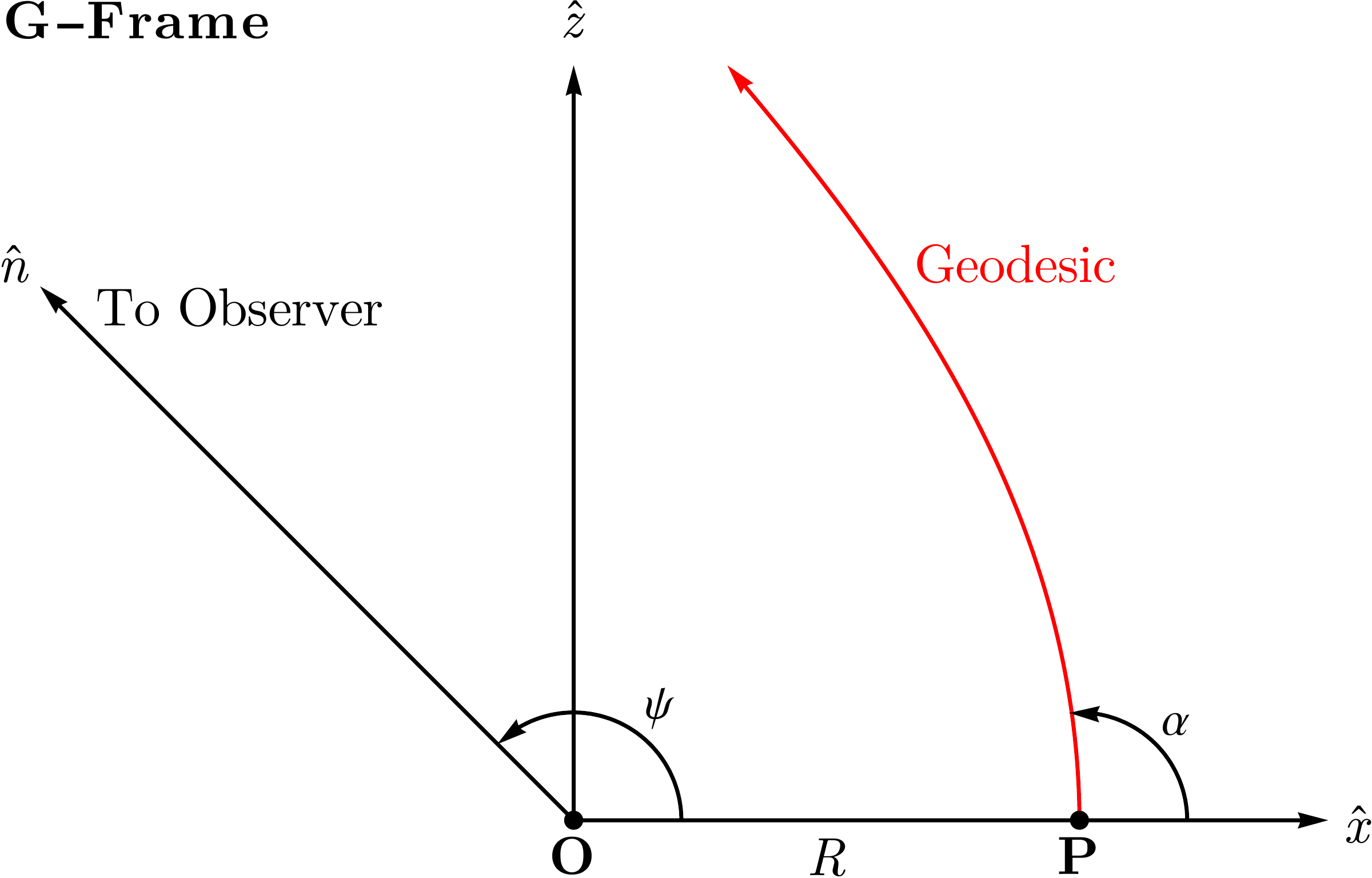}
\caption{Geometry in the geodesic frame, or G-frame. In the Schwarzschild metric, each null geodesic is confined to a plane that intersects the black hole. The G-frame, defined for photons emitted at point P and reaching a distant observer at relative angle $\psi$, corresponds to Cartesian axes centered on the black hole, with $\hat{x}$ in the direction of P and the $\hat{x}$-$\hat{z}$~plane given by the geodesic plane. We approximate the emission angle $\alpha$ in this frame using \autoref{belo}.
}
\label{fig:gframe}
\end{figure}

In the ring plane, we consider a fluid element P located at azimuthal
angle $\phi$ measured from the line-of-nodes. We are interested in a
null geodesic, a light ray, that travels from P to the observer.  This geodesic lies
in a plane that includes the line from the black hole O to the point P,
as well as the line from O to the observer (see Fig.~\ref{fig:gframe}). We set up
Cartesian coordinates in the geodesic plane so that the unit vector along the $x$-axis
$\hat{x}$ is oriented along OP and the observer lies on the
$\hat{x}$-$\hat{z}$~plane. We call this the geodesic frame, or G-frame. The
angle $\psi$ between $\hat{x}$ and the unit vector $\hat{n}$ towards the
observer satisfies
\begin{align}
\label{eq:psi}
\cos\psi &= -\sin\theta_{\rm o}\,\sin\phi,
\nonumber\\ \sin\psi &= (1-\cos^2\psi)^{1/2}.
\end{align}

We consider a null geodesic with conserved energy\footnote{This is the photon energy measured by an observer at infinity, and we normalize it to unity.} $k_t=-1$ traveling
from P to the observer.  At the location P, the orthonormal time
component $k_{({\rm G})}^{\hat{t}}$ of its 4-wavevector is given by (the redshift factor here is calculated using the Schwarzschild metric, as appropriate for the assumed non-spinning black hole)
\begin{equation}
k_{({\rm G})}^{\hat{t}} = -\frac{k_t}{\sqrt{-g_{tt}}} =
\frac{1}{\left(1-\frac{2}{R}\right)^{1/2}},
\end{equation}
where the subscript `$(G)$' indicates that this quantity is
measured in the G-frame. Also, since the geodesic lies in the
$xz$-plane, we have $k_{(G)}^{\hat{y}}=0$. To determine the other two
components of $k$, we need the angle $\alpha$ in Fig.~\ref{fig:gframe}, in terms of which we can write
\begin{equation}
k_{({\rm G})}^{\hat{x}} = k_{({\rm G})}^{\hat{t}}\cos\alpha, \qquad
k_{({\rm G})}^{\hat{z}}=k_{({\rm G})}^{\hat{t}}\sin\alpha.
\end{equation}
Instead of attempting to calculate $\alpha$  precisely, which would require
a numerical integration of the geodesic equation, we use the
following approximate formula obtained by \citet{Beloborodov_2002},
\begin{align}
\label{belo}
\cos\alpha &= \cos\psi + \frac{2}{R}\,(1-\cos\psi),
\nonumber\\ \sin\alpha &= (1-\cos^2\alpha)^{1/2}.
\end{align}
This approximation is surprisingly accurate even for values of $R$ of
order a few (see sec.~\ref{sec:numerical} and \autoref{sec:errorAnalysis}).

\begin{figure}[t]
\includegraphics[width=\columnwidth]{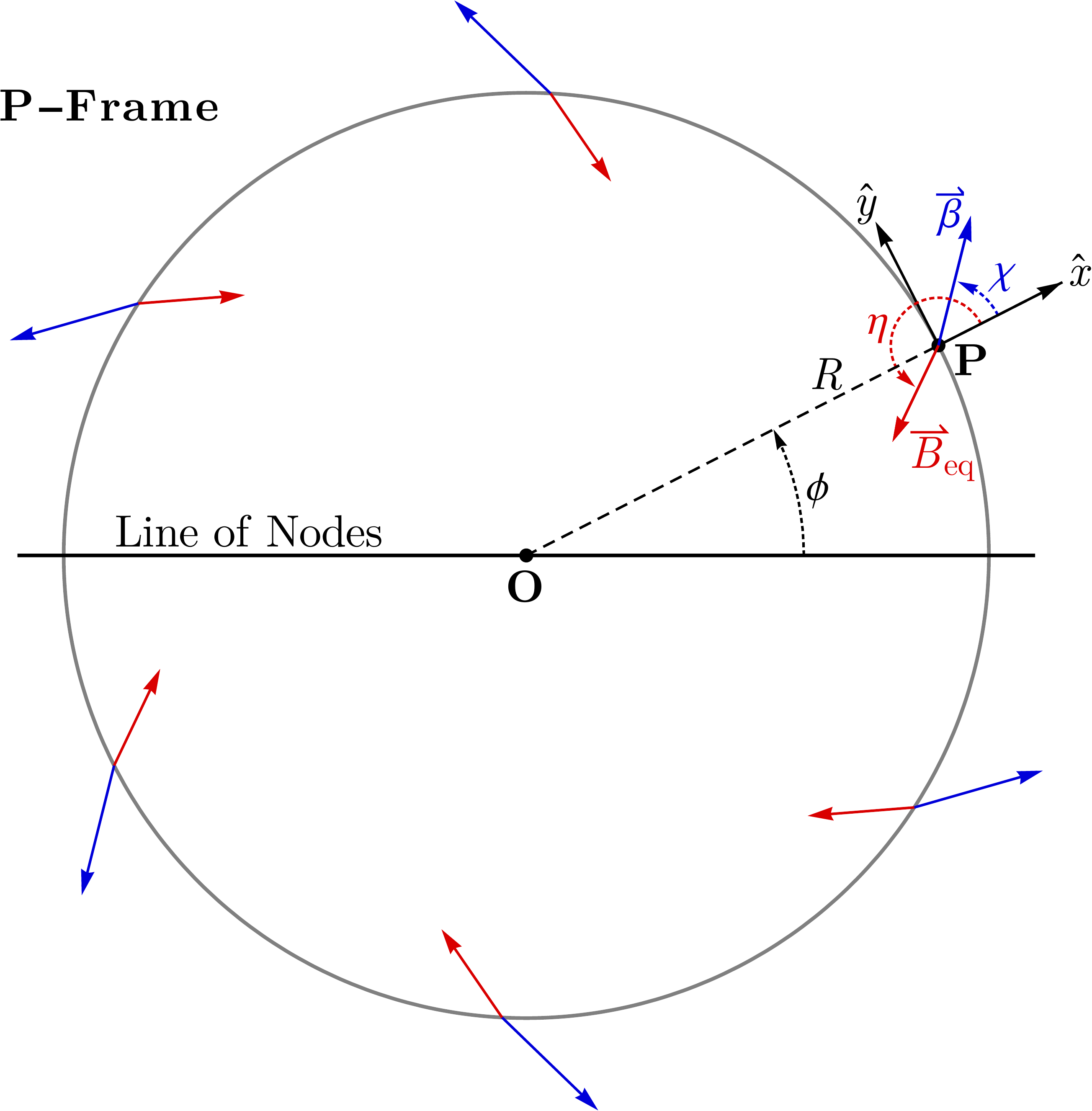}
\caption{Geometry in the P-frame. This frame is aligned with the rotating gas at emission radius $R$ and emission azimuth $\phi$. The $\hat{x}$ direction lies along the radial line from the black hole at O to the emission point P, and $\hat{y}$ is the azimuthal direction. The equatorial magnetic field $\vec{B}_{\rm eq}$ and fluid velocity $\vec{\beta}$ lie at angles $\eta$ and $\chi$ to $\hat{x}$  in the $x$-$y$ plane, respectively. Our model allows these angles to be specified independently, but we will later focus on the physically motivated choices of $\eta = \chi$ and $\eta = \chi+\pi$ (see \autoref{sec:examples}).}
\label{fig:pframe}
\end{figure}

We now switch to a Cartesian frame that is aligned with the orbiting
fluid ring. We take $\hat{x}$ along OP, $\hat{y}$ in the azimuthal
direction at P parallel to $\hat{\phi}$, and $\hat{z}$ perpendicular to the orbital plane. We
call this the P-frame (see Fig.~\ref{fig:pframe}). The G-frame and P-frame have a
common $\hat{x}$-axis. Therefore, transforming from one to the other
involves rotation by some angle $\xi$ around the $x$-axis. To
determine $\xi$, we note that the unit vector
$\hat{n}$ from the black hole O towards the observer has Cartesian
components $(\cos\psi, ~0, ~\sin\psi)$ in the G-frame, and Cartesian
components $(-\sin\theta_{\rm o} \sin\phi, ~-\sin\theta_{\rm o}
\cos\phi, ~\cos\theta_{\rm o})$ in the P-frame. Since a rotation by
angle $\xi$ transforms one set of components to the other, we obtain
\begin{equation}
\cos\xi = \frac{\cos\theta_{\rm o}}{\sin\psi}, \qquad
\sin\xi = \frac{\sin\theta_{\rm o}\,\cos\phi}{\sin\psi}.
\end{equation}
Applying this rotation to the orthonormal components of $k_{\rm (G)}$,
we obtain the corresponding orthonormal components in the P-frame,
\begin{alignat}{3}
k_{({\rm P})}^{\hat{t}} &= \frac{1}{\left(1-\frac{2}{R}\right)^{1/2}},\qquad &
k_{({\rm P})}^{\hat{x}} &= \frac{\cos\alpha}{\left(1-\frac{2}{R}\right)^{1/2}},\\
k_{({\rm P})}^{\hat{y}} &=-\frac{\sin\xi\,\sin\alpha}{\left(1-\frac{2}{R}\right)^{1/2}},\qquad &
k_{({\rm P})}^{\hat{z}} &=\frac{\cos\xi\,\sin\alpha}{\left(1-\frac{2}{R}\right)^{1/2}}.
\end{alignat}

The fluid at the point P moves in the $xy$-plane of the local P-frame with a velocity $\vec{\beta}$, 
which we write in the local Cartesian coordinate frame as (see Fig.~\ref{fig:pframe})
\begin{equation}
    \vec{\beta} = \beta \left(\cos\chi\,\hat{x} + \sin\chi\,\hat{y}\right).
    \label{velocity2}
\end{equation}
Our sign convention is that radial motion towards the black hole corresponds to $\cos\chi<0$, and clockwise rotation on the sky corresponds to $\sin\chi<0$. In the case of M87*, the rotation
is clockwise. The velocity $\vec{\beta}$ describes motion of the fluid through the ring; the ring model itself is not expanding or contracting.

We now transform to the fluid frame --- the F-frame --- by applying a
Lorentz boost with velocity $\vec{\beta}$. This gives the following orthonormal components of $k$,
\begin{eqnarray}
k_{({\rm F})}^{\hat{t}} &=& \gamma\, k_{({\rm P})}^{\hat{t}} -\gamma\beta\cos\chi\, k_{({\rm P})}^{\hat{x}}-\gamma\beta\sin\chi\, k_{({\rm P})}^{\hat{y}}, \nonumber \\
k_{({\rm F})}^{\hat{x}} &=& -\gamma\beta\cos\chi\, k_{({\rm P})}^{\hat{t}} +(1+(\gamma-1)\cos^2\chi)\, k_{({\rm P})}^{\hat{x}} \nonumber \\
&~& \qquad\qquad\qquad +(\gamma-1)\cos\chi\sin\chi\, k_{({\rm P})}^{\hat{y}}, \nonumber \\
k_{({\rm F})}^{\hat{y}} &=& -\gamma\beta\sin\chi\, k_{({\rm P})}^{\hat{t}} +(\gamma-1)\cos\chi\sin\chi\, k_{({\rm P})}^{\hat{x}} \nonumber \\
&~& \qquad\qquad\qquad +(1+(\gamma-1)\sin^2\chi)\, k_{({\rm P})}^{\hat{y}}, \nonumber \\
k_{({\rm F})}^{\hat{z}} &=& k_{({\rm P})}^{\hat{z}}.
\end{eqnarray}

\subsection{Transformation of Polarized Intensity}

Any radiation emitted along $k_{\rm(F)}^{\hat{\mu}}$ in the F-frame is
Doppler-shifted by the time it reaches the observer. Since $k_{\rm
  (O)}^{\hat{t}}$ in the observer frame is equal to unity, the Doppler
factor $\delta$ is
\begin{equation}
\delta = \frac{k_{\rm (O)}^{\hat{t}}}{k_{\rm (F)}^{\hat{t}}} =
\frac{1}{k_{\rm (F)}^{\hat{t}}}.
\end{equation}
This includes both gravitational redshift and Doppler shift from velocity.

In the fluid frame, there is a magnetic field which we write as\footnote{Because the emission of synchrotron radiation is best described in the fluid frame, we find it convenient to specify the magnetic field components in this frame. The $\hat{x}$, $\hat{y}$, $\hat{z}$ axes in the fluid frame are related to the corresponding axes in the P-frame (equivalently, the Schwarzschild frame, e.g., eq \ref{Schwarzschild}), via a Lorentz transformation with velocity $\vec{\beta}$. The transformation of field components between the two frames is worked out in Appendix~\ref{sec:transformations}.}
\begin{eqnarray}
    \vec{B} &=& B_r\hat{x} + B_\phi \hat{y} + B_z \hat{z} \nonumber \\
    &=& B_{\rm eq}\,(\cos\eta\,\hat{x} +  \sin\eta\,\hat{y}) + B_z\, \hat{z} \\
    &\equiv& \vec{B}_{\rm eq} + B_z\hat{z}, \nonumber
    \label{field_comps}
\end{eqnarray}
where the second line describes the field components in the equatorial plane in terms of a magnitude $B_{\rm eq}$ and an orientation $\eta$ (see Fig.~\ref{fig:pframe}).
The intensity of synchrotron radiation emitted along the 3-vector $\vec{k}_{\rm (F)}$ depends on $\sin\zeta$,
where $\zeta$ is the angle between $\vec{k}_{\rm (F)}$ and the magnetic field $\vec{B}$:
\begin{equation}
\sin\zeta = \frac{|\vec{k}_{\rm (F)} \times \vec{B}|}{|\vec{k}_{\rm
    (F)}|\,\,|\vec{B}|}.
\end{equation}

In the case of thermal synchrotron emission, the intensity also depends on the ratio of the emitted photon energy $h\nu$ to the electron temperature $kT_e$. At low frequencies $h\nu \ll kT_e$, the intensity is proportional to $\sin^{2/3}\zeta$ \citep[e.g.,][]{Mahadevan_1996}, whereas in the opposite limit $h\nu \gg kT_e$, the intensity varies as a very large positive power of $\sin\zeta$, because of the exponential cutoff of the particle energy distribution and the corresponding rapid decline of emissivity with increasing frequency. In general, if the emitted intensity varies as $I_\nu \sim \nu^{-\alpha_\nu}$, then the angle dependence goes as $(\sin\zeta)^{1+\alpha_\nu}$. In models of M87*, a dependence $\sim\sin^2\zeta$ is often obtained at 230\,GHz. 
This corresponds to $\alpha_\nu\sim1$, which is consistent with the synchrotron emission being close to its peak at this frequency ($\nu F_\nu$ roughly constant). In the analysis below, we explicitly retain the $\alpha_\nu$ dependence. However, we set $\alpha_\nu=1$ for the numerical calculations described in sec.~\ref{sec:examples}, and also when we series-expand the equations in Appendix~\ref{sec:series}.

The factor $(\sin\zeta)^{1+\alpha_\nu}$ discussed in the previous paragraph is the emission per unit volume. To convert this to the emerging intensity in the fluid frame we need to multiply by the geodesic path length $l_{\rm p}$ through the emitting region. We assume that the medium is optically thin to its own emission. If we model the emitting fluid as a thin disk of vertical thickness $H$, then the path length is
\begin{equation}
l_{\rm p} = \frac{k_{\rm (F)}^{\hat{t}}} {k_{\rm (F)}^{\hat{z}}}\,H. \label{eq:lp}
\end{equation}

So far, we have discussed the emitted intensity in the fluid frame. This intensity is Doppler-boosted
by a factor of $\delta^{3+\alpha_\nu}$ by the time it reaches the
observer.\footnote{In the context of a continuous relativistic jet, a Doppler boost factor of $\delta^{2+\alpha_\nu}$ is generally used \citep[e.g.,][]{Blandford_Konigl}. That corresponds to the combined quantity $l_p\delta^{3+\alpha_\nu}$, where for motion parallel to the jet axis, $l_p\propto \delta^{-1}$. Our formulation, with $l_p$ handled as a separate factor, is more general.}  Thus, the intensity
$|P|$ of linearly polarized synchrotron radiation that reaches the
observer from the location P is
\begin{eqnarray}
|P| &=& \delta^{3+\alpha_\nu}\,l_{\rm p}\,|\vec{B}|^{1+\alpha_\nu}\,\sin^{1+\alpha_\nu}\zeta  \label{absP0}\\ &\to& \delta^4\,l_{\rm p}\,|\vec{B}|^2\sin^2\zeta~~{\rm for~\alpha_\nu=1},
\label{absP}
\end{eqnarray} 
where we have omitted a proportionality constant. Since  $|\vec{B}|$ is  constant around the ring, the factors involving $|\vec{B}|$ could be eliminated from Equations \ref{absP0} and \ref{absP} and absorbed into the omitted proportionality constant. We retain these factors because keeping track of $|\vec{B}|^2$ and its components is convenient for much of the analysis in \autoref{sec:series}.\footnote{Alternatively, we could assume $|\vec{B}|=1$, as indeed we do in all the plots, eliminate $|\vec{B}|$ from Equations \ref{absP0} and \ref{absP}, but still keep track of the components of $\vec{B}$ in \autoref{sec:series}.}

\subsection{Transformation of Polarization Vector}

We next work on the polarization vector. In the
fluid frame, the $\vec{E}$-vector of the radiation is oriented along
$\vec{k}_{\rm (F)} \times \vec{B}$, i.e., perpendicular to both $\vec{k}_{\rm(F)}$ and $\vec{B}$. Therefore, we write the
orthonormal components of the polarization 4-vector $f^\mu$ as
\begin{eqnarray}
    f_{\rm (F)}^{\hat{t}} &=& 0,
    \quad\qquad\qquad f_{\rm (F)}^{\hat{x}} = \frac{\left(\vec{k}_{\rm (F)}\times\vec{B}\right)_{\hat{x}}}{|\vec{k}_{\rm (F)}|}, \nonumber \\
    f_{\rm (F)}^{\hat{y}} &=& \frac{\left(\vec{k}_{\rm (F)}\times\vec{B}\right)_{\hat{y}}}{|\vec{k}_{\rm (F)}|},
    \quad f_{\rm (F)}^{\hat{z}} = \frac{\left(\vec{k}_{\rm (F)}\times\vec{B}\right)_{\hat{z}}}{|\vec{k}_{\rm (F)}|}. \label{eq:fmuF}
\end{eqnarray}
By construction, this 4-vector satisfies
\begin{equation}
f^\mu k_\mu=0, \qquad f^\mu
f_\mu= \sin^2\zeta\,|\vec{B}|^2.
\label{fnorm}
\end{equation} 
An inverse Lorentz boost transforms the 4-vector $f_{\rm (F)}^{\hat\mu}$ back to
the P-frame:
\begin{eqnarray}
f_{({\rm P})}^{\hat{t}} &=& \gamma\, f_{({\rm F})}^{\hat{t}} +\gamma\beta\cos\chi\, f_{({\rm F})}^{\hat{x}}+\gamma\beta\sin\chi\, f_{({\rm F})}^{\hat{y}}, \nonumber \\
f_{({\rm P})}^{\hat{x}} &=& \gamma\beta\cos\chi\, f_{({\rm F})}^{\hat{t}} +(1+(\gamma-1)\cos^2\chi)\, f_{({\rm F})}^{\hat{x}} \nonumber \\
&~& \qquad\qquad\qquad +(\gamma-1)\cos\chi\sin\chi\, f_{({\rm F})}^{\hat{y}}, \nonumber \\
f_{({\rm P})}^{\hat{y}} &=& \gamma\beta\sin\chi\, f_{({\rm F})}^{\hat{t}} +(\gamma-1)\cos\chi\sin\chi\, f_{({\rm F})}^{\hat{x}} \nonumber \\
&~& \qquad\qquad\qquad +(1+(\gamma-1)\sin^2\chi)\, f_{({\rm F})}^{\hat{y}}, \nonumber \\
f_{({\rm P})}^{\hat{z}} &=& f_{({\rm F})}^{\hat{z}}.
\end{eqnarray}

Since the Cartesian unit vectors $\hat{x}, ~\hat{y}, ~\hat{z}$ in the
P-frame are oriented along the spherical polar unit vectors $\hat{r},
~\hat{\phi}, ~-\hat{\theta_{\rm o}}$ of the Schwarzschild frame, the orthonormal
components of $k$ and $f$ in Schwarzschild coordinates are
\begin{equation}
k^{\hat{t}}=k^{\hat{t}}_{\rm (P)}, \quad 
k^{\hat{r}}=k^{\hat{x}}_{\rm (P)}, \quad
k^{\hat{\theta}}=-k^{\hat{z}}_{\rm (P)}, \quad
k^{\hat{\phi}}=k^{\hat{y}}_{\rm (P)},
\label{Schwarzschild}
\end{equation}
\begin{equation}
f^{\hat{t}}=f^{\hat{t}}_{\rm (P)}, \quad 
f^{\hat{r}}=f^{\hat{x}}_{\rm (P)}, \quad
f^{\hat{\theta}}=-f^{\hat{z}}_{\rm (P)}, \quad
f^{\hat{\phi}}=f^{\hat{y}}_{\rm (P)}.
\end{equation}
The photon geodesic emitted at P has three conserved quantities (see for instance \citealt{Bardeen1973b}): its
energy $k_t=-1$, its angular momentum around the $\hat{z}$ axis $k_\phi = Rk^{\hat{\phi}}$, and
the \cite{Carter1968} constant $C$, which is the square of the total angular momentum of the photon for the Schwarzschild metric. In the P-frame  the Carter constant is
\begin{equation}
C = R^2 \left[\left(k^{\hat{\theta}}\right)^2 +
  \left(k^{\hat{\phi}}\right)^2\right].
\end{equation}
Using the conservation of $k_\phi$ and $C$, we find the
coordinates $x$ and $y$ of the geodesic at the observer sky plane (recall the orientation of the sky coordinates $x,y$ described at the top of section \ref{sec:model}) \citep{Bardeen1973b},
\begin{align}
\label{alphabeta}
x &= -\frac{k_\phi}{\sin\theta_{\rm o}}=-\frac{Rk^{\hat\phi}}{\sin\theta_{\rm o}},
\nonumber\\ y  &= k_\theta = R\,
\left[\left(k^{\hat{\theta}}\right)^2 -
  \cot^2\theta_{\rm o}\,\left(k^{\hat{\phi}}\right)^2\right]^{1/2} {\rm
  sgn}(\sin\phi).
\end{align}

To compute the polarization vector at the observer, we make use of the
Walker-Penrose constant $K_1+iK_2$ \citep{Walker_Penrose_1970}, which takes a simple form for
a Schwarzschild spacetime. At the position P, we have (using the sign convention in \citealt{Himwich_2020}),
\begin{equation}
K_1 = R(k^tf^r-k^rf^t), \qquad K_2 = -R^3 (k^\phi f^\theta - k^\theta f^\phi).
\end{equation}
Both $K_1$ and $K_2$ are conserved along the
geodesic. Therefore, knowing their values, we can evaluate the two
transverse components of the polarization electric field $\vec{E}$ at the
observer. If we use the normalization used in \citet{Himwich_2020}, the field components are
\begin{align}
    \label{Enorm}
E_{x,\rm norm} &= \frac{y K_2 + x K_1} {[(K_1^2+K_2^2)\,
    (x^2+y^2)]^{1/2}},
\nonumber\\ E_{y,\rm norm} &= \frac{y K_1 - x K_2} {[(K_1^2+K_2^2)\,
    (x^2+y^2)]^{1/2}},
\nonumber\\ E_{x,\rm norm}^2 + E_{y,\rm norm}^2 &= 1,
\end{align}
which is normalized to unity. This normalization is suitable for plotting the orientation of polarization vectors in the $xy$-plane. 

An alternative normalization is
\begin{align}
E_{x} &= \frac{y K_2 + x K_1} {
    x^2+y^2},
\nonumber\\ E_{y} &= \frac{y K_1 - x K_2} {
    x^2+y^2},
\nonumber\\ E_{x}^2 + E_{y}^2 &= \sin^2\zeta\,|\vec{B}|^2.
    \label{Enormzeta}
\end{align}
This retains the original normalization of $f^\mu$ in the fluid frame (eq \ref{fnorm}), hence the electric field is proportional to $\sin\zeta\,|\vec{B}|$.

For computing the observed polarized intensity, we need to include the dependence on the Doppler factor $\delta$ and path length $l_{\rm p}$, and must also ensure the correct powers of $\sin\zeta$ and $|\vec{B}|$ as given in Equations \ref{absP0} and \ref{absP}. Since the intensity is proportional to $|\vec{E}|^2$, we therefore write the observed electric field components as
\begin{align}
E_{x,\rm obs} &= \delta^{(3+\alpha_\nu)/2}\,l_{\rm p}^{1/2}\,(\sin\zeta)^{(1+\alpha_\nu)/2}\,|\vec{B}|^{(1+\alpha_\nu)/2}\,E_{x,\rm norm}
\nonumber\\ &= \delta^{(3+\alpha_\nu)/2}\,l_{\rm p}^{1/2}\,(\sin\zeta)^{(\alpha_\nu -1)/2}|\vec{B}|^{(\alpha_\nu-1)/2}\,E_{x}, \label{ealphaobs}\\ 
\nonumber E_{y,\rm obs} &= \delta^{(3+\alpha_\nu)/2}\,l_{\rm p}^{1/2}\,(\sin\zeta)^{(1+\alpha_\nu)/2}\,|\vec{B}|^{(1+\alpha_\nu)/2}\,E_{y,\rm norm} 
\nonumber\\ &= \delta^{(3+\alpha_\nu)/2}\,l_{\rm p}^{1/2}\,(\sin\zeta)^{(\alpha_\nu -1)/2}|\vec{B}|^{(\alpha_\nu-1)/2}\,E_{y}, \label{ebetaobs}\\
\nonumber E_{x,\rm obs}^2 + E_{y,\rm obs}^2 &= |P(\phi)|,
\end{align}
where $P(\phi)$ is the observed linear polarized intensity of radiation that is originally emitted by a fluid element at ring azimuthal angle $\phi$.

We need one more transformation:
we must convert the coordinates $(R,\phi)$ of the
emitting region in the fluid to the Cartesian sky coordinates $(x,y)$, or equivalently the polar sky coordinates $(\rho,\varphi)$,  at which the radiation is observed,
\begin{equation}
x = \rho\cos\varphi, \quad y = \rho\sin\varphi.
\label{rho_varphi}
\end{equation}
The relation between $(R,\phi)$ and $(\rho,\varphi)$ is worked out in Appendix~\ref{sec:mapping}. The observed linear polarization $P(\phi)$ can then be described in image coordinates by the complex function $P(\varphi)$, 
\begin{align}
    P(\varphi) \equiv Q(\varphi) + i U(\varphi), \label{eq:PQU}
\end{align}
where the Stokes parameters $Q(\varphi)$ and $U(\varphi)$ are obtained from the electric field components $E_{x,\rm obs}$, $E_{y,\rm obs}$ using \autoref{eq:QU_Ealphabeta}. The electric vector position angle (or EVPA) is then
\begin{align}
    {\rm EVPA} \equiv \frac{1}{2}\arctan{\frac{U}{Q}}.
\end{align}

This completes the calculation of the intensities $Q$, $U$, $P$ on the image plane. If one wishes to calculate fluxes in the sky plane corresponding to specific source configurations in ring coordinates $(R,\phi)$, it would be necessary to apply the Jacobian of the transformation from $(R,\phi)$ to $(\rho,\varphi)$, \added{as in \autoref{fig:gravity_loops}}. The Jacobian determinant is evaluated in Appendix~\ref{sec:mapping}. \deleted{, but we do not use it in this paper}.

To summarize, in this section we showed how, given the position $(R,\phi$, Fig.~\ref{fig:pframe}) and velocity $(\beta,\chi$, eq.~\ref{velocity2}) of a synchrotron-emitting fluid element located on a tilted equatorial plane around a Schwarzschild black hole, and given also the magnetic field configuration $(B_{\rm eq},\eta,B_z$, eq.~\ref{field_comps}) in the frame of the fluid, one can calculate the sky coordinates $(x,y$, equivalently $\rho,\varphi$) of the image of this radiating element, and the linearly polarized intensity and position angle of the observed radiation. The mapping from the radiating element to the observer's image plane is written as a sequence of analytical calculations that do not require numerically integrating the geodesic equation or iteratively solving any equation. The equations are written in sufficient detail for easy incorporation into modeling calculations.

\section{Example Models}\label{sec:examples}

\begin{figure*}[t]
\includegraphics[width=9cm]{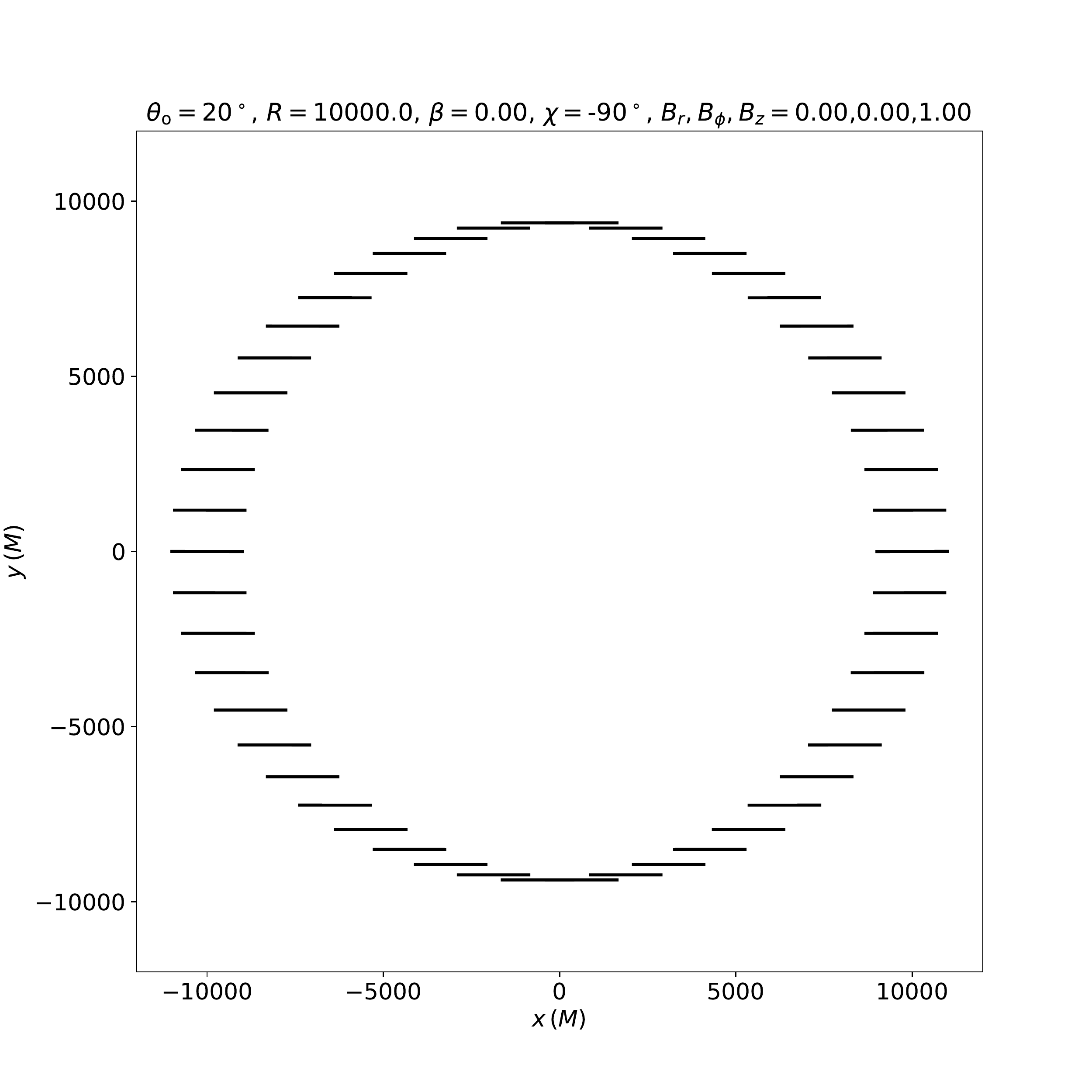}
\includegraphics[width=9cm]{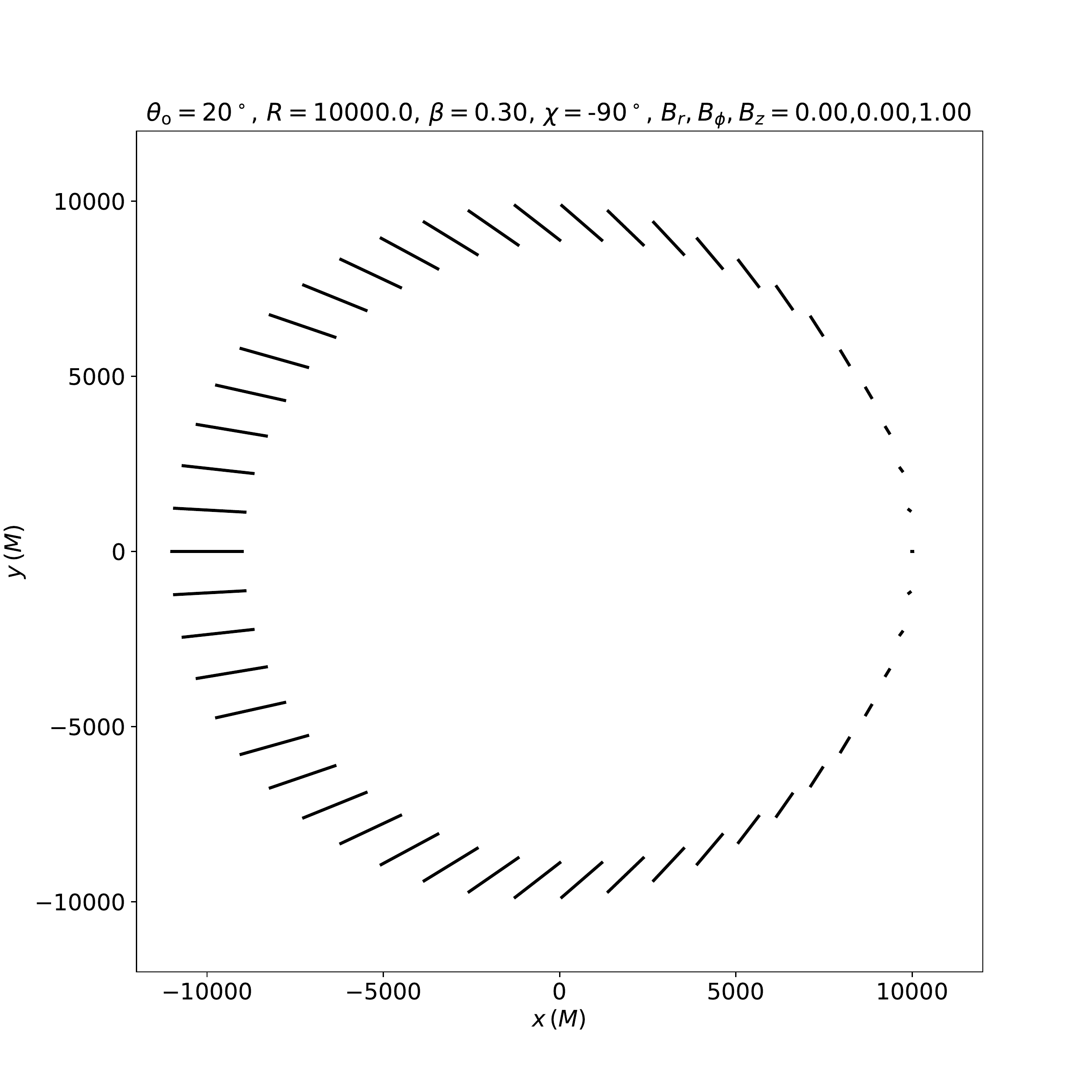}
\includegraphics[width=9cm]{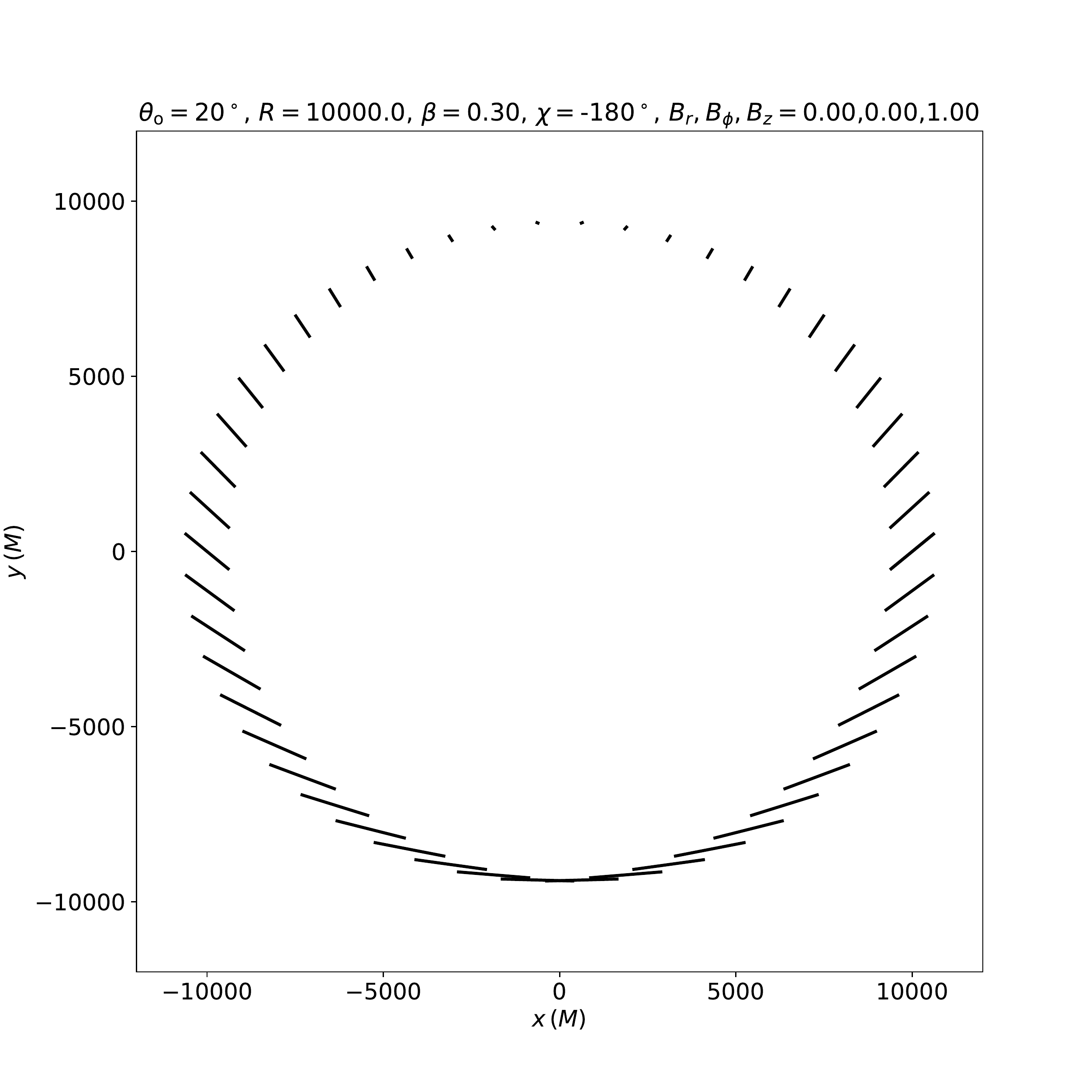}
\includegraphics[width=9cm]{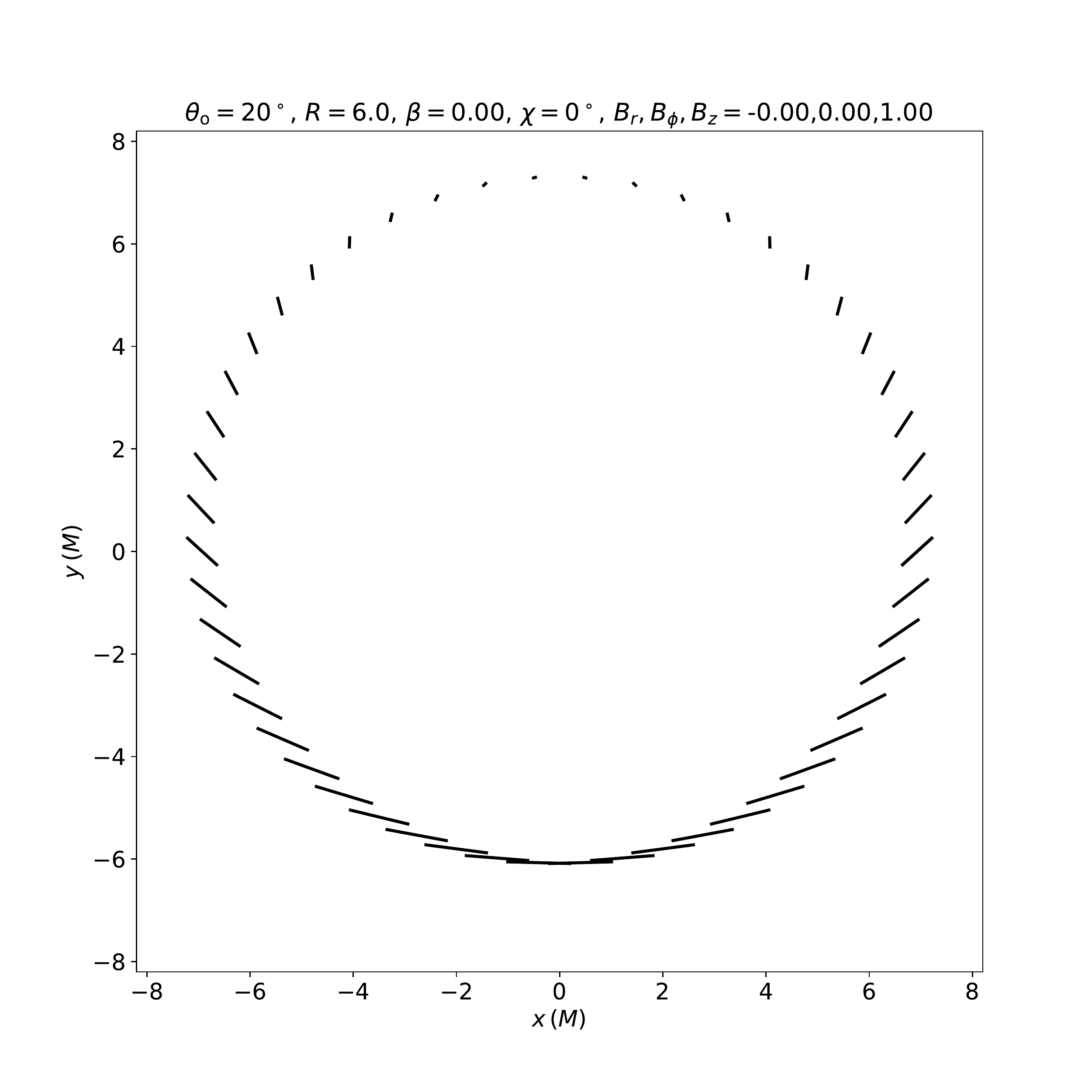}
\caption{Polarization patterns corresponding to models with a
  ``vertical'' magnetic field (non-zero $B_z$ in the fluid frame).  In
  each case, the directions of the ticks indicate the orientation of
  the polarization $\vec{E}$-vector around the ring as viewed on the
  sky. The lengths of the ticks are proportional to the polarized
  intensity. Top Left: Ring with a very large radius and no orbital
  velocity, so that neither velocity aberration nor lensing plays a role. Top
  Right: Large ring radius (i.e., no lensing), and fluid
  orbiting with a tangential velocity $\beta=0.3$ in the clockwise direction
  ($\chi=-90^\circ$). Bottom Left: Large ring radius (no lensing), and fluid flowing
  with velocity $\beta=0.3$ radially inward 
  ($\chi=-180^\circ$). Bottom Right: Ring
  with a small radius $R=6M$, hence strong gravitational lensing, but
  with no fluid velocity, hence no aberration.}
  \label{fig:Bz}
\end{figure*}

The simple model considered in the previous section has the following parameters: tilt angle
of the ring $\theta_{\rm o}$, ring radius $R$, velocity vector of the fluid $\vec{\beta}$, which is parameterized by $\beta=v/c$ and $\chi$ (eq~\ref{velocity2}), fluid frame magnetic
field $\vec{B}$, which is parameterized by either $B_r$, $B_\phi$, $B_z$, or $B_{\rm eq}$, $\eta$, $B_z$ (eq~\ref{field_comps}), and spectral index $\alpha_\nu$. Figures \ref{fig:Bz}--\ref{fig:Beq2} show the
polarization patterns produced by this model for selected values of
the parameters. In all these examples, we choose $\theta_{\rm o}=20^\circ$ and $\alpha_\nu=1$.

Before considering the examples, we briefly summarize the salient features of the polarized image of M87* obtained by the EHT \citepalias{PaperVII}. First, the linear polarized flux shows a pronounced asymmetry around the ring. The polarized flux is strong between PA (measured East of North) $\sim150^\circ$ and $\sim300^\circ$; the peak polarized intensity is around PA $200^\circ$ on April~5 and $240^\circ$ on April~11. The linear polarized flux is much weaker at other angles. The large scale jet in M87* is oriented towards PA $288^\circ$. Presumably, the accretion disk is also tilted toward this direction. Such a tilt is consistent with the EHT total intensity image shown in \citetalias{PaperIV}. Thus, if we measure angles counter-clockwise with respect to the presumed tilt direction in M87*, the polarized flux is strong between angles $\sim+10^\circ$ and $-140^\circ$, with peak at $-90^\circ$ and $-50^\circ$ on April~5 and April~11.

In our analytic model, the tilt and putative jet are toward the North. Thus, for a direct comparison of this model with the M87* image, we should rotate the calculated image clockwise by $72^\circ$. Alternatively, we could measure angles as offsets from the jet direction North. Thus, for a model to reproduce what is seen in M87*, it should have strong linearly polarized flux between $+10^\circ$ from the jet, i.e., just to the left of North, and $-140^\circ$ from the jet, which is located in the lower-right quadrant. That is, the polarized flux should concentrate in the right half of the panels in plots such as Figs.~\ref{fig:Bz}--\ref{fig:Beq2} below, shading towards the upper right quadrant. As we will see, this is a fairly strong constraint.

The second piece of information from the polarized image of M87* is that the polarization vectors show a twisting pattern that wraps around the black hole \citepalias{PaperVII, PaperVIII}. The twist is described quantitatively by the $\beta_2$ mode of the azimuthal decomposition of polarization described in \citet{PWP_2020}. The amplitude of $\beta_2$ describes the degree to which the EVPA obeys rotational symmetry and scales linearly with fractional polarization, while the phase of $\beta_2$ describes the twist angle between the EVPA and the local radial unit vector on the image. In the M87* image, the twist angle is fairly stable in the regions where the polarized flux is strong. With respect to the local radial direction, the EVPA of the polarization vector is rotated clockwise by $\sim70^\circ$. This too is a strong constraint on models, as discussed at length in \citetalias{PaperVIII}.

\subsection{Models with Pure Vertical Field}

\citet{Gravity_2018_orbit} reported observations of polarized flares in Sgr A$^*$ in near-IR, and showed that a model with a dominant vertical magnetic field can reproduce the observations. Motivated by this, we begin by studying the predictions of our toy model for a pure vertical field, oriented normal to the plane of the emitting ring. 

Figure \ref{fig:Bz} shows results from the analytical model for the case when $B_z=1$,
$B_r=B_\phi=0$. It explores the two primary
physical effects other than magnetic field direction that influence the observed polarization: (i) Doppler beaming and relativistic aberration caused by motion of the radiating fluid, and (ii) gravitational lensing caused by the gravity of the black hole. The Top Left panel in Fig.~\ref{fig:Bz} corresponds to a ring with a large radius ($R=10^4$) such that there is negligible gravitational lensing.We also set $\beta=0$, thereby eliminating Doppler beaming and aberration. The only remaining effect is the tilt of the ring, which causes the pure $B_z$ field in the ring frame to appear in projection on the sky as a vertically oriented (North-South) field. The polarized synchrotron emission from the ring has its EVPA perpendicular to the projected field, i.e., in the East-West direction. The observed polarized intensity, which is indicated by the sizes of the polarization ticks in the plot, is uniform around the ring. In this figure and all others shown in this section, ticks are shown at 50 equally spaced positions in $\phi$.

The Top Right panel in Fig.~\ref{fig:Bz} shows the effect of including an arbitrary relativistic velocity ($\beta=0.3$) for the fluid in the clockwise tangential direction ($\chi=-90^\circ$), but still keeping a large radius, hence no gravitational deflection. In this case, there is a strong asymmetry in the polarized flux around the ring. However, the bright region of the ring is in the left half of the plot, exactly the opposite of what we require to explain M87*. This contrary behavior is actually rather surprising. Given the direction of the tilt and the clockwise sense of rotation, the fluid in the right half of the plot has a component of its motion towards the observer, while the fluid on the left has a component away from the observer. Doppler beaming ought to favor the right side, yet we see the opposite. This paradoxical behavior is because of aberration, as we explain in sec.~\ref{sec:analytic}. 

The Bottom Left panel in Fig.~\ref{fig:Bz} shows the effect of a pure inward radial velocity ($\chi=-180^\circ$), again for a large ring radius. Once again, the bright region of the disk is on the wrong side compared to what is seen in M87*. It is also exactly the opposite of what we would expect from Doppler beaming, since the fluid in the upper half has a velocity component towards the observer, and ought to be bright. Once again, aberration is the explanation.

Finally, the bottom right panel considers a ring at small radius ($R=6$) such that gravitational deflection of light rays is important. For simplicity, we assume that there is no fluid velocity. In this case, the results are similar to the Bottom Left panel, and the strongest polarized flux is at the bottom, which does not match what is seen in M87*. 

We do not discuss the $\beta_2$ phase of the polarization patterns for models with pure vertical field, except to note that in the regions where M87* has its strongest polarized flux (upper right), the sense of the EVPA twist seen in all the examples in Fig.~\ref{fig:Bz} has the wrong sense.

The conclusion from these examples is the following. If the polarized emission that we see in M87* at 230\,GHz is from equatorial gas, and if the gas rotates in the clockwise direction, as \citetalias{PaperV} concluded, and/or flows radially inward, as is natural for accretion, then the magnetic field cannot be dominated by a pure vertical component. There must be substantial radial and tangential field components. 

\begin{figure*}[t]
\includegraphics[width=9cm]{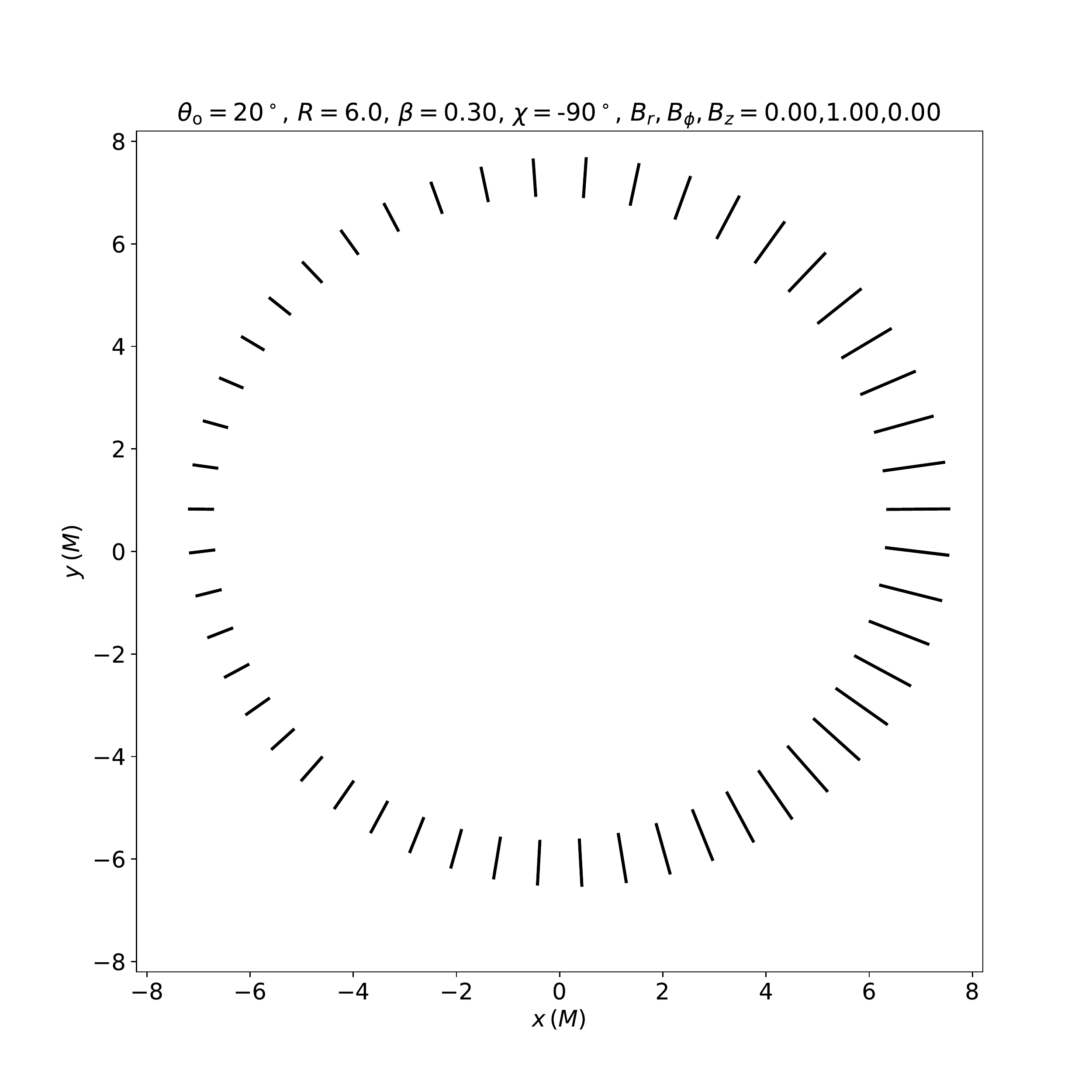}
\includegraphics[width=9cm]{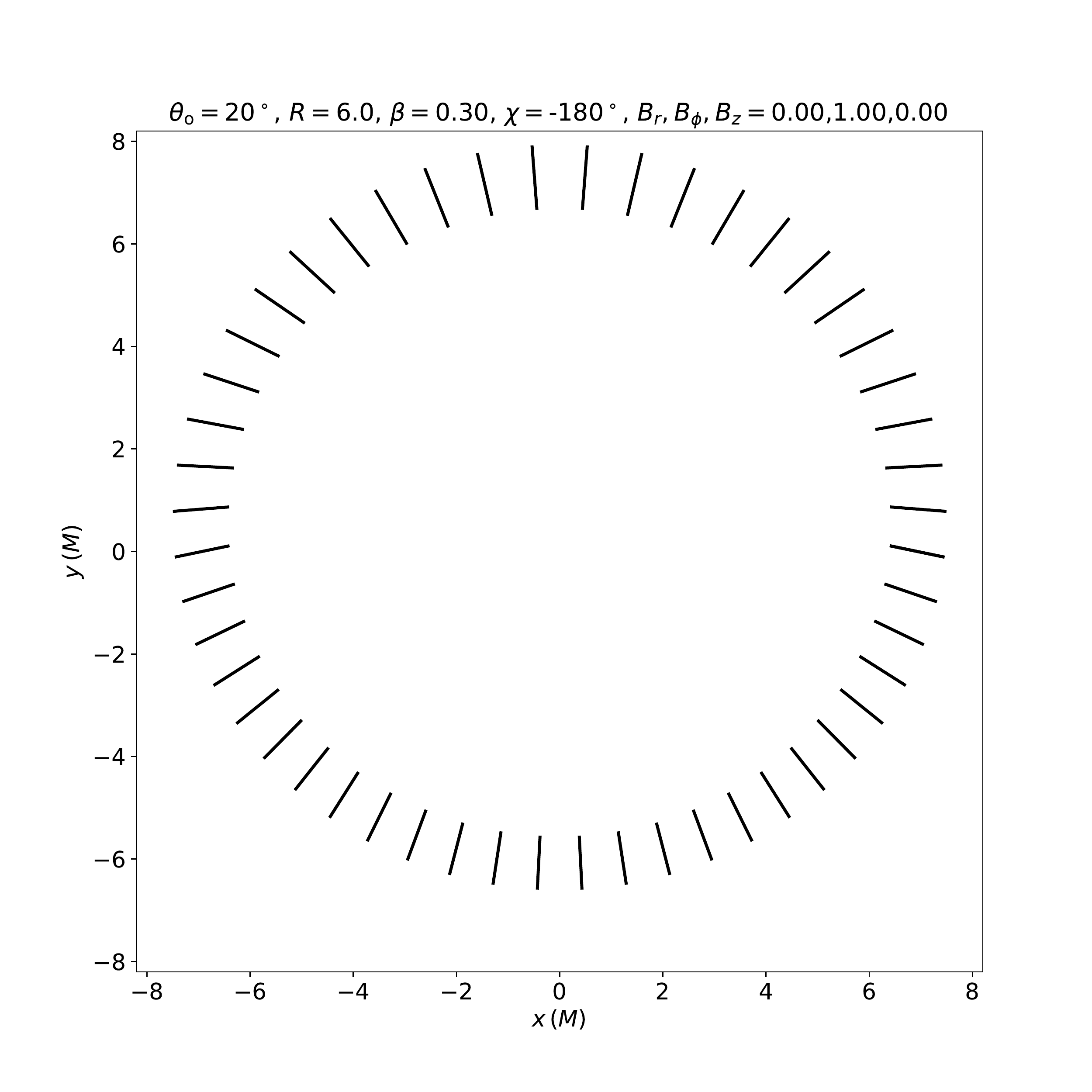}
\includegraphics[width=9cm]{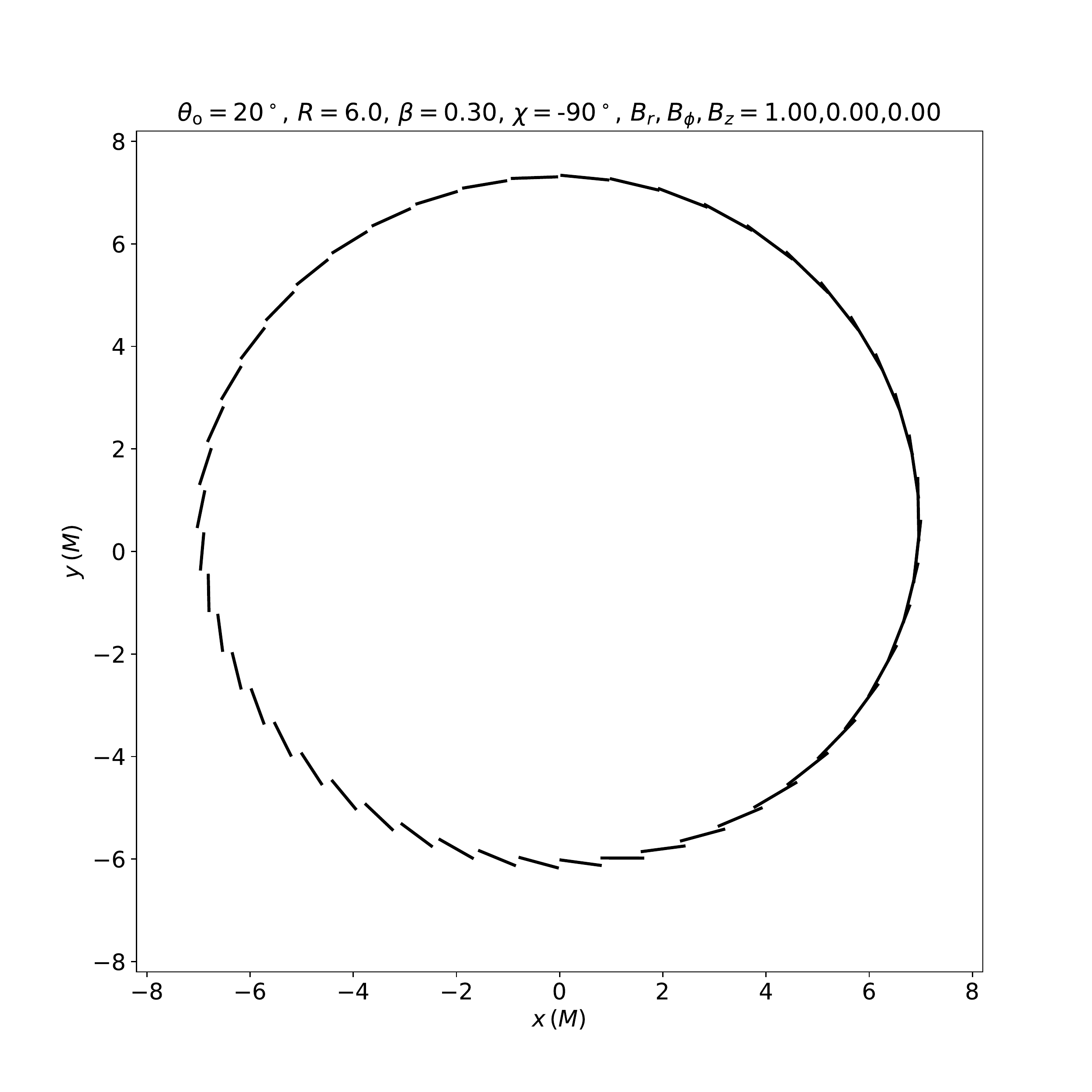}
\includegraphics[width=9cm]{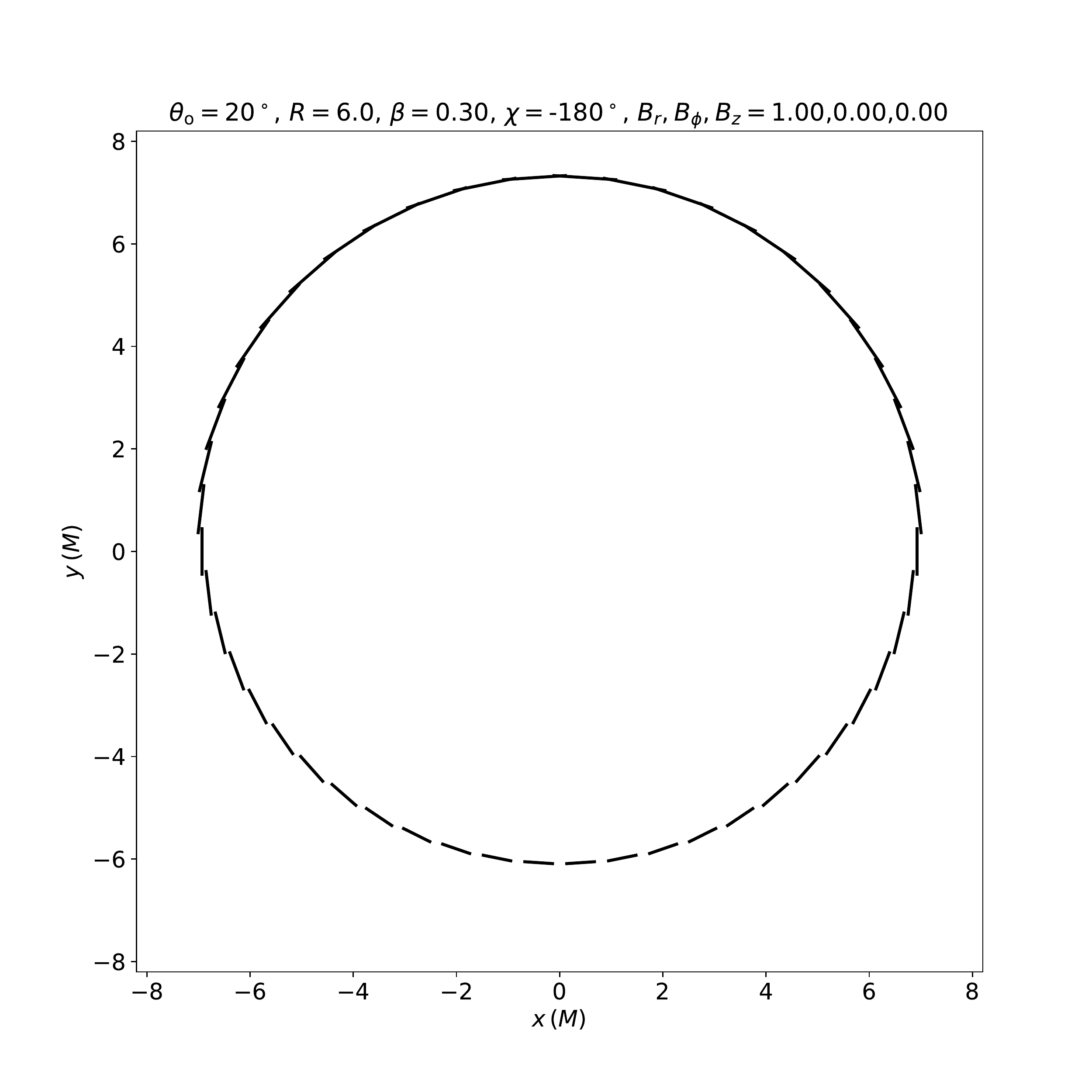}
\caption{Polarization patterns for models with magnetic field in the equatorial plane. Top Left: Azimuthal field ($\eta=90^\circ$) with azimuthal clockwise velocity ($\chi=-90^\circ$). Top Right: Azimuthal field ($\eta=90^\circ$) with radial inward velocity ($\chi=-180^\circ$). Bottom Left: Radial field ($\eta=0^\circ$) with azimuthal clockwise velocity ($\chi=-90^\circ$). Bottom Right: Radial field ($\eta=0^\circ$) with radial inward velocity ($\chi=-180^\circ$).}
\label{fig:Beq1}
\end{figure*}

\begin{figure*}[t]
\includegraphics[width=9cm]{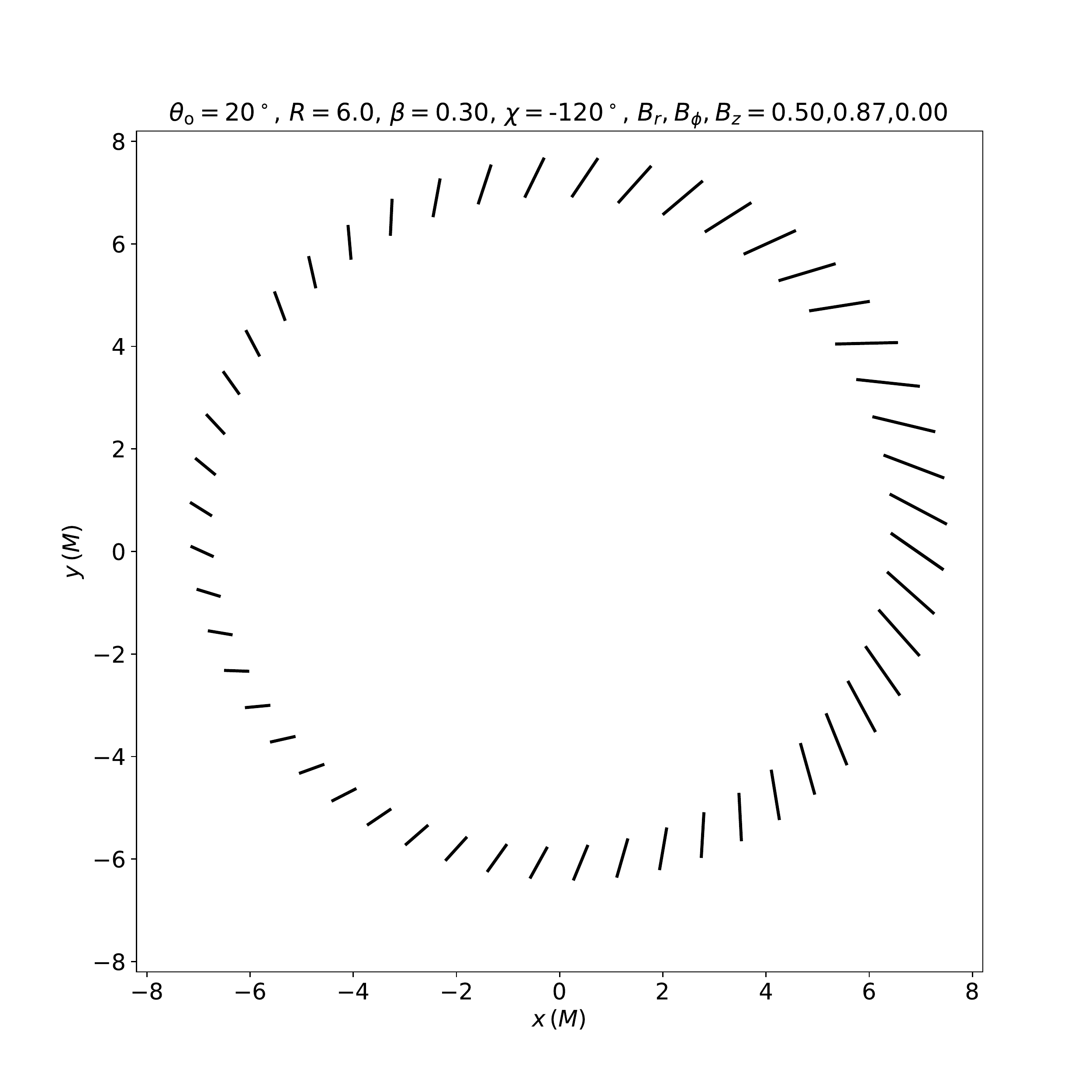}
\includegraphics[width=9cm]{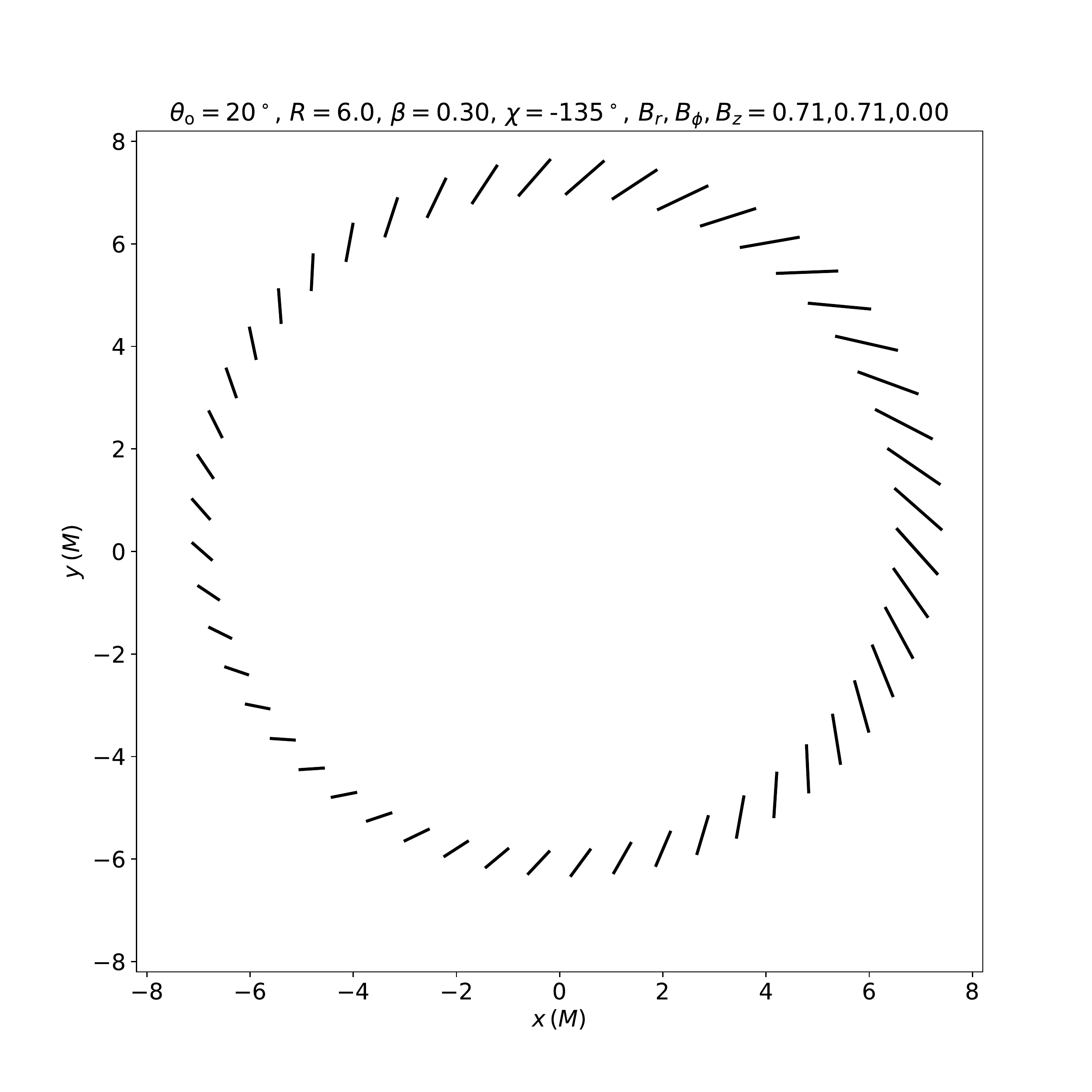}
\includegraphics[width=9cm]{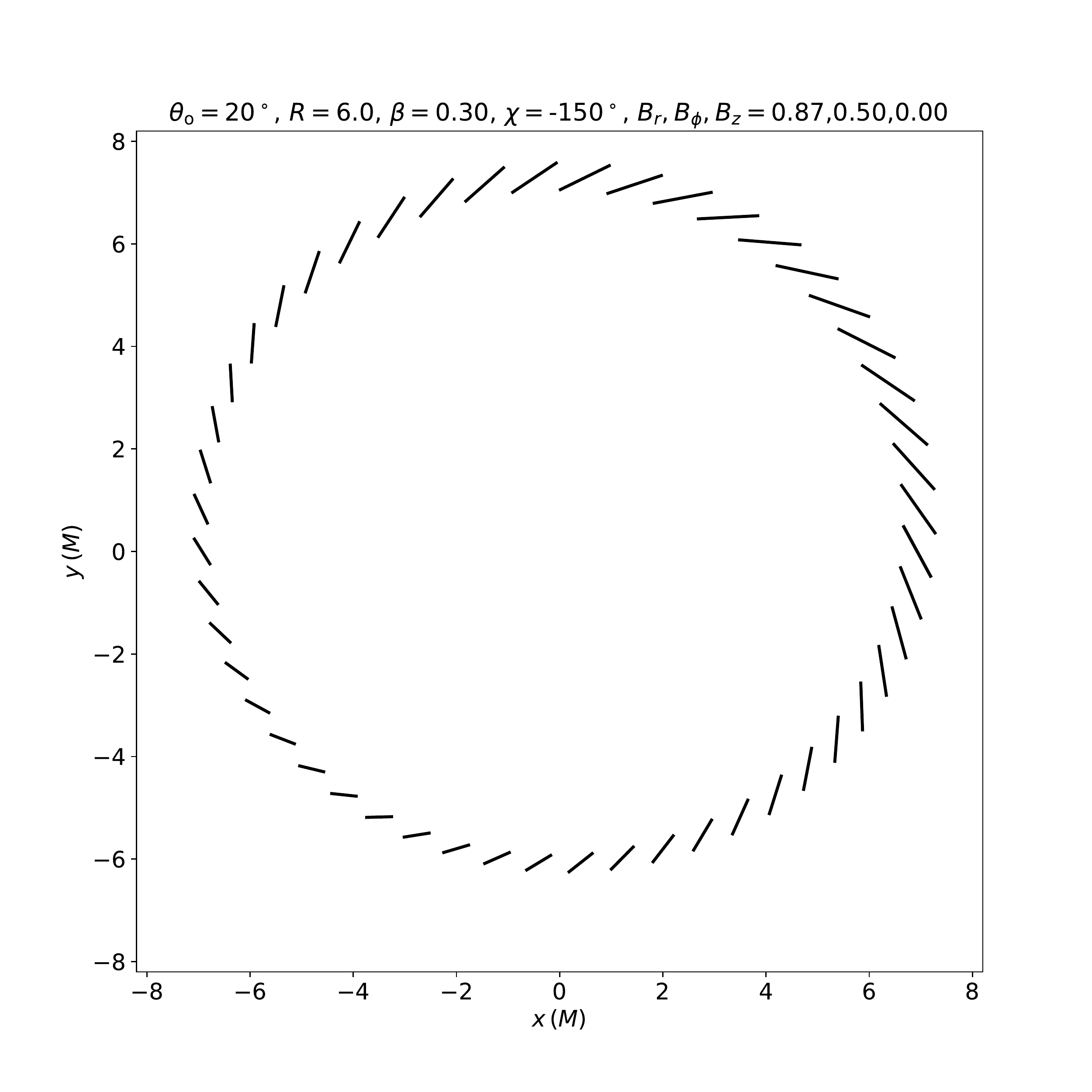}
\includegraphics[width=9cm]{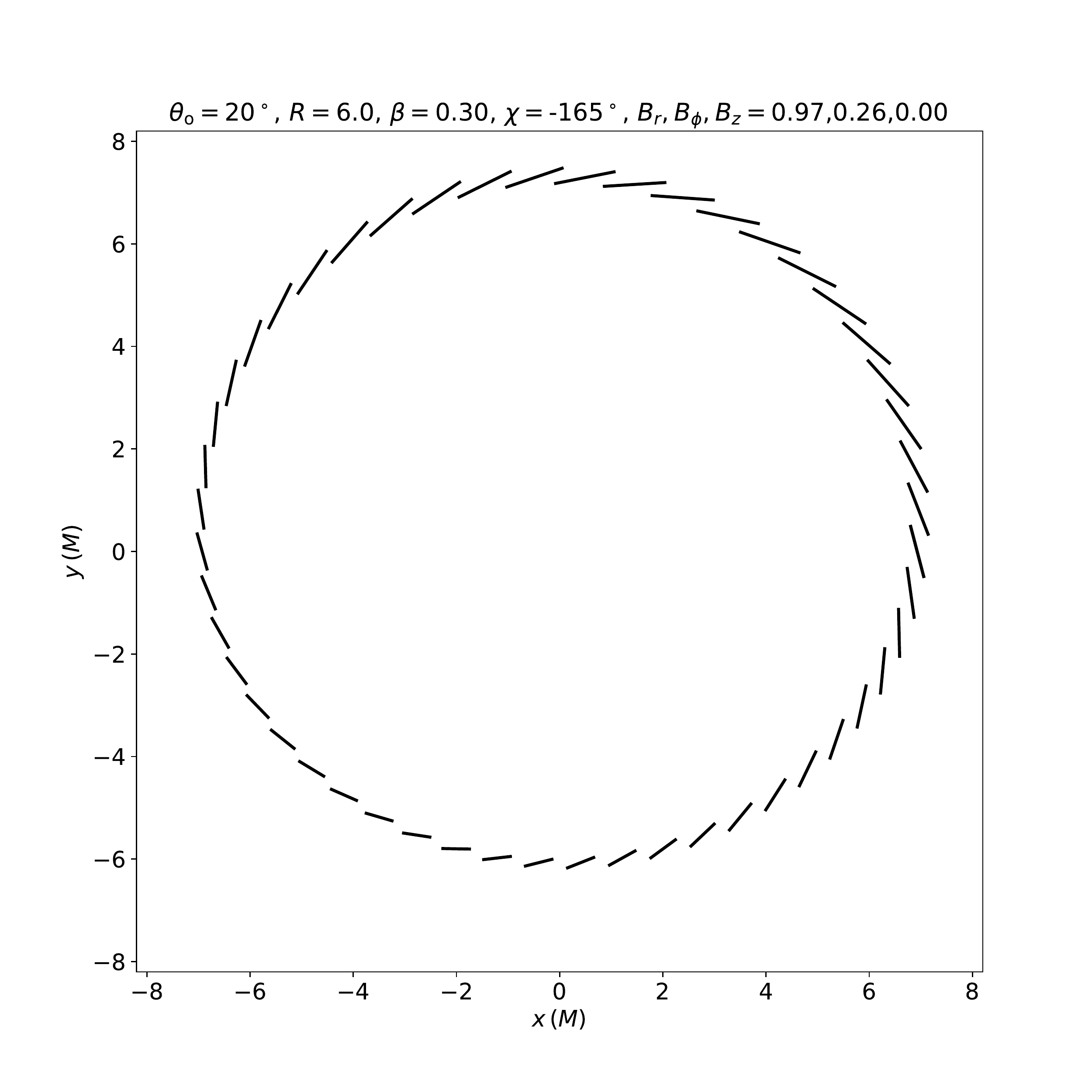}
\caption{Polarization patterns for four models that include both radial and azimuthal components of velocity and magnetic field. The models correspond to $\chi=-120^\circ$ (Top Left), $\chi=-135^\circ$ (Top Right), $\chi=-150^\circ$ (Bottom Left), $\chi=-165^\circ$ (Bottom Right), each with magnetic field trailing opposite to the velocity ($\eta=\chi+180^\circ$). The two models in the bottom row come closest to reproducing the polarization pattern seen in M87*.}\label{fig:Beq2}
\end{figure*}

Note that the observed ring in the Bottom Right panel in Fig.~\ref{fig:Bz} has a radius slightly larger than the original ring radius $R=6$. The ring is also shifted slightly upward relative to the origin. Both effects are the result of gravitational deflection, as we explain in sec.~\ref{sec:analytic}. The effect is seen only when $R$ is small (gravity is strong), which is the case in this panel of Fig.~\ref{fig:Bz}, and in all the panels in Figs.~\ref{fig:Beq1}, \ref{fig:Beq2}.

\subsection{Models with Pure Radial or Tangential Field}

We now turn our attention to models with magnetic field entirely in the equatorial plane, i.e., $B_z=0$, non-zero $B_r$ or $B_\phi$. We consider a ring with small radius ($R=6$) and include relativistic fluid motion; thus, lensing, Doppler and aberration are all included. Figure~\ref{fig:Beq1} shows four models, two with radial field ($\eta=0^\circ$) and two with tangential field ($\eta=90^\circ$). For each field configuration, we consider two velocity fields, either pure clockwise rotation ($\chi=-90^\circ$) or pure radial infall ($\chi=-180^\circ$).

Three of the four panels in Fig.~\ref{fig:Beq1} have their strongest polarized flux in the correct region of the ring (top and/or right) to match what is seen in M87*. Even the fourth (Top Right panel) has slightly stronger polarized flux at the top. The very different behavior of these models, compared to those in Fig.~\ref{fig:Bz}, is explained in detail in the next section. In brief, for models with magnetic field restricted to the equatorial plane, aberration induces the same sense of flux asymmetry as Doppler beaming and therefore enhances the effect of the latter, whereas in the pure $B_z$ models, aberration induces flux asymmetry with the opposite sign of that due to Doppler beaming, and in fact overwhelms the latter and reverses the sign of what is observed. In this sense, equatorial field-dominated models are more promising for M87*.

Considering the twist of the polarization pattern, as discussed in \citetalias{PaperVIII}, a pure tangential field is ruled out because the polarization ticks are predicted to be purely radial, which does not match M87. A pure radial field is also ruled out since it predicts polarization ticks entirely in the tangential direction. However, these models come closer to what is seen in M87*. It would appear that models in which $B_r>B_\phi$ are most suitable.

\subsection{Models with Both Radial and Tangential Field}

Figure~\ref{fig:Beq2} shows four models in which both $B_r$ and $B_\phi$ are non-zero, and $B_z=0$. All the models have fluid with clockwise rotation in the sky and radial infall, i.e., the angle $\chi$ of the vector $\vec{\beta}$ is in the lower left quadrant. Since the radial and tangential magnetic field components in the inner regions of an accretion disk are likely oriented parallel to the motion of the fluid -- the field is ``combed out" by the flow -- we simplify matters by assuming that the field is aligned with the velocity. Specifically, we choose
\begin{align}
    {\rm Pure}~B_{\rm eq}:\quad \eta=\chi ~~{\rm or} ~~\eta = \chi + \pi.
    \label{etachi}
\end{align}
For the specific case of a purely equatorial field, we can choose either of the two values of $\eta$ indicated above. The two choices correspond to oppositely oriented directions of the magnetic field lines; this ambiguity has no effect on the linear polarized emission. As we discuss in \autoref{subsec:allfield}, we need to be more careful about the choice of $\eta$ when we have both vertical and equatorial field components.

In \autoref{fig:Beq2}, the model in the Top Left panel has tangential velocity larger than radial velocity, and correspondingly $B_\phi>B_r$. In the Top Right panel, the radial and tangential components are equal, while in the lower two panels the radial components of velocity and magnetic field are larger than the respective tangential components. All four models have flux asymmetry that qualitatively matches M87*. All four models also have polarization patterns with the same sense of twist, or sign of $\beta_2$ phase, as observed in M87*. Among the four models, the ones in the bottom row come closest to M87*.

\subsection{Models with $R=4.5$ M and Varying Inclination}
\begin{figure*}[t]
\includegraphics[width=9cm]{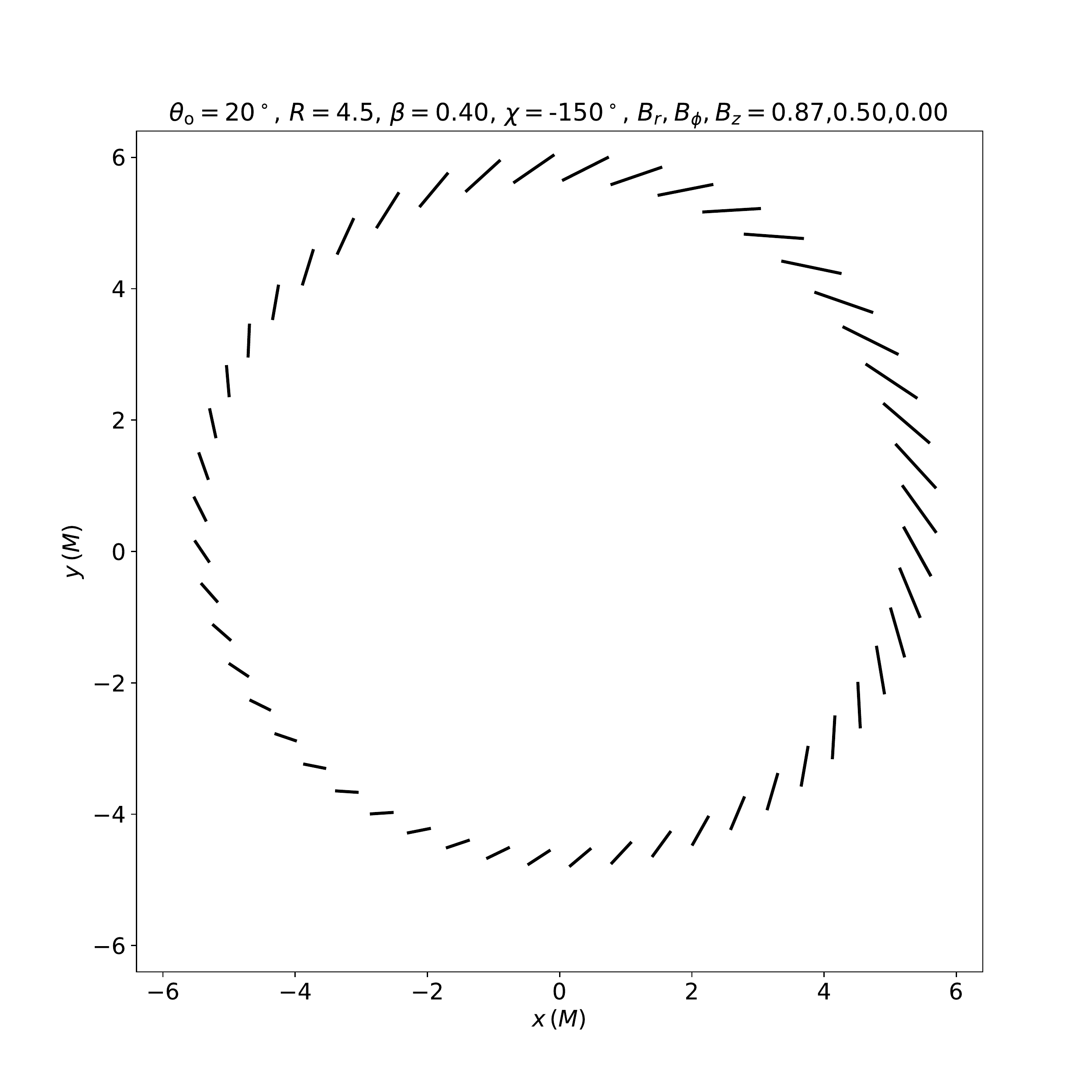}
\includegraphics[width=9cm]{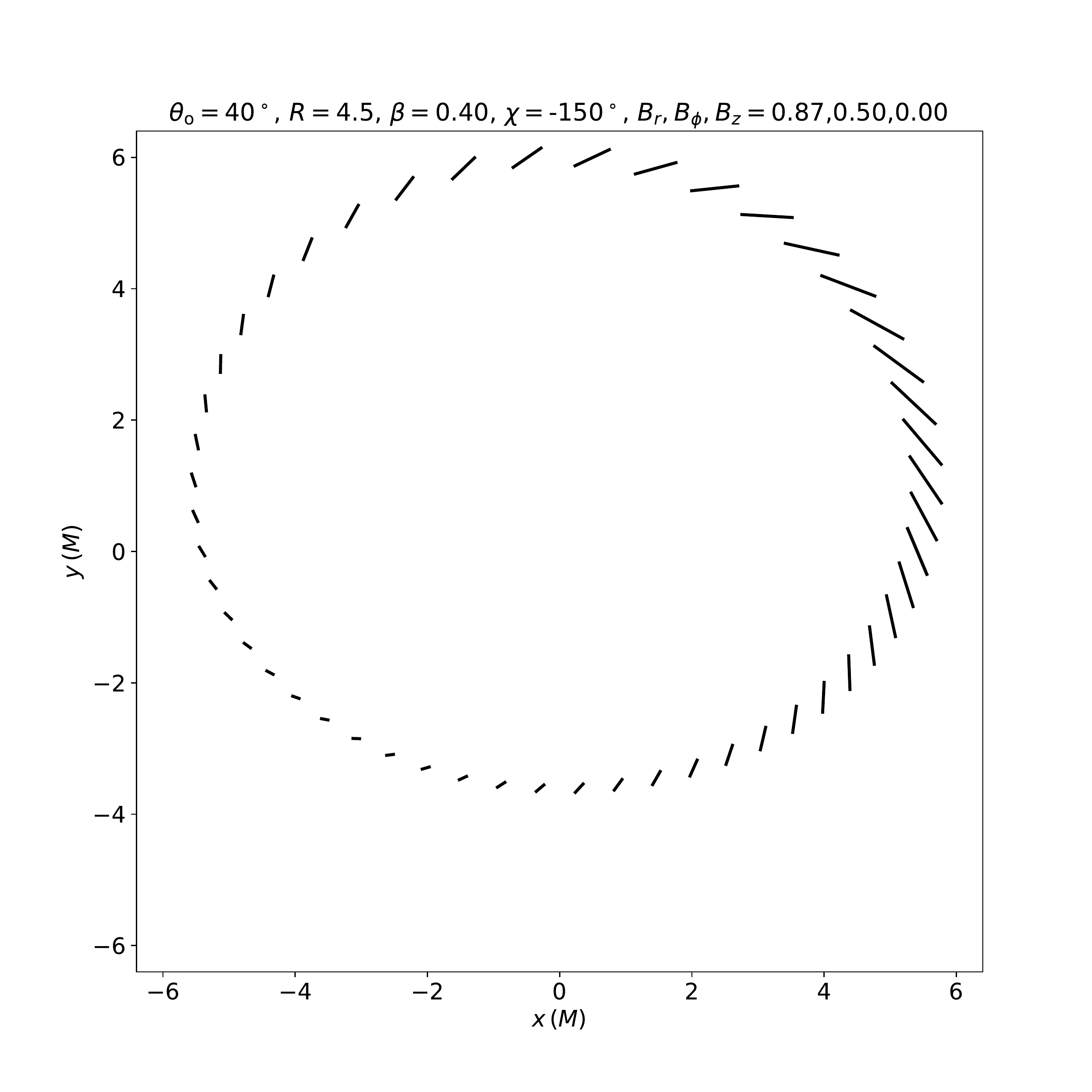}
\includegraphics[width=9cm]{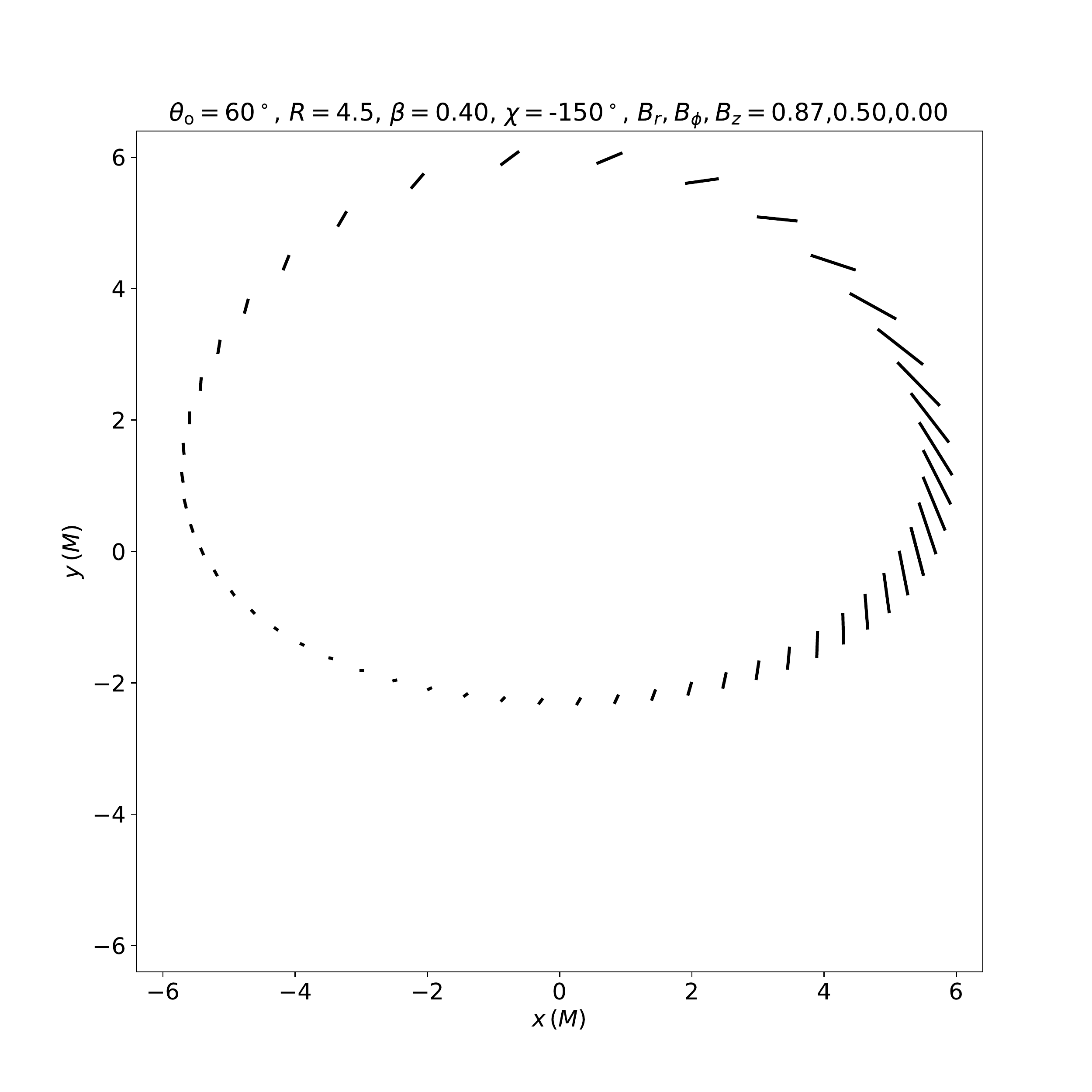}
\includegraphics[width=9cm]{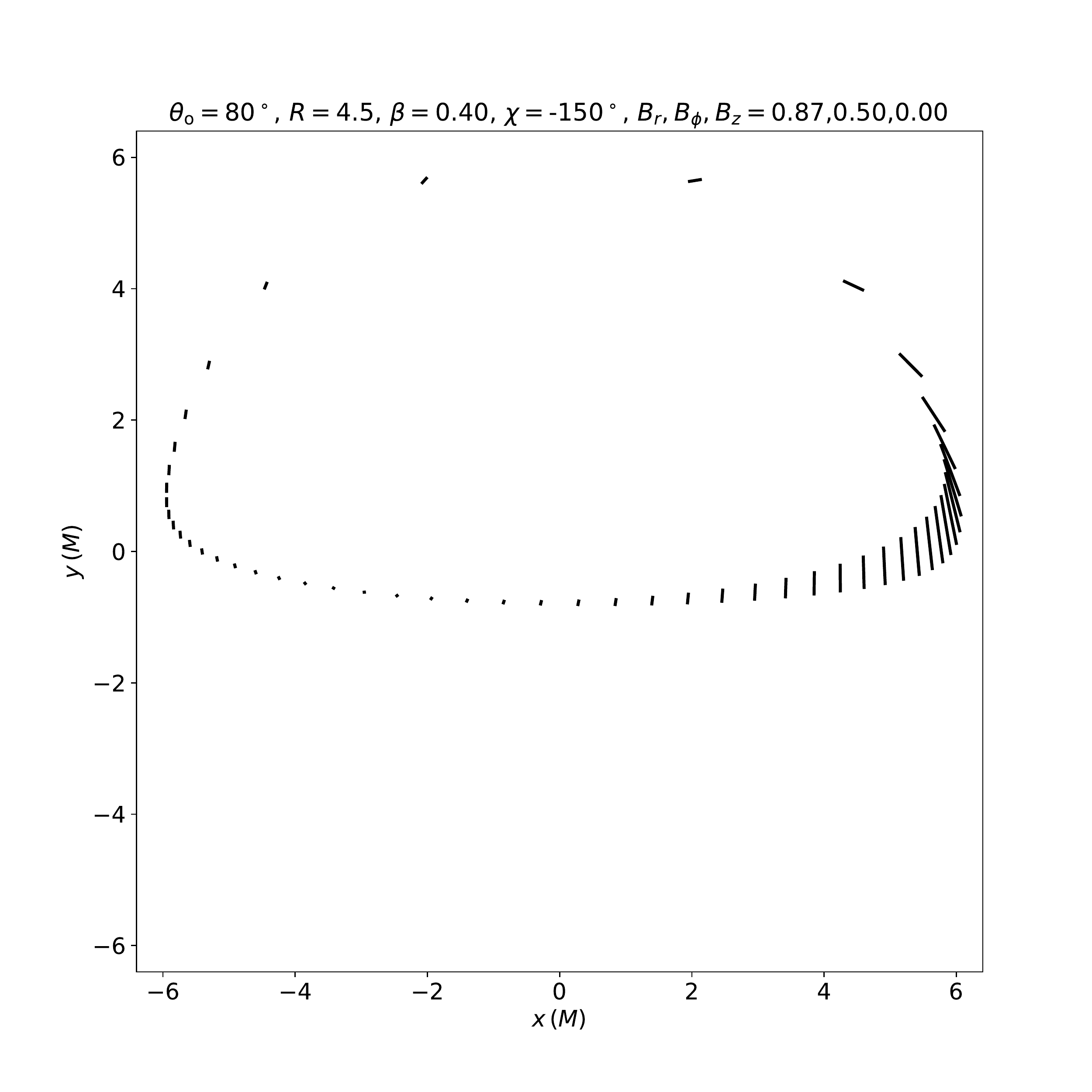}
\caption{Polarization patterns for four models with equatorial magnetic field and emission radius $R=4.5$, viewed at different inclination angles. Top left: $\theta_{\rm o}=20^\circ$. Top right: $\theta_{\rm o}=40^\circ$. Bottom left: $\theta_{\rm o}=60^\circ$. Bottom right: $\theta_{\rm o}=80^\circ$. All the models have velocity angle $\chi=-150^\circ$,  and magnetic field trailing opposite to the velocity ($\eta=\chi+180^\circ$). The model in the top left, rotated counter-clockwise by $288^\circ$ and with emission spread over a finite range of radii, is shown in \autoref{fig:M87_comparison} as a toy model of M87*.} \label{fig:incmodels}
\end{figure*}

We round out the discussion of examples by considering models with a smaller emission radius, $R=4.5$, which is better matched to M87*, and exploring the effect of varying the tilt angle $\theta_{\rm o}$.
\autoref{fig:incmodels} shows models with $\chi = -150^\circ$, $\eta = \chi + \pi = 30^\circ$, and four choices of $\theta_{\rm o}$: $20^\circ$, $40^\circ$, $60^\circ$, and $80^\circ$. 

The top left panel has $\theta_{\rm o}=20^\circ$ and is designed to resemble M87*. The polarized intensity asymmetry (relative to the direction of the jet), as well as the twist of the EVPA pattern, are similar to the EHT observations described in \citetalias{PaperVII} and \citetalias{PaperVIII}. This same model is shown again in \autoref{fig:M87_comparison} with the polarization pattern rotated counter-clockwise by $288^\circ$ to match the jet orentation in M87*, and with the emitting fluid spread out in radius with an exponential profile with scale width $2M$  (see \autoref{sec:M87_comparisons} for details), instead of the infinitely thin emitting ring assumed here.  

The remaining panels in \autoref{fig:incmodels} show the effect of increasing the tilt angle $\theta_{\rm o}$. The Doppler asymmetry in the polarized intensity increases rapidly since the fluid motion has a larger component parallel to the line-of-sight. The orientation of the asymmetry (bright on the right, dim on the left) as well as the twist of the polarization pattern qualitatively resemble what is seen in the $\theta_{\rm o}=20^\circ$ model. The ring appears increasingly flattened as $\theta_{\rm o}$ increases, but it also acquires an additional asymmetry such that, by $\theta_{\rm o}=80^\circ$ it looks more like a semi-circle than an ellipse. This is because of extreme lensing of radiation emitted from the far side of the ring. As in the previous Figures, ticks are equally spaced in $\phi$; the large gaps on the north side of the $\theta=80^\circ$ image indicate the relative stretching between $\varphi$ and $\phi$ at high inclination.

\subsection{Models with All Field Components}
\label{subsec:allfield}

We finally discuss models in which all three components of the magnetic field are non-zero. In this general case, we need to be careful about the geometry of the magnetic field. In a three-dimensional accretion flow in which magnetic field lines penetrate the disk from one side to the other, as for instance in a magnetically arrested disk (MAD) field geometry \citep{Narayan_et_al_2003,Igumenshchev_et_al_2003,Tchekhovskoy_et_al_2011,Bisnovatyi-Kogan_2019}, one expects a reflection antisymmetry in $B_{\rm eq}$ about the midplane. That is, $B_r$ and $B_\phi$ would flip sign when crossing the mid-plane, whereas $B_z$ would retain the same sign on the two sides. Let us assume, without loss of generality, that $B_z$ is positive, i.e., the $z$-component of the magnetic field line is pointed towards the observer, and let us also take $B_{\rm eq}$ to be positive. If the magnetic field is dragged and aligned with the flow, as we assumed in the previous two subsections, the field angle $\eta$ and the flow velocity angle $\chi$ must be related as follows on the two sides of the disk,
\begin{eqnarray}
&z>0 ~{\rm (near~side)}:\quad& \eta = \chi + \pi, \nonumber\\
&z<0 ~{\rm (far~side)}:\quad& \eta = \chi,
\end{eqnarray}
where ``near side" means the side of the disk facing the observer.
 
In the absence of Faraday rotation effects, the above antisymmetry affects emission only by changing the relative sign between $B_{\rm eq}$ and $B_z$, hence it is not relevant if either $B_{\rm eq}$ or $B_z$ is zero. However, when both $B_{\rm eq}$ and $B_z$ are non-zero, one should separately compute the polarized image produced by the near side and far side of the disk and add the resulting Stokes parameters.

If Faraday effects internal to the flow are strong enough to depolarize the emission from the far side, the polarized image seen by the observer will be dominated by the near side. The simulations considered in \citetalias{PaperVIII}, for instance, generally show large internal Faraday depths. In such cases, we need compute only a single image from the near side of the disk, setting $\eta=\chi+\pi$.

We do not show examples of models with both vertical and equatorial field since the parameter space is large. 

\subsection{Numerical Geodesics and Effect of Spin}\label{sec:numerical}

\begin{figure*}[t]
\includegraphics[width=\textwidth]{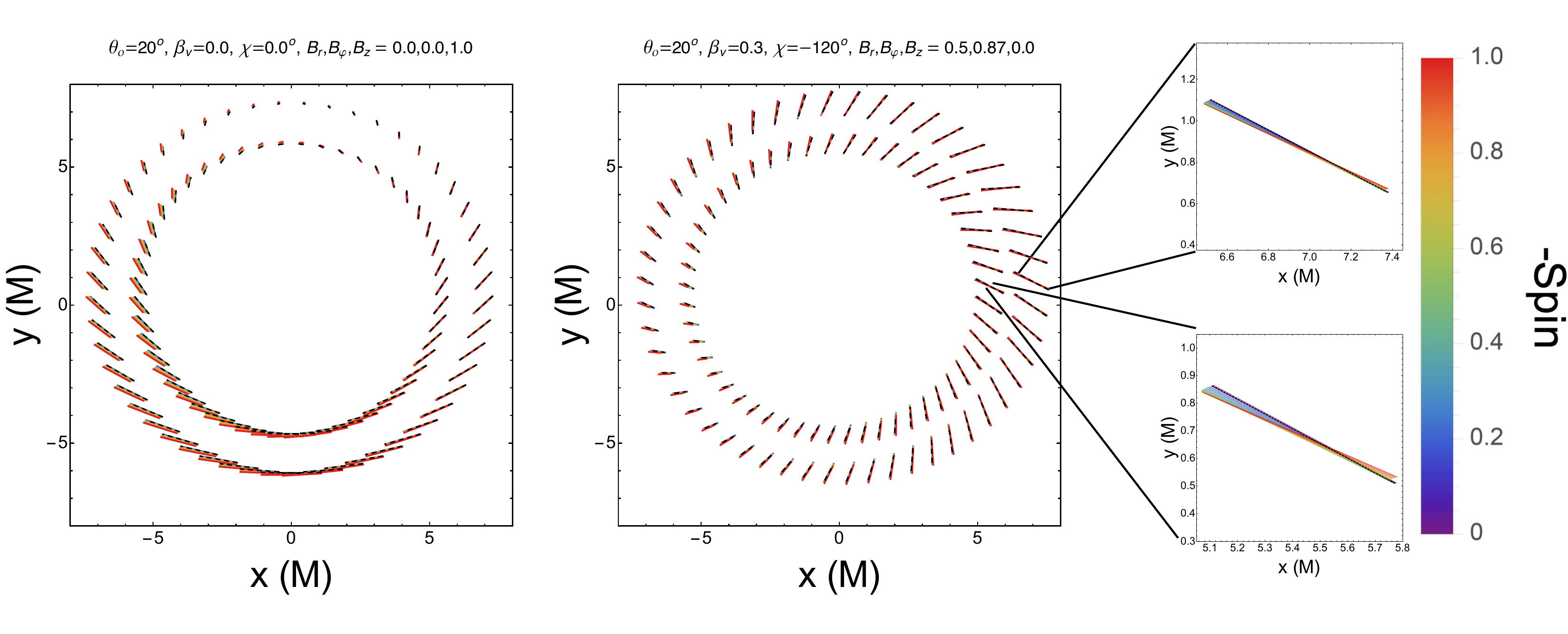}
\caption{The effects of spin on the observed polarization pattern. Each of the two main panels displays a different configuration of magnetized fluid. The first panel corresponds to the bottom right panel of Fig.~\ref{fig:Bz} and the second panel corresponds to the top left panel of Fig.~\ref{fig:Beq2}. Both panels show an inclination of $20^\circ$ and negative spin (i.e., clockwise rotation on the image). The inner and outer rings of polarization ticks correspond to emission from $R = 4.5$ and $R=6$, respectively. The color bar shows increasing spin from $a=0$ to $|a|=1$, and the Beloborodov approximation for Schwarzschild is shown in black overlaid dashes. The two small panels display a zoom-in of one set of ticks at $R =4.5$ (lower) and $R = 6$ (upper).  }\label{fig:withspincomparison}
\end{figure*}

A general Beloborodov-like analytic approximation for the emission angle of photons from equatorial matter around a spinning black hole is not known. However, it is possible to solve analytically for the observed polarization once the photon's arrival coordinates on the image are 
determined from a numerical solution to the geodesic equation; this relation can be explicitly expressed in terms of real elliptic integrals (\citealt{GL_lensing,GL_null}, see also  \citealt{Li_et_al_2005,Gates2020} for a calculation of images of an orbiting emitter in this formalism). 
For a spinning black hole, we generalize the P-frame to the ``zero-angular-momentum-observer" (ZAMO) frame, and then consider a boost $\vec{\beta}$ as in \eqref{velocity2} into the corresponding F-frame. The semi-analytic result for the polarized image of such a boosted fluid orbiting a spinning black hole is presented in Figure \ref{fig:withspincomparison}, in which changing spin is plotted by color. The inner and outer ring in the first two panels correspond to emission radii of $R = 4.5$ and $R = 6$, respectively. The results of the Beloborodov approximation are overlaid with black dashed lines and coincide with the low spin semianalytic solution from Kerr. The first and second panels of Fig.~\ref{fig:withspincomparison} generalize the scenarios from the bottom right panel of Fig.~\ref{fig:Bz} and the upper left panel of Fig.~\ref{fig:Beq2}, respectively. The small panels zoom in on one set of ticks from the second panel. 

\autoref{fig:withspincomparison} illustrates that for the idealized case of purely geometric and relativistic effects that we consider here, black hole spin has only a small effect on the observed EVPA and can be reasonably neglected for the purposes of the toy model. It also shows that the Beloborov approximation is fairly accurate even at radii as small as $R=4.5$. The effects of spin on observed polarization become more pronounced at very small radius and high observer inclination, neither of which are considered in this paper but will be the subject of future work.

\subsection{Generalizations}
\label{subsec:generalizations}

Although the examples presented in this paper are restricted to axisymmetric models with emission limited to a single radius, the underlying model is more general. The primary result of the analysis presented in sec.~\ref{sec:model} is an analytical method to map emission properties at a given $(R,\phi)$ in the emitting ring to the properties of the observed radiation in the sky plane. This transformation can be easily applied to models with non-axisymmetric emission, as well as to radially extended sources. In such models, $|\vec{B}|$ would be a function of location and this would need to be included in the calculations. Other quantities like the electron temperature and number density that affect the emissivity could also vary with position and will need to be accounted for.

Two other approximations in the model, both made in the interests of simplicity, deserve discussion: (1) We restricted the emitting gas to lie in a single equatorial plane. (2) We took the velocity to lie entirely within the same plane (though we did allow for a general magnetic field). Both limitations can be eliminated. 

The Beloborodov approximation can be applied at any emission location $(R,\phi,z)$, not just at equatorial locations. For non-equatorial locations, the geometry of the Geodesic Frame and the computation of $\alpha$ (Fig.~\ref{fig:gframe}) will differ. This will modify the result for the components of $k^{\hat\mu}_{\rm (P)}$. If a given null geodesic has contributions from several emission regions at different heights $z$ from the equatorial plane, one could compute their individual contributions to the Stokes parameters and add the contributions incoherently. 

Similarly, an off-plane velocity component will modify the Lorentz transformation coefficients between the P-Frame and the F-Frame, and will alter the geometrical factor that enters the path length calculation. The distinction between ``vertical" and ``in-plane" magnetic field components would become less clear, but this is merely a matter of definition.

The model discussed in this paper has been derived for a non-spinning (Schwarzschild) black hole. However, as shown in \autoref{sec:numerical}, and as discussed also in \citet{Gravity_2020} and \citetalias{PaperVIII}, black hole spin has very little effect on the polarized image, at least for the low inclination angles considered so far.

Finally, the analysis here is focused on optically thin synchrotron emission for which the polarization four-vector $f^\mu$ is given by equation~(\ref{eq:fmuF}) and the electric field is normalized as in equation~(\ref{Enormzeta}). For optically thick emission from a thin accretion disk, other prescriptions will need to be substituted, e.g., \cite{Li_et_al_2009} discuss polarization of X-rays emitted by the scattering atmosphere above a black hole X-ray binary disk. Except for this change, the rest of the analysis should remain the same.

\section{Analytical Understanding of the Results}\label{sec:analytic}

By Taylor-expanding the expressions given in sec.~\ref{sec:model} in suitably chosen ``small''
quantities, and keeping terms up to second order, we can obtain useful analytical approximations for various observables. This provides a physical understanding of the results shown in sec.~\ref{sec:examples}. 

In the present context of trying to understand M87* and Sgr A$^*$, we have three small quantities, $2/R \approx 1/3$ (lensing), $\beta\approx 1/3$ (Doppler and aberration), $\sin\theta_{\rm o} \approx 1/3$ (ring tilt\footnote{In the case of M87*, observations of the radio jet suggest a tilt $\theta_{\rm o} \sim17^\circ$ \citep{Walker_2018}, and in the case of Sgr~A$^*$, \cite{Gravity_2018_orbit} estimate $\theta_{\rm o} <30^\circ$ based on the polarization signatures of infrared flares.}), where the numerical values correspond to the models shown in sec.~\ref{sec:examples}. We treat all three quantities on an equal footing in the series expansions we carry out. The full results, with all terms up to quadratic order, are listed in Appendix~\ref{sec:series}. The reason for going up to quadratic order is explained below. Here we use the series expansion of the equations to interpret the numerical results presented in sec.~\ref{sec:examples}.

\subsection{Shape of the Observed Ring}\label{sec:ring_shape}

We begin with the shape of the ring as observed on the sky. To quadratic order, the result is
\begin{align}
\label{alphaseries}
x &= (R+1)\cos\varphi\\
\nonumber &\qquad + \left[ -\frac{1}{2R}\cos\varphi + \sin\theta_{\rm o}\sin 2\varphi - \frac{R}{2}\sin^2\theta_{\rm o}\sin^2\varphi \cos\varphi\right],\\ 
\label{betaseries}
y  &= (R+1)\sin\varphi\\
\nonumber &\qquad + \left[ -\frac{1}{2R}\sin\varphi + 2\sin\theta_{\rm o}\sin^2\varphi  -\frac{R}{2}\sin^2\theta_{\rm o}\sin^3\varphi  \right]. 
\end{align}
The first term in each expression gives the answer up to linear order, and the remaining terms inside the square brackets correspond to quadratic order. Up to linear order we see that the observed ring is circular, but with an apparent radius larger by unity (i.e., $GM/c^2$) than the radius of the source ring. The radial ``expansion" of the observed ring is caused by gravitational deflection (lensing) of geodesics. As shown in Fig.~\ref{fig:gframe}, lensing causes the geodesic to curve around the black hole such that the impact parameter is larger than the naive straight-line estimate $R\sin\psi$.

Among the quadratic terms in equations~(\ref{alphaseries}) and (\ref{betaseries}), the terms proportional to $1/R$ are second-order corrections to the ring radius, and the $\sin^2\theta_{\rm o}$ terms describe the flattening of the observed ring because of tilt. The latter is simple geometry: a tilted circular ring appears elliptical in shape, with a minor axis radius equal to $\cos\theta_{\rm o} \approx 1-(1/2)\sin^2\theta_{\rm o}$ times the original ring radius. The $\sin\theta_{\rm o}$ terms describe the effect of tilt on lensing. Geodesics reaching the observer from the upper half of the ring ($0<\phi<\pi$) travel a longer distance near the black hole and suffer more deflection (this is the case shown schemaically in Fig.~\ref{fig:gframe}), while geodesics from the lower half ($\pi<\phi<2\pi$) experience less deflection. This causes an upward shift of the observed ring, i.e., a net positive bias in $y$. The shift is of the order of $\sin\theta_{\rm o}$ in units of $GM/c^2$. The shift is seen in all the models in sec.~\ref{sec:examples} that have a smallish radius ($R=6$, Lower Right panel in Fig.~\ref{fig:Bz}, and all panels in Figs.~\ref{fig:Beq1}, \ref{fig:Beq2}, \ref{fig:incmodels}).

\subsection{Doppler Factor and $\sin\zeta$}

Expanding up to second order, we find for the Doppler factor $\delta$,
\begin{eqnarray}
\delta &=& \left(1-\frac{1}{R}\right) \nonumber \\
&~& -\left[\frac{\beta^2}{2} + \frac{1}{2R^2} - \frac{2\beta}{R}\cos\chi + \beta\sin\theta_{\rm o}\sin(\chi+\varphi)\right],
\label{delta}
\end{eqnarray}
where the second order terms are shown on the second line inside square brackets. The linear order term $-1/R$ describes deboosting of the observed  intensity by gravitational redshift, and the first three second-order terms describe various other deboosting effects such as second-order Doppler. Since $\cos\chi$ is negative for radial infall, all three terms have a positive magnitude for the inflowing models we have considered, causing uniform dimming all around the ring. 

Azimuthal modulation of the intensity from relativistic beaming is described by the final term, $\beta\sin\theta_{\rm o}\sin(\chi+\varphi)$, and this is the only term that varies as a function of $\varphi$. The fact that this important effect appears only at second order is a major reason for expanding the equations up to quadratic order rather than stopping at linear. Why is it second order? It is because azimuthal modulation from Doppler beaming requires both tilt and fluid velocity, each of which is treated as a small quantity in our analysis.\footnote{For the models considered in sec.~\ref{sec:examples}, where each of the three small quantities is $\approx 1/3$, one expects second-order terms to be of order 10\% of the leading-order terms. However, many second-order terms come with large coefficients, e.g., intensity is proportional to $\delta^4$ so Doppler boost goes like $-4\beta\sin\theta_{\rm o}\sin(\chi+\varphi)$. Hence the second-order contributions are often not small. The analysis in this section should thus be used only for qualitative understanding. For accurate results, it is necessary to evaluate numerically the full equations given in sec.~\ref{sec:model}.} 

Doppler beaming causes an increase in the observed polarized intensity when $\sin(\chi+\varphi)$ is negative, with the maximum boost occurring when $\chi+\varphi=-90^\circ$. For pure clockwise rotation ($\chi=-90^\circ$), the maximum boost is at $\varphi=0$. This is natural since, for a ring tilted towards the North, the fluid at $\varphi=0$ has the largest velocity component towards the observer and hence produces the most Doppler-boosted radiation. For pure radial infall ($\chi=-180^\circ$), the maximum boost is at $\varphi=90^\circ$, again because the fluid there has the maximum velocity towards the observer. Since we consider models that lie between these two extremes, we expect the polarized intensity to be maximum somewhere in the top right quadrant, $0<\varphi<90^\circ$ (for a tilt to the North). This agrees with what is observed in M87* (once we allow for the different tilt/jet direction). Surprisingly, it is not true for the models shown in Fig.~\ref{fig:Bz}. To understand the reason for this discrepancy, we need to consider a second effect.

From equation~(\ref{absP}), the observed polarized intensity depends on the Doppler factor $\delta$ as well as the path length $l_{\rm p}$ and the angle $\zeta$ between the photon wave-vector $\vec{k}_{\rm (F)}$ in the fluid frame and the local magnetic field $\vec{B}$. For small tilt angles, the variation in the path length is small and not very important. We ignore it in the discussion below. The angle $\zeta$, however, is crucial since synchrotron emission is maximum when $\vec{k}_{\rm (F)}$ and $\vec{B}$ are orthogonal to each other ($\zeta=\pm\,\pi/2$) and vanishes when they are parallel ($\zeta=0,~\pi$). Appendix~\ref{sec:series} evaluates $|\vec{B}|^2\sin^2\zeta$ up to quadratic order. We consider  in the following subsections the effect of various terms in the series expansion.

\subsection{Models with Pure Vertical Field}

We begin by considering a model with pure $B_z$ and consider the non-zero terms in $|\vec{B}|^2\sin^2\zeta$:
\begin{align}
\label{sinzeta1}
&B_z~{\rm Finite},~B_{\rm eq}=0:
\nonumber\\ &|\vec{B}|^2\sin^2\zeta = \biggl[ -\frac{4}{R}\sin\theta_{\rm o}\sin\varphi + \frac{4}{R^2} + \sin^2\theta_{\rm o} - \frac{4\beta}{R}\cos\chi
\nonumber\\ &\qquad \qquad \qquad +  2\beta\sin\theta_{\rm o}\sin(\chi+\varphi)  + \beta^2 \cdots \biggr]\,B_z^2. 
\end{align}
There are several interesting effects here. First, we have only second-order terms, no zeroth- or first-order terms (this is another reason for going up to second order in the analysis). It suggests that the observed flux should be strongly suppressed. This is not surprising since the emission towards the observer goes as $\sin^2\zeta \sim\sin^2\theta_{\rm o}$, which is small for models with small tilt. The lack of zeroth- and first-order terms also means that the importance of the second-order quantities in equation~(\ref{sinzeta1}) is enhanced.

Consider first the term $-(4/R)\sin\theta_{\rm o}\sin\varphi$, which describes the combined effect of lensing ($4/R$) and tilt ($\sin\theta_{\rm o}$). Figure~\ref{fig:gframe} shows the origin of this term. In the absence of lensing, a geodesic travels on a straight line to the observer and hence subtends an angle $\theta_{\rm o}$ to the (vertical) magnetic field. When gravitational ray deflection is included, the angle at the emission point is modified. For a point on the North or upper half of the ring (the case shown in Fig.~\ref{fig:gframe}), the deflection is such that the photon wave-vector becomes more nearly parallel to the $z$-axis, i.e., more parallel to the magnetic field. Thus $\zeta$ is reduced, and this causes the emissivity to go down. The decrease is largest when $\varphi=90^\circ$, as indeed we find in equation (\ref{sinzeta1}). If we consider instead a point on the South or lower half of the ring, e.g., $\varphi=-90^\circ$, the gravitational deflection works in the opposite sense and causes $\zeta$ to increase, and the emissivity to correspondingly increase. The net result is an asymmetry in the polarized flux around the ring such that the maximum flux is in the South and the minimum is in the North, precisely as seen in the Bottom Right panel in Fig.~\ref{fig:Bz}.

Consider next the term $2\beta\sin\theta_{\rm o}\sin(\chi+\varphi)$, which corresponds to the combined effect of tilt and relativistic motion. Here the relevant effect is aberration. Because of the motion of the fluid, the orientation of the wave-vector $\vec{k}_{\rm (F)}$ in the fluid frame is different from its orientation $\vec{k}_{\rm (P)}$ in the P-frame. The aberration effect is such that fluid that is moving towards the observer has $\vec{k}_{\rm (F)}$ rotated closer to the $z$-axis in the fluid frame, i.e., more nearly parallel to $\vec{B}$, while fluid that is moving away from the observer has the tilt of $\vec{k}_{\rm (F)}$ with respect to $\vec{B}$ increased. The former fluid element thus emits less and the latter more in the direction of the observer. This cancels the effect of Doppler beaming. Actually, since the constant $\varphi$-independent terms in equation~(\ref{sinzeta1}) are of the same order as the modulation term  $\sin(\chi+\varphi)$ (note that $2\beta\sin\theta_{\rm o}$ is almost equal to $4/R^2+\sin^2\theta_{\rm o}+\beta^2$), the cancellation tends to be quite pronounced when $\chi+\varphi \sim -90^\circ$. The net effect is that aberration overwhelms Doppler beaming and gives the patterns seen in the Top Right and Bottom Left panels in Fig.~\ref{fig:Bz}.

\subsection{ Models with Pure Equatorial Field}

When we consider models with pure equatorial field ($B_{\rm eq}$ finite, $B_z=0$), the situation is quite different. Focusing on $|\vec{B}|^2\sin^2\zeta$, we find
\begin{align}
    \label{sinzeta2}
    &B_{\rm eq}~{\rm Finite},~B_z=0,~\eta=\chi+\pi:\nonumber\\ 
&|\vec{B}|^2\sin^2\zeta \approx B_{\rm eq}^2 + \left[ - 2\beta\sin\theta_{\rm o}\sin(\chi+\varphi) \cdots  \right]\,B_{\rm eq}^2,
\end{align}
where we have written only one of the second-order terms. As in sec.~\ref{sec:examples}, we have simplified matters by assuming that the magnetic field is oriented anti-parallel with the velocity: $\eta=\chi+\pi$. 

The first thing to note is that in the case of an equatorial field there is a non-vanishing zero-order term. For small tilt, a magnetic field in the equatorial plane is almost orthogonal to the photon wave-vector, hence synchrotron emissivity in the direction of the observer is nearly maximum. Correspondingly, the second-order terms are less important. Moreover, the second order term in equation~\ref{sinzeta2} appears with the same sign as the corresponding term in $\delta$ (eq.~\ref{delta}), and the opposite sign as in equation~(\ref{sinzeta1}). The reason is simple. When aberration tilts the wavevector closer to the $z$-axis, the wavevector becomes more nearly orthogonal to $\vec{B}$, and hence the emissivity increases. Thus in equatorial field models, the second-order terms in $|\vec{B}|^2\sin^2\zeta$ cooperate with and enhance the effect of Doppler beaming, as seen in the panels in Figs~\ref{fig:Beq1} and \ref{fig:Beq2}. As an aside, when both $B_{\rm eq}$ and $B_z$ are non-zero, and if we assume as before that $\eta=\chi+\pi$, then there is a first order term $-2\sin\theta_{\rm o}\sin(\eta+\varphi)\,B_{\rm eq}B_z$, which again has the same sign as the corresponding term in $\delta$.

\subsection{Twist of the Polarization Pattern}
\label{sec:twist}

We now briefly discuss the twist of the polarization pattern around the ring. When the field is purely in the equatorial plane, the results are transparent. To zeroth order, the electric field in the sky plane is given by
\begin{alignat}{2}
&&E_{x,\rm obs} &= -\sin\varphi\, B_r -
\cos\varphi\, B_\phi =-\sin(\eta+\varphi)\,B_{\rm eq}^2,\nonumber\\
\label{Ebetaobs}
&&E_{y,\rm obs} &= 
\cos\varphi\, B_r -
\sin\varphi\, B_\phi = \cos(\eta+\varphi)\,B_{\rm eq}^2. 
\end{alignat}
That is, the electric field is oriented perpendicular to the projected magnetic field, as one would expect.

Instead of considering the electric field, one could consider the Stokes parameters $Q$ and $U$ and look at their Fourier coefficients $\beta_m$ \citep{PWP_2020}, as described in Appendix~\ref{sec:series}. The most useful coefficient is $\beta_2$, whose complex phase directly gives the orientation of the twist. If the electric field is radial, the phase of $\beta_2$ is zero, if it is rotated clockwise from radial by $45^\circ$, the phase is $-90^\circ$, and if the electric field is tangential, the phase is $-180^\circ$. The EHT observations of M87* give a phase $\sim -130^\circ \equiv +230^\circ$. From Appendix~\ref{sec:series}, the leading order term in $\beta_2$ in the case of a pure equatorial magnetic field is
\begin{equation}
    \beta_2 \approx e^{i(\pi+2\eta)} B_{\rm eq}^2.
    \label{beta2eq}
\end{equation}
The phase of this quantity will match the phase observed in M87* if $\eta\sim 25^\circ$. Hence, the magnetic field must be mostly radial.

When $B_{\rm eq}=0$ and we have a purely vertical field, the phase of $\beta_2$ is determined by the coefficient of $B_z^2$, which consists entirely of second-order terms:
\begin{equation}
    B_{\rm eq}=0:\quad \beta_2= \left[\left(-\frac{4}{R^2}+\frac{4\beta}{R}e^{i\chi}-\beta^2e^{2i\chi}\right) B_z^2\right].
    \label{beta2z}
\end{equation}
If lensing is unimportant, i.e., $R$ is large, then $\beta^2$ dominates and the phase of $\beta_2$ is determined by the orientation angle $\chi$ of the fluid velocity. For a radial velocity ($\chi=\pi$), the phase of $\beta_2$ is $\pi$, i.e., the polarization vectors should be tangentially oriented. This is indeed seen in the brightest part of the ring in the Bottom Left panel in Fig.~\ref{fig:Bz}. Similarly, for a tangential velocity ($\chi=-\pi/2$), the phase of $\beta_2=0$ and the polarization ticks should be radial, as seen in the Top Right panel of Fig.~\ref{fig:Bz}. Finally, if there is no velocity but we consider strong lensing (small $R$), then equation~(\ref{beta2z}) shows that $\beta_2$ has phase $=\pi$ and the polarization should be tangential, as in the Bottom Right panel.

\begin{figure*}
    \centering
    \includegraphics[width=0.9\textwidth]{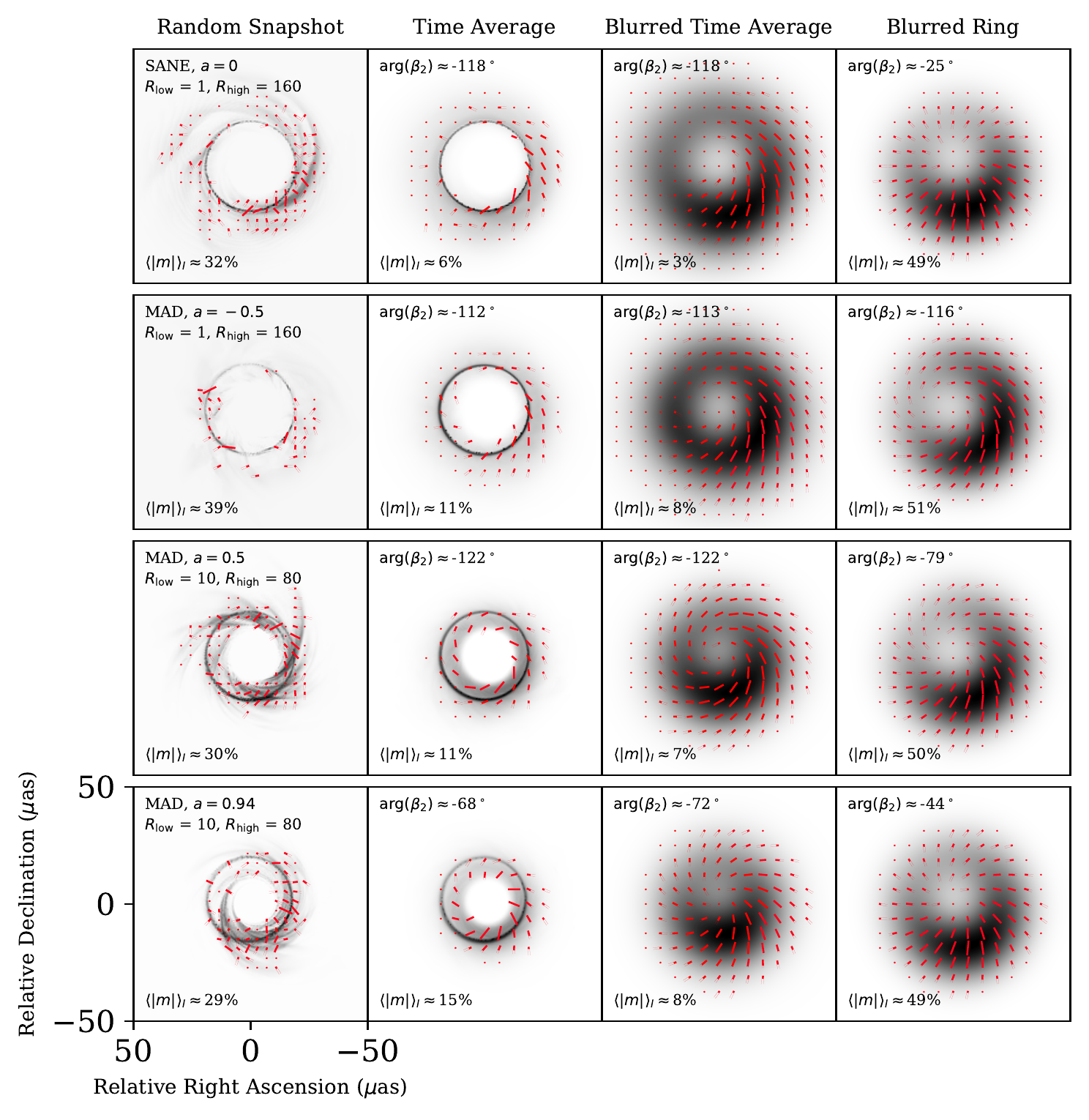}
    \caption{Comparison of GRMHD simulations to images of the ring model for simulation parameters favored in \citetalias{PaperVIII}. 
    The left three columns show random snapshots, time averaged images, and blurred time averages of each GRMHD simulation; the right column shows the image generated by the simple ring model when evaluated for magnetic field and fluid velocity values taken from the simulations at $R=4.5$ after azimuthal and temporal averaging. 
    Ticks show polarization magnitude and position angle where total intensity exceeds 5\% of the maximum. Grayscale shows total intensity in linear scale (directly proportional to polarization magnitude for the ring model). The total intensity and polarization magnitude are separately normalized in each panel. Panels show the average fractional polarization weighted by total intensity at bottom left; note that the GRMHD images are heavily depolarized, whereas the ring model images are not. The ring model and averaged images show the argument of the $\beta_2$ PWP mode at top left.}
    \label{fig:GRMHD}
\end{figure*}

\begin{figure*}[t]
    \centering
\includegraphics[height=7.7cm]{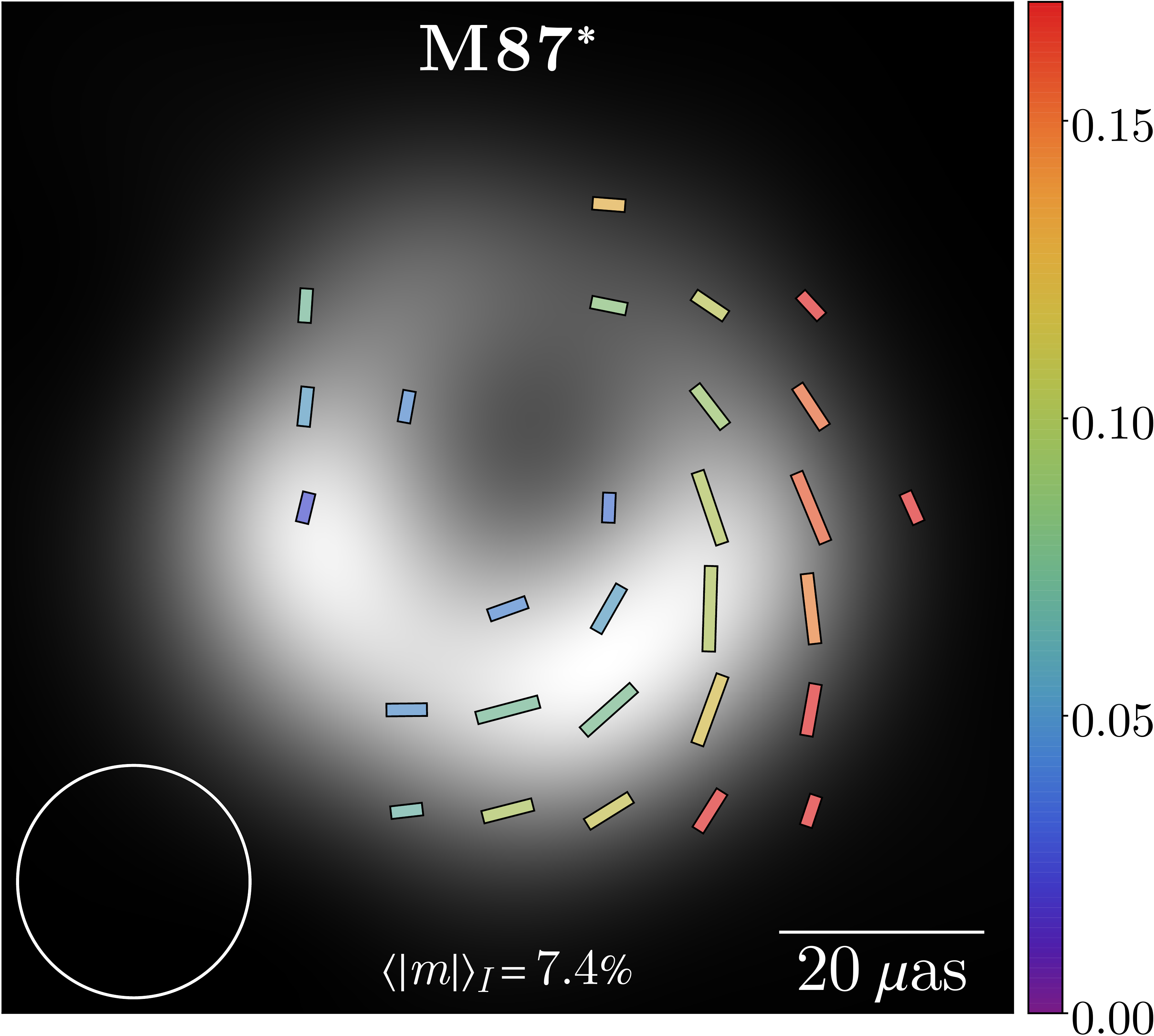}
\hfill\includegraphics[height=7.7cm]{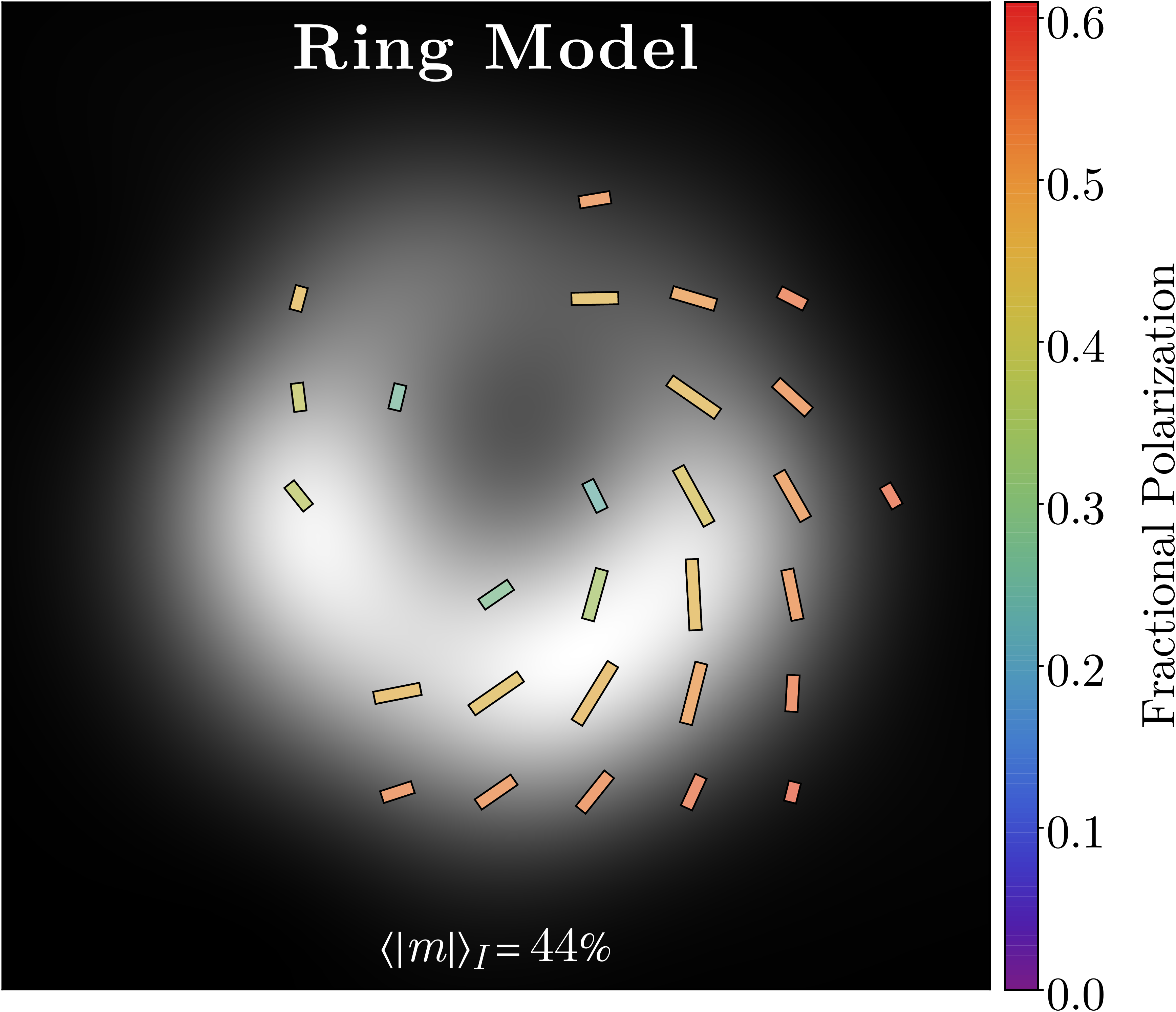}
    \caption{Comparison of the EHT polarimetric image of M87* on 2017 April~11 (left) with a representative ring model (right). Ticks show polarization fraction (color), magnitude  (length), and position angle  (direction); grayscale is identical for the two panels and shows total intensity of the EHT image of M87*. Ticks are only plotted where the M87* polarization exceeds 2\% of the maximum intensity. All images are shown after convolution with a circular beam of FWHM $23\,\mu{\rm as}$ (shown in the left panel). 
    As in \autoref{fig:GRMHD}, the total intensity and polarization are individually normalized for each panel.  
    The ring model has clockwise rotation with radial inflow, corresponding to the top left model in \autoref{fig:incmodels} after counterclockwise rotation by $288^\circ$. For complete model details, see \autoref{sec:M87_comparisons}. The fractional polarization of the resolved ring model is set to 70\%; the fractional polarization is reduced only through beam depolarization. Even after blurring, the ring model has significantly higher fractional polarization than the M87* image, although the relative variation in fractional polarization is similar across both images.
    }
    \label{fig:M87_comparison}
\end{figure*}

\section{Comparison to Observations}
\label{sec:observation_comparisons}

Our ring model provides a convenient framework for direct comparison with a variety of polarimetric observations of near-horizon emission. We now discuss two specific cases of particular interest: polarimetric imaging with the EHT and infrared flares of Sgr~A*. 

\subsection{Comparison to the M87 Polarized Image}
\label{sec:M87_comparisons}

Recent EHT observations produced  polarized images of M87* \citepalias{PaperVII}. \added{As reported in the one-zone model comparisons performed in \citetalias{PaperV} and \citetalias{PaperVIII}, the brightness, angular size, and expectation of significant Faraday effects coarsely constrain the magnetic field strength $B$, electron number density $n_e$, and electron temperature $T_e$ in the flow imaged by the EHT. The \citetalias{PaperVIII} results suggest that $B \lesssim 30$ G, ${10^4<n_e<10^7{\rm cm^{-3}}}$, and {$10^{10} < T_e < 1.2\times10^{11} \rm{K}$}.} \replaced{These images}{The reconstructed images in \citetalias{PaperVII}} were compared to general relativistic magnetohydrodynamic (GRMHD) simulations to identify a space of favored model parameters \citepalias{PaperVIII}. We will now explore whether our ring model can reproduce the polarization structure in these favored GRMHD simulations and in EHT images of M87*. 

For the GRMHD comparison, we first perform an azimuthal and temporal averaging in the fluid domain to approximate a stationary axisymmetric flow.  In the fluid frame, the magnetic field in each cell is decomposed in Cartesian Kerr-Schild coordinates, which are then recast into cylindrical coordinates and then azimuthally averaged.  These azimuthally averaged magnetic field decompositions are then further averaged over time between $7500\leq t/(GM/c^3) \leq 10000$ (the final quarter of these simulations). We then sample values of the fluid velocity and magnetic field vectors from the averaged simulations and use these values to generate ring models at $\theta_0 = 17^\circ$. To avoid sampling near where the tangential and radial field directions tend to abruptly flip sign, we use $z=1 M$, just above the midplane. We use $R=4.5 M$, corresponding to the apparent lensed size of the emission ring in EHT images of M87* (see the later discussion of the observed image). To create an image from the one-dimensional ring model, we adopt a radial profile that decays symmetrically in $R$ about $R=4.5$ as an exponential with a scale width of $2M$ \citepalias[EHT images only constrain this width to be ${<}\,5M$;][]{PaperVI}. 
We take a pixel-wise fractional polarization $|m|$ of 0.7 before blurring in the ring model.
Finally, we convolve both the ring model image and the GRMHD image with a $20\,\mu{\rm as}$ Gaussian kernel.

Using this approach, \autoref{fig:GRMHD} compares four favored GRMHD models to the corresponding ring models. In each case, the ring model reproduces the sense of EVPA twist and relative polarized intensity of the averaged and blurred GRMHD image, although discrepancies in $\arg ( \beta_2 )$ suggest contributions from emission away from the midplane or from other effects that are not included in the ring model (e.g., black hole spin or Faraday effects). The $R_{\rm low}$ and $R_{\rm high}$ parameters adapted from \citet{Mosci_2016} for use in \citetalias{PaperV} tune the ratio of electron to ion temperatures depending on the magnetic energy density of the plasma; large values of $R_{\rm high}$ tend to produce significant emission far from the midplane, particularly in SANE models. Also, Faraday effects in MAD models can produce significant coherent rotation of the EVPA and, hence, in $\arg ( \beta_2 )$ \citepalias{PaperVIII}.

\autoref{fig:M87_comparison} compares a representative ring model to the ``consensus'' EHT polarimetric image for 11 April, 2017 (i.e., the method-averaged image, see \citetalias{PaperVII}). The ring model parameters are chosen based on the observed image and a priori expectations for M87*. For simplicity, we take $B_z=0$, although non-zero values of $B_z/B_{\rm eq}$ over a modest range also give similar results. We use $\chi=-150^\circ$, to roughly match the observed $\beta_2$ for M87* (see \autoref{sec:twist}). We take $R = d/(2\theta_{\rm g}) - 1 \approx 4.5$ (\autoref{sec:ring_shape} explains the $-1$ factor), where $d \approx 42\,\mu{\rm as}$ is the observed ring diameter and $\theta_g \approx 3.8\,\mu{\rm as}$ is the angular gravitational radius \citepalias{PaperVI}. We use $\beta = 0.4$, which is comparable to the equatorial velocity seen in GRMHD simulations \citep[see][]{Ricarte_2020}. We use $\theta_0 = 20^\circ$ to match the jet inclination of M87*. Thus, this model has a modestly relativistic fluid with clockwise rotation and predominantly radial infall. This model corresponds to the top left panel of \autoref{fig:incmodels} after rotation to match the jet position angle of M87*, $288^\circ$. As with the GRMHD comparison, the ring model is evaluated over an exponential profile with a scale width of 2\,M centered at $R=4.5$\,M. The resulting ring model image is broadly consistent with the polarization morphology of the EHT image.

Although the qualitative agreement in \autoref{fig:M87_comparison} is encouraging, our simple ring model fundamentally fails to reproduce all the features in the M87* image. Namely, our simplest model would produce a high fractional polarization (${\gsim}\,60\%$), while the M87* image has a low resolved fractional polarization ${\lsim}\,20\%$. This suggests that significant depolarization from internal Faraday effects are essential when modeling and interpreting the M87* image. Nevertheless, the success of the ring model in reproducing the structure of some GRMHD images that have significant Faraday effects is encouraging for the prospects of physical inference from this simple model. 

One possibility for using our model for a more complex emission scenario is to combine multiple ring models that correspond to different emission regions. Specifically, the assumption $\eta = \chi + \pi$ corresponds to emission sourced by entrained magnetic field lines on the near side of the accretion flow (see \autoref{subsec:allfield}). 
The far side of the flow would instead have $\eta = \chi$, flipping $\vec{B}_{\rm eq}$. Ignoring that contribution is equivalent to assuming that 
Faraday depolarization effects in the midplane are strong, so that the far-side emission is fully depolarized (as indicated in many models considered in \citetalias{PaperVIII}; see \citealt{Ricarte_2020}). Our ring model could also be adapted to the case of weak Faraday rotation in the midplane; the resulting image would be the sum of two ring models, one with $\eta = \chi$ and the other with $\eta = \chi + \pi$. Both cases would reduce the image polarization substantially and may give better agreement with the M87* image, but we defer a full analysis to a future paper.

\subsection{Comparison to Sgr~A* Polarization}

The polarization of Sgr~A* shows continuous variability in the submillimeter \citep{Marrone_2006,Johnson_2015,Bower_2018} and also shows rapid variability during near-infrared (NIR) flares \citep{Eckart_2006,Trippe_2007,Zamaninasab_2010,Gravity_2018}. The variability often appears as ``loops'' in Stokes $Q$-$U$, and is frequently attributed to localized emission from an orbiting ``hotspot'' \citep{Broderick_Loeb_2005,Broderick_Loeb_2006,Fish_2009}. For the case of NIR flares, Faraday effects, absorption, and background emission are insignificant, so we can directly compare observed values of polarization and centroid motion with a simulated hotspot-only model.

\begin{figure}[t]
    \centering
    \includegraphics[width=\columnwidth]{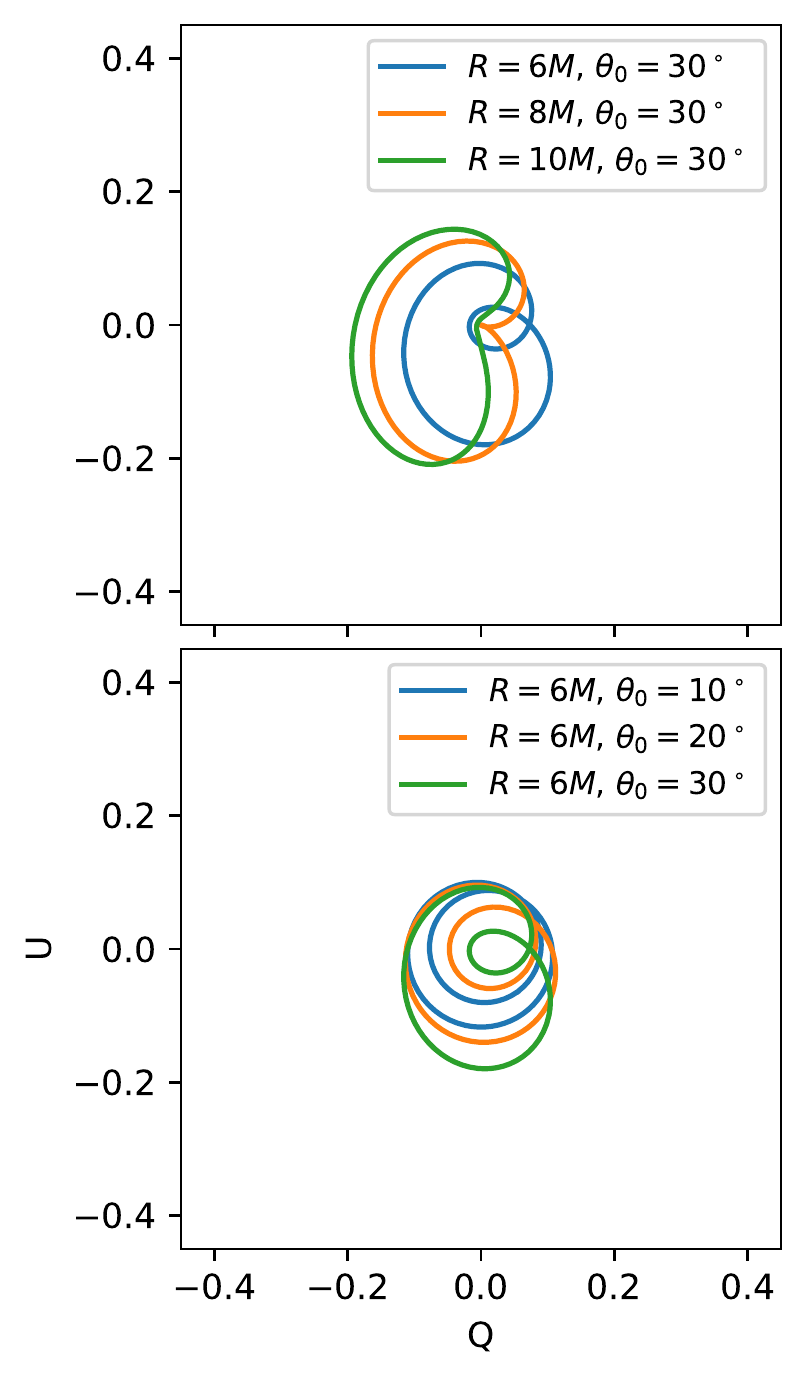}
    \caption{Polarization signatures for a vertically magnetized hotspot on a circular, relativistic Keplerian orbit. Each curve shows the \replaced{polarization}{polarized flux} for a full orbit. Different curves correspond to varying the hotspot radius (top) and viewing inclination (bottom). 
    \added{Note that we use radio astronomy conventions for $Q$ and $U$ here, distinct from those in \autoref{eq:QU_Ealphabeta} by an overall sign.}
    }
    \label{fig:gravity_loops}
\end{figure}

\autoref{fig:gravity_loops} shows a representative example. In this figure, we compute the hotspot \replaced{polarization}{polarized flux in the} $(Q,U)$ plane over a full period for a set of orbits with varying emission radius and inclination. We hold the underlying magnetic field structure to be vertical and constant, and adopt a relativistic Keplerian velocity for the hotspot: $\beta = 1/\sqrt{r-2}$. 
Our results are similar to previous studies with fully numerical calculations \citep[see, e.g.][]{Fish_2009, Gravity_2018_orbit, Gravity_2020}; lensing and aberration compress the image of azimuthal evolution of polarization on one side of the flow and expand it on the other. In the formalism of azimuthal Fourier modes on the ring \citep{PWP_2020}, power is shifted from the $m=2$ mode to the $m=1$ mode.

\section{Summary}
\label{sec:summary}

We have developed an analytical method for computing the polarized image of a synchrotron-emitting fluid ring orbiting a Schwarzchild black hole. Given simple assumptions for the magnetic field geometry and fluid velocity, this model allows us to generate predictions of EVPA and relative polarized intensity as a polar function in the observed image at arbitrary viewing inclination. We explored the main features of the model through a number of representative examples and by further expansion in the inverse emission radius (lensing), fluid velocity (Doppler and aberration), and observer inclination (ring tilt). These reveal how the various physical effects influence the polarized image. 

\added{In its simplest form, the fractional polarization of our model is significantly higher than that seen in EHT images of M87* \citepalias{PaperVII}. This may indicate significant sub-beam depolarization, potentially from strong internal Faraday effects \citepalias{PaperVIII}. If so, observations at higher frequencies, where Faraday effects are suppressed, may show significantly higher image polarizations, while observations at lower frequencies are expected to show a heavily depolarized ``core.''}

Our polarized ring model provides intuition and insights about how a black hole's accretion flow and spacetime combine to produce a polarized image. It also provides a pathway to constrain these physical properties through direct comparisons with data and images from the EHT, GRAVITY, and future X-ray polarimetry studies. Extensions such as non-axisymmetric structure and non-equatorial emission will provide an expanded class of geometrical models to complement the growing library of GRMHD simulations \citepalias{PaperV}. The inclusion of black hole spin will be necessary for rigorous understanding of M87* polarization, particularly if emission at small radii is significant. Further studies which examine the capability of the model in matching snapshots of GRMHD simulations with similar magnetic field and flow conditions will elucidate how readily field geometries may be directly inferred from polarized images.

\acknowledgments{We thank the National Science Foundation (awards OISE-1743747, AST-1816420, AST-1716536, AST-1440254, AST-1935980) and the Gordon and Betty Moore Foundation (GBMF-5278) for financial support of this work. This work was supported in part by the Black Hole Initiative, which is funded by grants from the John Templeton Foundation and the Gordon and Betty Moore Foundation to Harvard University. \added{Support for this work was also provided by the NASA Hubble Fellowship grant
HST-HF2-51431.001-A awarded by the Space Telescope
Science Institute, which is operated by the Association
of Universities for Research in Astronomy, Inc.,
for NASA, under contract NAS5-26555.}

\replaced{The authors of the present paper thank}{The Event Horizon Telescope Collaboration thanks} the following
organizations and programs: the Academy
of Finland (projects 274477, 284495, 312496\added{, 315721}); \deleted{the
Advanced European Network of E-infrastructures for
Astronomy with the SKA (AENEAS) project, supported
by the European Commission Framework Programme
Horizon 2020 Research and Innovation action
under grant agreement 731016;} \added{the Agencia Nacional de Investigación y Desarrollo (ANID), Chile via NCN$19\_058$ (TITANs) and Fondecyt 3190878}, the Alexander
von Humboldt Stiftung; an Alfred P. Sloan Research Fellowship;
Allegro, the European ALMA Regional Centre node in the Netherlands, the NL astronomy
research network NOVA and the astronomy institutes of the University of Amsterdam, Leiden University and Radboud University;
the Black Hole Initiative at
Harvard University, through a grant (60477) from
the John Templeton Foundation; the China Scholarship
Council; \deleted{Comisión Nacional de Investigación
Científica y Tecnológica (CONICYT, Chile, via PIA
ACT172033, Fondecyt projects 1171506 and 3190878,
BASAL AFB-170002, ALMA-conicyt 31140007);} Consejo
Nacional de Ciencia y Tecnolog\'{\i}a (CONACYT,
Mexico, projects  U0004-246083, U0004-259839, F0003-272050, M0037-279006, F0003-281692,
104497, 275201, 263356);
the Delaney Family via the Delaney Family John A.
Wheeler Chair at Perimeter Institute; Dirección General
de Asuntos del Personal Académico-—Universidad
Nacional Autónoma de México (DGAPA-—UNAM,
projects IN112417 and IN112820); the European Research Council Synergy
Grant "BlackHoleCam: Imaging the Event Horizon
of Black Holes" (grant 610058); the Generalitat
Valenciana postdoctoral grant APOSTD/2018/177 and
GenT Program (project CIDEGENT/2018/021); MICINN Research Project PID2019-108995GB-C22;
the
Gordon and Betty Moore Foundation \replaced{(grants GBMF-
3561, GBMF-5278)}{(grant GBMF-3561}; the Istituto Nazionale di Fisica
Nucleare (INFN) sezione di Napoli, iniziative specifiche
TEONGRAV; the International Max Planck Research
School for Astronomy and Astrophysics at the
Universities of Bonn and Cologne; \deleted{the Jansky Fellowship
program of the National Radio Astronomy Observatory
(NRAO);}
Joint Princeton/Flatiron and Joint Columbia/Flatiron Postdoctoral Fellowships, research at the Flatiron Institute is supported by the Simons Foundation; 
the Japanese Government (Monbukagakusho:
MEXT) Scholarship; the Japan Society for
the Promotion of Science (JSPS) Grant-in-Aid for JSPS
Research Fellowship (JP17J08829); the Key Research
Program of Frontier Sciences, Chinese Academy of
Sciences (CAS, grants QYZDJ-SSW-SLH057, QYZDJSSW-
SYS008, ZDBS-LY-SLH011); the Leverhulme Trust Early Career Research
Fellowship; the Max-Planck-Gesellschaft (MPG);
the Max Planck Partner Group of the MPG and the
CAS; the MEXT/JSPS KAKENHI (grants 18KK0090,
JP18K13594, JP18K03656, JP18H03721, 18K03709,
18H01245, 25120007); the Malaysian Fundamental Research Grant Scheme (FRGS)\\ FRGS/1/2019/STG02/UM/02/6; the MIT International Science
and Technology Initiatives (MISTI) Funds; the Ministry
of Science and Technology (MOST) of Taiwan (105-
2112-M-001-025-MY3, 106-2112-M-001-011, 106-2119-
M-001-027, 107-2119-M-001-017, 107-2119-M-001-020,
\deleted{and }107-2119-M-110-005\added{, 108-2112-M-001-048, and 109-2124-M-001-005}); the National Aeronautics and
Space Administration (NASA grant NNX17AL82G, Fermi Guest Investigator
grant \replaced{80NSSC17K0649}{80NSSC20K1567}, NASA Astrophysics Theory Program grant 80NSSC20K0527 \deleted{, and Hubble Fellowship grant
HST-HF2-51431.001-A awarded by the Space Telescope
Science Institute, which is operated by the Association
of Universities for Research in Astronomy, Inc.,
for NASA, under contract NAS5-26555,}, \added{NASA NuSTAR award 80NSSC20K0645}); the National
Institute of Natural Sciences (NINS) of Japan; the National
Key Research and Development Program of China
(grant 2016YFA0400704, 2016YFA0400702); the National
Science Foundation (NSF, grants AST-0096454,
AST-0352953, AST-0521233, AST-0705062, AST-0905844, AST-0922984, AST-1126433, AST-1140030,
DGE-1144085, AST-1207704, AST-1207730, AST-1207752, MRI-1228509, OPP-1248097, AST-1310896, \deleted{AST-1337663, AST-1440254,} AST-1555365, AST-1615796, AST-1715061, AST-1716327, \deleted{AST-1716536,
OISE-1743747, AST-1816420,} AST-1903847\deleted{,
AST-1935980}, \added{AST-2034306}); the Natural Science
Foundation of China (grants 11573051, 11633006,
11650110427, 10625314, 11721303, 11725312, 11933007, 11991052\added{, 11991053}); a fellowship of China Postdoctoral Science Foundation (2020M671266); the Natural
Sciences and Engineering Research Council of
Canada (NSERC, including a Discovery Grant and
the NSERC Alexander Graham Bell Canada Graduate
Scholarships-Doctoral Program); the National Research
Foundation of Korea (the Global PhD Fellowship
Grant: grants NRF-2015H1A2A1033752, 2015-
R1D1A1A01056807, the Korea Research Fellowship Program:
NRF-2015H1D3A1066561, Basic Research Support Grant 2019R1F1A1059721); the Netherlands Organization
for Scientific Research (NWO) VICI award
(grant 639.043.513) and Spinoza Prize SPI 78-409; the New Scientific Frontiers with Precision Radio Interferometry Fellowship awarded by the South African Radio Astronomy Observatory (SARAO), which is a facility of the National Research Foundation (NRF), an agency of the Department of Science and Innovation (DSI) of South Africa; the South African Research Chairs Initiative of the Department of Science and Innovation and National Research Foundation; the Onsala Space Observatory
(OSO) national infrastructure, for the provisioning
of its facilities/observational support (OSO receives
funding through the Swedish Research Council under
grant 2017-00648) the Perimeter Institute for Theoretical
Physics (research at Perimeter Institute is supported
by the Government of Canada through the Department
of Innovation, Science and Economic Development
and by the Province of Ontario through the
Ministry of Research, Innovation and Science); \deleted{the Russian
Science Foundation (grant 17-12-01029);} the Spanish
Ministerio de Economía y Competitividad (grants
\replaced{AYA2015-63939-C2-1-P}{PGC2018-098915-B-C21}, AYA2016-80889-P, PID2019-108995GB-C21); the State
Agency for Research of the Spanish MCIU through
the "Center of Excellence Severo Ochoa" award for
the Instituto de Astrofísica de Andalucía (SEV-2017-
0709); the Toray Science Foundation; the Consejería de Economía, Conocimiento, Empresas y Universidad of the Junta de Andalucía (grant P18-FR-1769), the Consejo Superior de Investigaciones Científicas (grant 2019AEP112);
the US Department
of Energy (USDOE) through the Los Alamos National
Laboratory (operated by Triad National Security,
LLC, for the National Nuclear Security Administration
of the USDOE (Contract 89233218CNA000001);
\deleted{the Italian Ministero dell’Istruzione Università e Ricerca
through the grant Progetti Premiali 2012-iALMA (CUP
C52I13000140001);} the European Union’s Horizon 2020
research and innovation programme under grant agreement
No 730562 RadioNet; ALMA North America Development
Fund; the Academia Sinica; Chandra \added{DD7-18089X and}TM6-
17006X; the GenT Program (Generalitat Valenciana)
Project CIDEGENT/2018/021. This work used the
Extreme Science and Engineering Discovery Environment
(XSEDE), supported by NSF grant ACI-1548562,
and CyVerse, supported by NSF grants DBI-0735191,
DBI-1265383, and DBI-1743442. XSEDE Stampede2 resource
at TACC was allocated through TG-AST170024
and TG-AST080026N. XSEDE JetStream resource at
PTI and TACC was allocated through AST170028.
The simulations were performed in part on the SuperMUC
cluster at the LRZ in Garching, on the
LOEWE cluster in CSC in Frankfurt, and on the
HazelHen cluster at the HLRS in Stuttgart. This
research was enabled in part by support provided
by Compute Ontario (http://computeontario.ca), Calcul
Quebec (http://www.calculquebec.ca) and Compute
Canada (http://www.computecanada.ca). We thank
the staff at the participating observatories, correlation
centers, and institutions for their enthusiastic support.
This paper makes use of the following ALMA data:
ADS/JAO.ALMA\#2016.1.01154.V. ALMA is a partnership
of the European Southern Observatory (ESO;
Europe, representing its member states), NSF, and
National Institutes of Natural Sciences of Japan, together
with National Research Council (Canada), Ministry
of Science and Technology (MOST; Taiwan),
Academia Sinica Institute of Astronomy and Astrophysics
(ASIAA; Taiwan), and Korea Astronomy and
Space Science Institute (KASI; Republic of Korea), in
cooperation with the Republic of Chile. The Joint
ALMA Observatory is operated by ESO, Associated
Universities, Inc. (AUI)/NRAO, and the National Astronomical
Observatory of Japan (NAOJ). The NRAO
is a facility of the NSF operated under cooperative agreement
by AUI. APEX is a collaboration between the
Max-Planck-Institut f{\"u}r Radioastronomie (Germany),
ESO, and the Onsala Space Observatory (Sweden). The
SMA is a joint project between the SAO and ASIAA
and is funded by the Smithsonian Institution and the
Academia Sinica. The JCMT is operated by the East
Asian Observatory on behalf of the NAOJ, ASIAA, and
KASI, as well as the Ministry of Finance of China, Chinese
Academy of Sciences, and the National Key R\&D
Program (No. 2017YFA0402700) of China. Additional
funding support for the JCMT is provided by the Science
and Technologies Facility Council (UK) and participating
universities in the UK and Canada. The LMT is a project operated by the Instituto Nacional de Astrofísica, Óptica, y Electrónica (Mexico) and the University of Massachusetts at Amherst (USA), with financial support from the Consejo Nacional de Ciencia y Tecnología and the National Science Foundation. The
IRAM 30-m telescope on Pico Veleta, Spain is operated
by IRAM and supported by CNRS (Centre National de
la Recherche Scientifique, France), MPG (Max-Planck-
Gesellschaft, Germany) and IGN (Instituto Geográfico
Nacional, Spain). The SMT is operated by the Arizona
Radio Observatory, a part of the Steward Observatory
of the University of Arizona, with financial support of
operations from the State of Arizona and financial support
for instrumentation development from the NSF.
The SPT is supported by the National Science Foundation
through grant PLR- 1248097. Partial support is
also provided by the NSF Physics Frontier Center grant
PHY-1125897 to the Kavli Institute of Cosmological
Physics at the University of Chicago, the Kavli Foundation
and the Gordon and Betty Moore Foundation grant
GBMF 947. The SPT hydrogen maser was provided on
loan from the GLT, courtesy of ASIAA. The EHTC has
received generous donations of FPGA chips from Xilinx
Inc., under the Xilinx University Program. The EHTC
has benefited from technology shared under open-source
license by the Collaboration for Astronomy Signal Processing
and Electronics Research (CASPER). The EHT
project is grateful to T4Science and Microsemi for their
assistance with Hydrogen Masers. This research has
made use of NASA’s Astrophysics Data System. We
gratefully acknowledge the support provided by the extended
staff of the ALMA, both from the inception of
the ALMA Phasing Project through the observational
campaigns of 2017 and 2018. We would like to thank
A. Deller and W. Brisken for EHT-specific support with
the use of DiFX. We acknowledge the significance that
Maunakea, where the SMA and JCMT EHT stations
are located, has for the indigenous Hawaiian people.}

\appendix
\numberwithin{equation}{section}
\section{Accuracy of The Beloborodov Approximation}
\label{sec:errorAnalysis}

The model developed in \autoref{sec:model} relies on the approximate formula \autoref{belo} derived by \cite{Beloborodov_2002}. This approximation provides an estimate for $\alpha$ (and, equivalently, for $\rho$; \autoref{rhosqr}) for given emission coordinates $R$ and $\phi$.  We now quantify the accuracy of this approximation.

\begin{figure*}[t]
    \centering
    \includegraphics[width=1.0\textwidth]{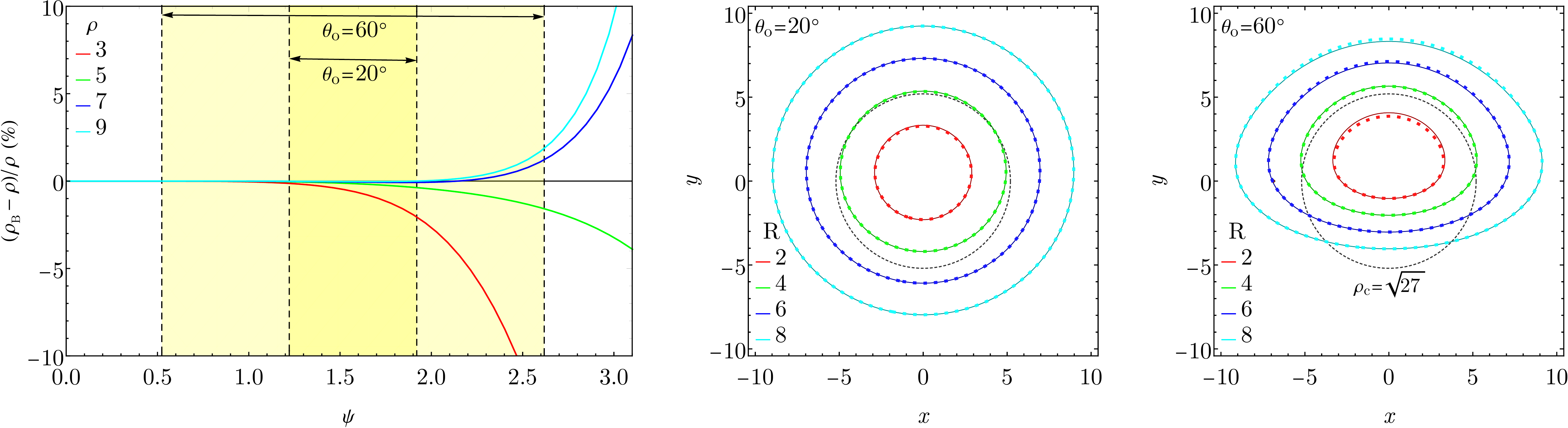}
    \caption{Testing the accuracy of the Beloborodov approximation. The left panel shows fractional error in $\rho=\sqrt{x^2+y^2}$ as a function of $\psi$ for $\rho = 3$, 5, 7, and 9. Yellow ranges denote values of $\psi$ relevant for observer inclinations $\theta_{\rm o}=20^\circ$ and $60^\circ$. The center and right panels show the image coordinates for rings with emission radius $R = 2$ (red), 4 (green), 6 (blue), and 8 (cyan) viewed at inclinations of $20^\circ$ and $60^\circ$, respectively. For each ring, the solid line shows the exact calculation, while the dotted line shows the Beloborodov approximation (see \autoref{rhosqr}). The black dotted line shows the critical curve, $\rho = \rho_{\rm c} \equiv \sqrt{27}$. 
    }
    \label{fig:Beloborodov_Fractional_Error}
\end{figure*}

Emission from the equatorial plane arriving at a given observer inclination angle $0\leq\theta_{\rm o}\leq \pi/2$ will sweep through $\psi \in \left[\pi/2 \pm \theta_{\rm o}\right]$ as the azimuthal angle $\phi$ varies (see \autoref{eq:psi}). 
In particular, all emission from a face-on disk has $\psi = \pi/2$, while emission from an edge-on disk samples angles $0\leq\psi\leq\pi$. As the left panel in \autoref{fig:Beloborodov_Fractional_Error} shows, the error in the Beloborodov approximation increases with $\psi$. In the context of the ring model, the approximation is most accurate at small inclinations. For $\theta_{\rm o} = 17^\circ$ for example (relevant for M87*), the approximation for $\rho$ has a fractional error smaller than $2\%$ for all values of $R$. 
This error decreases rapidly as $\rho$ grows; e.g., for $\rho=9$, the fractional error in $\rho$ is smaller than $0.03\%$. In general, for emission on the side of the accretion disk closer to the observer (i.e., $\pi<\phi<2\pi$, $\psi< \pi/2$), the approximation for $\rho$ will have fractional error smaller than $0.6\%$ for all $\rho\geq 3$ and any inclination. The error is larger for points on the far side of the ring ($0<\phi<\pi$, $\psi>\pi/2$). However, even at an inclination angle of $60^\circ$, the accuracy is quite adequate, as shown by the right panel in \autoref{fig:Beloborodov_Fractional_Error}.

\section{Transformations of Field Components}\label{sec:transformations}

In the analysis given in the main text, we assumed that the magnetic field components $B_r$, $B_\phi$, $B_z$ are specified in the fluid frame. Under the usual assumptions of ideal MHD, the electric field vanishes in this frame: $E_r=E_\phi=E_z=0$. Alternatively, we might wish to work with field components in the P-frame: $B_r^{\rm (P)}$, $B_\phi^{\rm (P)}$, $B_z^{\rm (P)}$, $E_r^{\rm (P)}$, $E_\phi^{\rm (P)}$, $E_z^{\rm (P)}$ (the electric field does not vanish in this frame). 

The two frames are related by a Lorentz transformation with velocity $\vec{\beta}$ (expressed in terms of $\beta$ and $\chi$, see eq \ref{velocity2}). The transformation is most transparent when we rewrite the radial and tangential field components in terms of ``parallel" and ``perpendicular" field components relative to the velocity:
\begin{equation}
    B_\parallel^{\rm (P)} = \cos\chi\, B_r^{\rm (P)}+\sin\chi\, B_\phi^{\rm (P)}, \qquad B_\perp^{\rm (P)} = -\sin\chi\, B_r^{\rm (P)}+\cos\chi\, B_\phi^{\rm (P)},
    \end{equation}
    \begin{equation}
    B_r^{\rm (P)} = \cos\chi\,B_\parallel^{\rm (P)} -\sin\chi\,B_\perp^{\rm (P)}, \qquad B_\phi^{\rm (P)}=\sin\chi\,B_\parallel^{\rm (P)} + \cos\chi\,B_\perp^{\rm (P)},
\end{equation}
with similar expressions for $\vec{E}^{\rm (P)}$ and $\vec{B}$. The transformation rules are then
\begin{equation}
    B_\parallel = B_\parallel^{\rm (P)}, \qquad E_\parallel = E_\parallel^{\rm (P)},
\end{equation}
\begin{equation}
    B_\perp = \gamma\, B_\perp^{\rm (P)}+\beta\gamma\, E_z^{\rm (P)}, \qquad B_z=\gamma\, B_z^{\rm (P)}-\beta\gamma\, E_\perp^{\rm (P)},
\end{equation}
\begin{equation}
    B_\perp^{\rm (P)} = \gamma\, B_\perp - \beta\gamma\, E_z, \qquad B_z^{\rm (P)}=\gamma\, B_z + \beta\gamma\, E_\perp,
\end{equation}
\begin{equation}
    E_\perp = \gamma\, E_\perp^{\rm (P)}-\beta\gamma\, B_z^{\rm (P)}, \qquad E_z=\gamma\, E_z^{\rm (P)}+\beta\gamma\, B_\perp^{\rm (P)},
\end{equation}
\begin{equation}
    E_\perp^{\rm (P)} = \gamma\, E_\perp + \beta\gamma\, B_z, \qquad E_z^{\rm (P)}=\gamma\, E_z - \beta\gamma\, B_\perp,
\end{equation}
where, as usual, $\gamma=(1-\beta^2)^{-1/2}$.

Using the above transformations, if we are given $B_r$, $B_\phi$, $B_z$ in the fluid frame, we can solve for $\vec{B}^{\rm (P)}$ and $\vec{E}^{\rm (P)}$ in the P-frame:
\begin{eqnarray}
    B_r^{\rm (P)}&=&(\cos^2\chi+\gamma\sin^2\chi)\,B_r-(\gamma-1)\cos\chi \sin\chi\, B_\phi, \\
    B_\phi^{\rm (P)} &=& -(\gamma-1)\cos\chi \sin\chi\, B_r+(\sin^2\chi+\gamma\cos^2\chi)\,B_\phi, \\
    B_z^{\rm (P)} &=&\gamma\, B_z, \\
    E_r^{\rm (P)} &=& -\beta\gamma\sin\chi\, B_z, \\
    E_\phi^{\rm (P)} &=& \beta \gamma \cos\chi\, B_z, \\
    E_z^{\rm (P)} &=& \beta\gamma\sin\chi\, B_r-\beta\gamma\cos\chi\, B_\phi.
\end{eqnarray}
Similarly, if we are given the magnetic field components in the P-frame, we can solve for the other field components:
\begin{eqnarray}
B_r &=& [\cos^2\chi+(1/\gamma)\sin^2\chi]B_r^{\rm (P)} +((\gamma-1)/\gamma)\cos\chi\sin\chi B_\phi^{\rm (P)}, \\
B_\phi &=& ((\gamma-1)/\gamma)\cos\chi\sin\chi B_r^{\rm (P)}+[\sin^2\chi+(1/\gamma)\cos^2\chi]B_\phi^{\rm (P)}, \\
B_z &=& (1/\gamma) B_z^{\rm (P)}, \\
E_r^{\rm (P)} &=& -\beta\sin\chi B_z^{\rm (P)}, \\
E_\phi^{\rm (P)} &=& \beta\cos\chi B_z^{\rm (P)}, \\
E_z^{\rm (P)} &=& \beta\sin\chi B_r^{\rm (P)} - \beta\cos\chi B_\phi^{\rm (P)}.
\end{eqnarray}
These transformations are provided here for the convenience of readers who might prefer to work with field components in the Schwarzschild frame. 

\section{EMISSION LOCATION VS. OBSERVED COORDINATES}\label{sec:mapping}

The radiation emitted by the point P in the ring at $(R,\phi)$ reaches the observer at sky coordinates $(x,y)$, which we can write in terms of polar coordinates $(\rho,\varphi)$  as described in equation~(\ref{rho_varphi}). Here we work out the relation between these two coordinates. 

The relation between $\varphi$ and $\phi$ is straightforward. Since
the observer frame is tilted with respect to the ring plane by a rotation angle
$\theta_{\rm o}$ around the line of nodes, and since the geodesic lies
entirely on a plane (because we have limited our analysis to the Schwarzschild spacetime), we find
\begin{equation}
\tan\varphi = \tan\phi\cos\theta_{\rm o}.
\end{equation}
This relation can be used to translate $\phi$ to $\varphi$ and vice
versa. For the analysis in Appendix~\ref{sec:series}, it is useful to express $\varphi$ in terms of $\phi$ up to quadratic order. The corresponding relations are
\begin{equation}
    \sin\phi \to \sin\varphi +(1/2)\sin^2\theta_{\rm o}\sin\varphi\cos^2\varphi, \quad \cos\phi \to \cos\varphi - (1/2)\sin^2\theta_{\rm o}\cos\varphi\sin^2\varphi.
    \label{phitovarphi}
\end{equation}

To calculate the mapping between $R$ and $\rho$, consider the G-frame (Fig.~\ref{fig:gframe}), where the geodesic lies
in the $xz$-plane. At the emission point $(x,y,z)=(R,0,0)$, the geodesic
makes an angle $\alpha$ with respect to the $x$-axis, where
$\alpha$ is given by the Beloborodov approximation (\ref{belo}).  Since the
angular momentum around the $y$-axis in the G-frame is conserved, we have
\begin{equation}
\rho = k_\phi = Rk^{\hat{\phi}} =
\frac{R\sin\alpha}{\left(1-\frac{2}{R}\right)^{1/2}}.
\end{equation}
Squaring both sides,
\begin{equation}
\rho^2 = \frac{R^2 (1-\cos^2\alpha)}{\left(1-\frac{2}{R}\right)}
=R^2(1-\sin^2\theta_{\rm o}\sin^2\phi)
+2R\,(1+\sin^2\theta_{\rm o}\sin^2\phi+2\sin\theta_{\rm o}\sin\phi).
\label{rhosqr}
\end{equation}
This directly gives $\rho$ in terms of $R$ and $\phi$; conversely, the quadratic
equation can be solved to obtain $R$ for a given $\rho$ and
$\phi$. Equation (\ref{rhosqr}) is exact, except for the fact that we used the
Beloborodov approximation (\ref{belo}) for $\cos\alpha$.

Since $(\partial\varphi/\partial R)_{\phi}=0$, the Jacobian determinant $|J|$, which describes the transformation of differential area elements between $(R,\phi)$ and $(\rho,\varphi)$, is given by
\begin{equation}
|J| = \left(\frac{\partial\rho}{\partial R}\right)_{\phi}  \left(\frac{\rho\partial\varphi}{R\partial\phi}\right)_{R} = \frac{1}{R}\left[(R+1) - (R-1)\sin^2\theta_{\rm o}\sin^2\phi +2\sin\theta_{\rm o}\sin\phi\right] \left(\frac{\sec^2\phi\cos\theta_{\rm o}}{1+\tan^2\phi\cos^2\theta_{\rm o}}\right).
\end{equation}

\section{Series Expansion to Quadratic Order}\label{sec:series}

The analysis in sec.~\ref{sec:model} is exact, modulo the Beloborodov approximaton, and is convenient for numerical calculations. However, for analytical studies, we need simpler relations. For this, we expand all the equations up to second order, treating the quantities $\sin\theta_{\rm o}$, $\beta$ and $2/R$, which describe tilt, relativistic velocity and gravity, as being small.\footnote{Because the solution for the coordinate $x$ involves a division by $\sin\theta_{\rm o}$, it is necessary to keep terms up to $\sin^3\theta_{\rm o}$ in the expressions leading up to this quantity.} The relevant series expansion results are given below. In each equation, the second-order terms are shown inside square brackets. 

The observed coordinates $(x,y)$ of the geodesic emitted at location $(R,\phi)$ in the ring are given by
\begin{eqnarray}
x &=& (R+1)\cos\varphi + \left[ -\frac{1}{2R}\cos\varphi + 2\sin\theta_{\rm o}\sin \varphi \cos\varphi -\frac{R}{2}\sin^2\theta_{\rm o}\sin^2\varphi \cos\varphi \right], \\ 
y &=& (R+1)\sin\varphi + \left[ -\frac{1}{2R}\sin\varphi +2\sin\theta_{\rm o}\sin^2\varphi -\frac{R}{2}\sin^2\theta_{\rm o}\sin^3\varphi \right].
\end{eqnarray}
In deriving these results, we first evaluated equation~(\ref{alphabeta}) and then made the substitutions given in equation~(\ref{phitovarphi}). The latter substitution is made in all the subsequent results presented in this Appendix; thus the results are expressed in terms of the observed azimuthal angle $\varphi$.

To quadratic order, the Doppler factor $\delta$ is
\begin{equation}
\delta = 1-\frac{1}{R} -\left[\frac{\beta^2}{2} + \frac{1}{2R^2} - \frac{2\beta}{R}\cos\chi + \beta\sin\theta_{\rm o}\sin(\chi+\varphi)\right]. \label{delta2}
\end{equation}
Note that Doppler boost due to azimuthal velocity is described by the last term, $\beta\sin\theta_{\rm o}\sin(\chi+\varphi)$, which appears only at second order in the small quantities $\sin\theta_{\rm o}$ and $\beta$. This is one of the reasons for expanding the equations to quadratic order.

Assuming that the spectral index $\alpha_\nu=1$, the intensity of the linear polarized radiation at the observer is given by equation~(\ref{absP}):
\begin{equation}
  |P| = \delta^4 \,l_{\rm p}\, |\vec{B}|^2\sin^2\zeta.
  \label{absP2}
\end{equation}
Expanding to quadratic order, the term $|\vec{B}|^2\sin^2\zeta$ is given by
\begin{eqnarray}
|\vec{B}|^2\sin^2\zeta &=& B_{\rm eq}^2 + \left(2\sin\theta_{\rm o}\sin(\eta+\varphi)-\frac{4}{R}\cos\eta+2\beta\cos(\chi-\eta)\right)B_{\rm eq}B_z \nonumber \\
&~~& +\left[-\left(\sin\theta_{\rm o}\sin(\eta+\varphi)-\frac{2}{R}\cos\eta+\beta\cos(\chi-\eta)\right)^2B_{\rm eq}^2 \right. \nonumber \\
&~~& ~~+\left(-\frac{4}{R}\sin\theta_{\rm o}\sin\varphi + \frac{4}{R^2} + \sin^2\theta_{\rm o} + 2\beta\sin\theta_{\rm o}\sin(\chi+\varphi) - \frac{4\beta}{R}\cos\chi + \beta^2\right)B_z^2 \nonumber \\
&~~& ~~\left. -\frac{4}{R}\sin\theta_{\rm o}\cos\eta\sin\varphi B_{\rm eq}B_z\right].
\end{eqnarray}
We have written the result in terms of the parameters $B_{\rm eq}$, $\eta$, $B_z$ of the magnetic field in the fluid frame (see eq \ref{field_comps}). This is helpful for the discussion in sec.~\ref{sec:analytic}. Note that, in the absence of any equatorial magnetic field, the only contributions are at the second order (because the only terms with $B_z^2$ are inside the square brackets). Since the observed intensity is directly proportional to $|\vec{B}|^2\sin^2\zeta$, we need to expand to quadratic order to handle models with  pure $B_z$.

To quadratic order, the path length $l_{\rm p}$ in equation (\ref{eq:lp}) is
\begin{equation}
\frac{l_{\rm p}}{H} = 1 + \frac{1}{2}\left[\beta^2+\frac{4}{R^2}+\sin^2\theta_{\rm o} +2\beta\sin\theta_{\rm o}\sin(\chi+\varphi) - \frac{4\beta}{R}\cos\chi -\frac{4}{R}\sin\theta_{\rm o}\sin\varphi \right].
\end{equation}

We calculate the linear polarized intensity $|P|$ as the product of the three terms, $\delta^4$, $l_{\rm p}$ and $|\vec{B}|^2\sin^2\zeta$ (see equation~\ref{absP2}). This gives
\begin{eqnarray}
|P(\varphi)| &=& \left(1-\frac{4}{R}\right)\,\left(B_r^2+B_\phi^2\right) + 2\left(\sin\theta_{\rm o}\cos\varphi + \beta\sin\chi\right)B_\phi B_z +
2\left(- \frac{2}{R}+\beta\cos\chi+\sin\theta_{\rm o}\sin\varphi  \right)B_z B_r \nonumber \\
&~~& ~+\left[ \left( \frac{2}{R}\sin\theta_{\rm o}\sin\varphi+\frac{2}{r^2}+\frac{1}{2}\sin^2\theta_{\rm o}\cos 2\varphi +\frac{10\beta}{R}\cos\chi +\beta\sin\theta_{\rm o}\left(\sin(\chi-\varphi)-4\sin(\chi+\varphi)\right) -\frac{\beta^2}{2}(4+\cos2\chi)  \right)\,B_r^2 \right. \nonumber \\
&~~& \quad +
\left( -\frac{2}{r}\sin\theta_{\rm o} \sin\varphi+ \frac{6}{R^2} -\frac{1}{2}\sin^2\theta_{\rm o}\cos 2\varphi -\beta\sin\theta_{\rm o}\left(4 \sin(\chi+\varphi)+\sin(\chi-\varphi)\right) +\frac{6\beta}{R}\cos\chi-\frac{\beta^2}{2}(4-\cos2\chi)\right)\,B_\phi^2 \nonumber \\
&~~& \quad +\left(  -\frac{4}{R}\sin\theta_{\rm o}\sin\varphi+\frac{4}{R^2} +\sin^2\theta_{\rm o} +2\beta\sin\theta_{\rm o}\sin(\chi+\varphi) -\frac{4\beta}{R}\cos\chi +\beta^2 \right)\, B_z^2 \nonumber \\
&~~& \quad + \left(  \frac{4}{R}\sin\theta_{\rm o}\cos\varphi -\sin^2\theta_{\rm o}\sin 2\varphi-2\beta\sin\theta_{\rm o}\cos(\chi-\varphi)  +\frac{4\beta}{R}\sin\chi -\beta^2\sin2\chi\right)\, B_rB_\phi \nonumber \\
&~~& \quad\left. +\left(  -\frac{8}{R}\sin\theta_{\rm o}\cos\varphi -\frac{8\beta}{R}\sin\chi \right)\,B_\phi B_z + \left(-\frac{12}{R}\sin\theta_{\rm o}\sin\varphi+\frac{16}{R^2}-\frac{8\beta}{R}\cos\chi  \right)\,B_z B_r\right],
\label{absP3}
\end{eqnarray}
where we have written the answer in terms of $B_r$, $B_\phi$, $B_z$ in the fluid frame.

The electric field components $E_x$, $E_y$, which are normalized such that they are proportional to $\sin\zeta\,|\vec{B}|$ (see eq~\ref{Enormzeta}), are
\begin{eqnarray}
E_{x} &=& -\sin\varphi\, B_r -
\cos\varphi\, B_\phi -\left(\sin\theta_{\rm o} -\frac{2}{R}\sin\varphi
+\beta\sin(\chi+\varphi)\right) B_z \nonumber \\
&~~& ~+\left[ \left( -\frac{2}{R}\sin\theta_{\rm o}\sin^2\varphi+\frac{2}{R^2}\sin\varphi+ \frac{1}{2}\sin^2\theta_{\rm o}\sin^3\varphi    +\frac{\beta}{2}\sin\theta_{\rm o}(\cos\chi-\cos(\chi+2\varphi)) \right.\right.\nonumber \\
&~~&\qquad \left. -\frac{2\beta}{R}\sin(\chi+\varphi) +\frac{\beta^2}{4}\left(\sin\varphi+\sin(2\chi+\varphi)\right) \right)\,B_r  \nonumber \\
&~~& \quad+\left( -\frac{1}{R}\sin\theta_{\rm o}\sin 2\varphi +\frac{1}{8}\sin^2\theta_{\rm o}(5\cos\varphi - \cos3\varphi) +\frac{\beta}{2}\sin\theta_{\rm o}(\sin\chi+\sin(\chi+2\varphi)) \right.\nonumber\\
&~~&\qquad \left.\frac{\beta^2}{4}(\cos\varphi-\cos(2\chi+\varphi))     \right)B_\phi \nonumber \\
&~~& \quad \left.+\frac{2}{R}\sin\theta_{\rm o}\sin^2\varphi \,B_z\right], \\
E_{y} &=& \cos\varphi\, B_r -
\sin\varphi\, B_\phi +\left( -\frac{2}{R}\cos\varphi
+\beta\cos(\chi+\varphi)\right) B_z \nonumber \\
&~~& ~+\left[ \left( \frac{1}{R}\sin\theta_{\rm o}\sin 2\varphi-\frac{2}{R^2}\cos\varphi- \frac{1}{8}\sin^2\theta_{\rm o}(\cos\varphi-\cos3\varphi)    +\frac{\beta}{2}\sin\theta_{\rm o}(\sin\chi-\sin(\chi+2\varphi)) \right. \right.\nonumber \\
&~~&\qquad \left.+\frac{2\beta}{R}\cos(\chi+\varphi) -\frac{\beta^2}{4}\left(\cos\varphi+\cos(2\chi+\varphi)\right) \right)\,B_r  \nonumber \\
&~~& \quad+\left( \frac{2}{R}\sin\theta_{\rm o}\cos^2\varphi -\frac{1}{8}\sin^2\theta_{\rm o}(\sin\varphi + \sin3\varphi) -\frac{\beta}{2}\sin\theta_{\rm o}(\cos\chi+\cos(\chi+2\varphi)) \right.\nonumber\\
&~~&\qquad \left.+\frac{\beta^2}{4}(\sin\varphi-\sin(2\chi+\varphi))     \right)B_\phi \nonumber \\
&~~& \quad \left. -\frac{1}{R}\sin\theta_{\rm o}\sin2\varphi \,B_z\right].
\end{eqnarray}

From $E_x$, $E_y$, we can obtain the observed field components, $E_{x,\rm obs}$, $E_{y,\rm obs}$, from equations~(\ref{ealphaobs}), (\ref{ebetaobs}). We can then compute the Stokes parameters $Q$ and $U$ via 
\begin{equation}
    Q = E_{x,\rm obs}^2-E_{y,\rm obs}^2 = (E_{x}^2-E_{y}^2)\,\delta^2\,l_{\rm p}^{1/2}, \qquad U = 2 E_{x, \rm obs} E_{y,\rm obs} = 2 E_{x} E_{y}\,\delta^2\,l_{\rm p}^{1/2}.
    \label{eq:QU_Ealphabeta}
\end{equation}
We can also calculate $|P|= E_{x,\rm obs}^2+E_{y,\rm obs}^2$, but this will simply reproduce the answer given in equation~(\ref{absP3}). We do not write down the results for $Q$ and $U$ as the expressions are large. Instead we define the complex polarization $P(\varphi)$ in the usual way (see eq.~\ref{eq:PQU}),
and expand it in a Fourier series as described in \citet{PWP_2020},
\begin{equation}
P(\varphi) \equiv Q(\varphi) + i U(\varphi)
= \frac{1}{2\pi}\sum_{m=-\infty}^{\infty}
\beta_m\,e^{im\varphi}.
\label{eq:P_QU}
\end{equation}
To zeroth and linear order there are only two non-zero coefficients, $\beta_1$ and $\beta_2$, and to quadratic order, there are five non-zero coefficients, $\beta_0 - \beta_4$. The expressions for these coefficients are given below (second-order contributions are shown inside square brackets): 
\begin{eqnarray}
\beta_0 &=& \left[-\frac{1}{4}\sin^2\theta_{\rm o} \left(B_r^2 +3B_\phi^2 -4B_z^2  -2iB_r B_\phi\right) \right] \\
&=& \left[\frac{1}{4}\sin^2\theta_{\rm o}\left(e^{2i\eta}-2\right)B_{\rm eq}^2+\sin^2\theta_{\rm o}B_z^2\right], \\
\beta_1 &=& 2\sin\theta_{\rm o} \left(-iB_r + B_\phi\right)\,B_z +\left[ \left(-\frac{i}{R} +i\beta\left(\frac{3}{2}e^{-i\chi}+e^{i\chi}\right)\right)\sin\theta_{\rm o}\,B_r^2 +\left(-\frac{3i}{R} +i\beta\left(-\frac{3}{2}e^{-i\chi}+e^{i\chi}\right)\right) \sin\theta_{\rm o}\,B_\phi^2 \right.
\nonumber \\
&~~& \quad\left. +\left(\frac{4i}{R} -2i\beta e^{i\chi} \right)\sin\theta_{\rm o}\,B_z^2
-\left( \frac{2}{R}+3\beta e^{-i\chi} \right) \sin\theta_{\rm o}\,B_rB_\phi -\frac{10}{R}\sin\theta_{\rm o}\,B_\phi B_z +\frac{10i}{R}\sin\theta_{\rm o}\,B_zB_r \right] \\
&=& -2i\sin\theta_{\rm o}e^{i\eta}B_{\rm eq}B_z + \left[\left(-\frac{i}{R}\left( 2-e^{2i\eta} \right)\sin\theta_{\rm o}+i\beta\sin\theta_{\rm o}\left(e^{i\chi}+\frac{3}{2}e^{i(2\eta-\chi)}\right) \right)B_{\rm eq}^2 \right. \nonumber \\
&~~&\qquad\qquad\qquad\qquad\left. +\left(\frac{4i}{R}-2i\beta e^{i\chi}\right)\sin\theta_{\rm o}B_z^2 +\frac{10i}{R}\sin\theta_{\rm o} e^{i\eta}B_{\rm eq}B_z \right], \\
\beta_2 &=& -\left(1-\frac{4}{R}\right)\,\left(B_r +i B_\phi\right)^2 
 -2\left(\beta e^{i\chi}-\frac{2}{R}\right)\,(B_r+iB_\phi)B_z \nonumber \\
&~~& ~+ \left[ \left(-\frac{2}{R^2}-\frac{i\beta}{R}(4\sin\chi-10i\cos\chi) +\frac{\beta^2}{2}\left(4+e^{2I\chi}\right)  \right)\,B_r^2 \right. \nonumber \\
&~~& \quad +\left( \frac{6}{R^2}  +\frac{6\beta}{R}\cos\chi +\frac{\beta^2}{2}\left(-4+e^{2i\chi}\right)\right)\,B_\phi^2  \nonumber \\
&~~& \quad+\left(-\frac{4}{R^2}+\frac{4\beta}{R}e^{i\chi}-\beta^2 e^{2i\chi} \right) B_z^2
+\left(-\frac{8i}{R^2}  +\frac{4\beta}{R}(\sin\chi-4i\cos\chi) +4i\beta^2  \right)\,B_r B_\phi \nonumber \\
&~~& \quad\left. +\left(-\frac{16i}{R^2}+\frac{8i\beta}{R}e^{i\chi} \right)\,B_\phi B_z +\left( -\frac{16}{R^2} +\frac{8\beta}{R}e^{i\chi} \right)\,B_z B_r \right] \\
&=& -\left(1-\frac{4}{R}\right)e^{2i\eta}B_{\rm eq}^2 + \left(\frac{4}{R}e^{i\eta}-2\beta e^{i(\chi+\eta)}\right)B_{\rm eq}B_z \nonumber \\
&~~& ~+\left[\left(\frac{2}{R^2}\left(1-2e^{2i\eta}\right)+\frac{\beta^2}{2}\left(e^{2i\chi}+4e^{2i\eta}\right)-\frac{\beta}{R}\left(e^{2i\eta}\left(6\cos\chi+2e^{i\chi}\right)+2e^{i\chi}\right)  \right) B_{\rm eq}^2\right. \nonumber \\
&~~& \quad \left. +\left(-\frac{4}{R^2}+\frac{4\beta}{R}e^{i\chi}-\beta^2e^{2i\chi}\right)B_z^2 + \left(-\frac{16}{R^2}e^{i\eta}+\frac{8\beta}{R}e^{i(\chi+\eta)}\right)B_{\rm eq}B_z\right], \\
\beta_3 &=& \left[ \left(\frac{i}{R}-\frac{5i\beta}{2} e^{i\chi}\right)\sin\theta_{\rm o} \left(B_r +iB_\phi \right)^2 -\frac{2i}{R}\sin\theta_{\rm o} \left(B_r +iB_\phi \right)\,B_z \right] \\
&=& \left[\left(\frac{i}{R}-\frac{5i\beta}{2}e^{i\chi}\right)\sin\theta_{\rm o}e^{2i\eta}B_{\rm eq}^2 -\frac{2i}{R}\sin\theta_{\rm o}e^{i\eta}B_{\rm eq}B_z\right], \\
\beta_4 &=& \left[-\frac{1}{4}\sin^2\theta_{\rm o}\, \left(B_r +iB_\phi\right)^2\right] \\
&=& \left[-\frac{1}{4}\sin^2\theta_{\rm o}e^{2i\eta}B_{\rm eq}^2\right].
\end{eqnarray}
For each $\beta_m$ coefficient, we give the result both in terms of $B_r$, $B_\phi$, $B_z$, and in terms of $B_{\rm eq}$, $\eta$, $B_z$.

\bibliography{ring_pol}

\end{document}